\documentclass[12pt]{article}
\usepackage{amsmath,amssymb}
\usepackage{latexsym,graphicx,color,verbatim}

\def\nn{\nonumber\\}

\def\6#1{{\underline{#1}}}
\def\m6#1{{\underline{#1}\,}}

\newdimen\Tdim
\def\ispan{{\setbox0=\hbox{i}%
\Tdim\ht0\advance\Tdim\dp0\rule[-\dp0]{0pt}{\Tdim}}}
\def\jspan{{\setbox0=\hbox{j}%
\Tdim\ht0\advance\Tdim\dp0\rule[-\dp0]{0pt}{\Tdim}}}
\def\Tspan#1{{\setbox0=\hbox{#1}%
\Tdim\ht0\advance\Tdim\dp0\advance\Tdim.55ex\rule[-\dp0]{0pt}{\Tdim}\box0}}

\def\be{\begin{eqnarray}}
\def\ben{\begin{eqnarray*}}
\def\ee{\end{eqnarray}}
\def\een{\end{eqnarray*}}
\def\Tr{{\rm Tr}}

\def\p{\partial}
\def\ra{\rightarrow}
\def\D{\mathcal{D}}
\def\O{\mathcal{O}}

\def\=:{=\hspace{-.7em}\raisebox{1.1ex}{.}\hspace{.1em}\raisebox{-0.2ex}{.} }

\newcommand{\NF}{N_{\rm F}}
\newcommand{\NC}{N_{\rm C}}
\newcommand{\hs}[1]{\hspace{#1 mm}}

\newcommand {\beq}{\begin{eqnarray}}
\newcommand {\eeq}{\end{eqnarray}}
\newcommand {\non}{\nonumber\\}

\makeatletter
\@addtoreset{equation}{section}

\renewcommand{\thefootnote}{\fnsymbol{footnote}}
\makeatother

\makeatletter
\newcommand{\thetablename}{Table}
\def\fnum@table{\thetablename\ \thetable}
\makeatother

\begin{document}
\thispagestyle{empty}
\begin{flushright}
DAMTP-2008-91, IFUP-TH/2008-31, TIT/HEP-585, \\
{\tt arXiv:} \\
October, 2008 \\
\end{flushright}
\vspace{3mm}
\begin{center}
{\LARGE \bf 
Dynamics of Strings 
between Walls
} \\ 
\vspace{3mm}

{\normalsize\bfseries
Minoru~Eto$^{a,b}$, 
Toshiaki~Fujimori$^c$, 
Takayuki Nagashima$^c$, \\
Muneto~Nitta$^d$,
Keisuke~Ohashi$^e$, 
and
 Norisuke~Sakai$^f$}
\footnotetext{
e-mail~addresses: \tt
minoru(at)df.unipi.it;
fujimori,
nagashi(at)th.phys.titech.ac.jp;\\
nitta(at)phys-h.keio.ac.jp;
K.Ohashi(at)damtp.cam.ac.uk;
sakai(at)lab.twcu.ac.jp,
}

\vskip 1.5em
$^a$ {\it INFN, Sezione di Pisa,
Largo Pontecorvo, 3, Ed. C, 56127 Pisa, Italy
}
\\
$^b$ {\it Department of Physics, University of Pisa
Largo Pontecorvo, 3,   Ed. C,  56127 Pisa, Italy
}
\\
$^c$ {\it Department of Physics, Tokyo Institute of
Technology, Tokyo 152-8551, Japan
}
\\
$^d$ 
{\it Department of Physics, Keio University, Hiyoshi, Yokohama,
Kanagawa 223-8521, Japan
}
\\
$^e$
{\it Department of Applied Mathematics and Theoretical Physics, \\
University of Cambridge, CB3 0WA, UK}
\\
$^f$
{\it Department of Mathematics, Tokyo Woman's Christian University, 
Tokyo 167-8585, Japan }

\abstract{
Configurations of 
vortex-strings stretched between or ending on domain walls 
were previously found to be 
1/4 Bogomol'nyi-Prasad-Sommerfield(BPS) states 
in ${\cal N}=2$ supersymmetric gauge theories 
in $3+1$ dimensions. 
Among zero modes of string positions, 
the center of mass of strings in each region 
between two adjacent domain walls is shown to be non-normalizable 
whereas the rests are normalizable. 
We study dynamics of vortex-strings stretched 
between separated domain walls by using two methods, 
the moduli space (geodesic) approximation of full 1/4 BPS states 
and the charged particle approximation for 
string endpoints in the wall effective action. 
In the first method 
we explicitly obtain the effective Lagrangian,  
in terms of hypergeometric functions, 
and 
find the 90 degree scattering for head-on collision.
In the second method 
the domain wall effective action is assumed to be 
$U(1)^N$ gauge theory, and 
we find a good agreement between two methods 
for 
well separated strings. 
}

\end{center}

\vfill
\newpage
\setcounter{page}{1}
\setcounter{footnote}{0}
\renewcommand{\thefootnote}{\arabic{footnote}}

\section{Introduction}\label{sec:intro}

Dirichlet(D-)branes \cite{Polchinski:1995mt}
have been necessary ingredients to study non-perturbative 
dynamics of string theory since their discovery. 
They are defined as endpoints of open strings. 
The low-energy effective theory on a D-brane is 
described by the Dirac-Born-Infeld(DBI) action.
String ending on a D-brane can be realized as solitons 
or solutions with a source term 
in the DBI action. 
These solitons are called BIons \cite{Callan:1997kz}. 
Usually these solitons are constructed as 
deformations of the D-brane surface 
such as a spike. 
It is not easy to construct a string stretched between 
D-branes as a soliton of the DBI theory. 
One reason of difficulty is that 
no DBI action for multiple D-branes 
is known so far.

\begin{figure}[htb]
\begin{center}
\includegraphics[width=120mm]{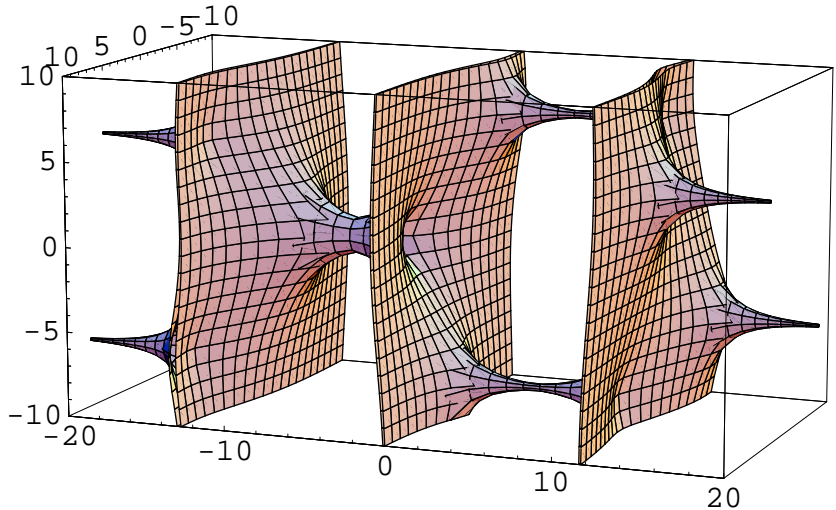}
\caption{
An example of {\it exact} solution of the D-brane soliton. 
A same energy surface is plotted. 
A figure taken from \cite{Isozumi:2004vg}.
}
\label{fig:example}
\end{center}
\end{figure}

Solitons resembling with 
strings ending on D-branes 
have been found in a field theory framework 
\cite{Gauntlett:2000de}. 
They have given an exact solution of 
vortex-strings ending on a domain wall 
in a ${\bf C}P^1$ nonlinear sigma model. 
This theory or the ${\bf C}P^N$ extention 
was known to admit single or multiple 
domain wall solutions \cite{Abraham:1992vb,U(1)walls,Tong:2002hi}.
Assuming the DBI action 
on 
the effective action on a single domain wall, 
they have further shown that that soliton  
can be identified with a BIon, 
a soliton on a D2-brane \cite{Gauntlett:2000de}, 
and so have called it a ``D-brane soliton".
Later it has been extended to a solution 
in $U(1)$ gauge theory coupled to 
two charged Higgs fields \cite{Shifman:2002jm}. 
Exact solutions of multiple domain walls have been 
constructed in $U(N)$ gauge theory in strong coupling limit,
by introducing the ``moduli matrix" \cite{Isozumi:2004jc}.
By extending this method, 
the most general solutions of 
D-brane solitons have been constructed \cite{Isozumi:2004vg} 
which offers {\it exact} ({\it analytic}) solutions of 
multiple domain walls 
with arbitrary number of vortex-strings 
stretched between (ending on) domain walls, 
see Fig.~\ref{fig:example}.
Some aspects of these solitons have been studied. 
We have found 
an object with a monopole charge which contributes 
negatively to the total energy of composite solitons 
in $U(1)$ gauge theory \cite{Isozumi:2004vg}. 
It has later been called a ``boojum" and studied 
extensively \cite{Sakai:2005sp}.
In the case of $U(N)$ gauge theory 
a monopole confined by vortices is also admitted 
\cite{Tong:2003pz,Hanany:2004ea,Shifman:2004dr,Eto:2004rz}, 
which can be understood as a kink in non-Abelian vortices 
found earlier \cite{Hanany:2003hp}. 
The moduli space has been studied \cite{Eto:2006pg} 
for composite solitons consisting of 
domain walls, vortex-strings and monopoles. 
See review papers \cite{Tong:2005un,Eto:2006pg,Konishi:2007dn,Shifman:2007ce} 
for recent developments of BPS composite solitons. 
It has been proposed that 
domain walls actually can be regarded as 
D-branes after taking into account quantum corrections by 
loop effect of vortex-endpoints 
\cite{Tong:2005nf} (see also \cite{Shifman:2006ea}). 
It has been proposed that this provides some field theoretical model 
of the open-closed string duality.

In this paper we study classical dynamics of D-brane solitons 
in $3+1$ dimensions
by using two methods and compare their results. 
One is to use the moduli space (geodesic) approximation 
found by Manton in 
studying  
the monopole dynamics 
\cite{Manton:1981mp,Manton:2004tk}.
In this approximation, geodesics on the moduli space 
of solitons correspond to dynamics or scattering of solitons.  
So far the moduli approximation has been used to describe 
classical scattering of particle-like solitons 
such as monopoles in three space-dimensions,
vortices in two space-dimensions \cite{Samols:1991ne,Eto:2005yh}, 
and kinks in one space-dimension 
\cite{Shifman:1997hg,Tong:2002hi},
which are 1/2 {Bogomol'nyi-Prasad-Sommerfield (BPS)} states. 
As the first example of composite solitons, 
it has been recently applied to dynamics of domain wall networks 
\cite{Eto:2006bb,Eto:2007uc}, 
which are 1/4 BPS states \cite{Eto:2005cp}.  
Here we apply it to dynamics of D-brane solitons, 
vortex-strings stretched between domain walls. 

The other is to use a charged particle approximation of solitons 
and a domain wall effective action. 
It was suggested by Manton that 
monopoles can be regarded as particles with magnetic 
and scalar charges \cite{Manton:1985hs,Manton:2004tk}. 
It was used to derive an asymptotic metric on 
the moduli space of well-separated BPS monopoles
\cite{Gibbons:1995yw}. 
On the other hand, 
the effective action on a single domain wall is a free Lagrangian 
of the $U(1)$ Nambu-Goldstone zero mode 
and the translational zero mode. 
This $U(1)$ zero mode can be dualized to a $U(1)$ gauge field 
in the 2+1 dimensional 
world volume of the wall \cite{Gauntlett:2000de}, 
then the effective Lagrangian becomes 
a dual $U(1)$ gauge theory plus one neutral scalar field. 
It has been found by Shifman and Yung \cite{Shifman:2002jm} 
that string endpoints can be regarded as electrically charged particles 
in a {dual} $U(1)$ gauge theory 
of the domain wall effective action.

We generalize this discussion to $N$ parallel domain walls. 
As effective theory of 
well separated $N$ domain walls, 
we propose $U(1)^N$ gauge theory and 
$N$ scalar fields corresponding to wall positions. 
Then we use the particle approximation for 
endpoints of strings on the domain walls. 
By comparing the moduli metric derived by the moduli approximation 
for the full 1/4 BPS configurations, 
we find a good agreement in  
the asymptotic 
metric. 

This is instructive for clarifying similarity or difference 
between D-branes and field theory solitons. 
The BPS monopoles can be realized as a D1-D3 bound state, 
D1-branes stretched between separated D3-branes.  
The endpoints of the D1-branes at the D3-branes 
can be regarded as BPS monopoles in 
the D3-brane effective action \cite{Diaconescu:1996rk}. 
The monopole (or D1-brane) dynamics by the particle approximation 
in the D3-brane effective theory 
is parallel to our second derivation 
of the vortex-string dynamics as the charged particles 
in the domain wall effective theory. 
The only difference is the 
number of codimensions of string-endpoints, 
which is three for D1-D3 and two for vortex-strings on walls.
In fact our asymptotic metric is similar to that of 
monopoles \cite{Manton:1985hs,Gibbons:1995yw} by replacing 
$1/r$ by $\log r$, 
where $r$ is the 
distance between solitons. 
However there exists a crucial difference 
when $N$ host 
branes (D3-branes or walls) coincide. 
The effective theory of D3-branes is in fact 
$U(N)$ gauge theory with 
several adjoint Higgs fields, 
reducing to $U(1)^N$ gauge group 
only when eigenvalues of the adjoint Higgs field (positions of $N$ 
D3-branes) are 
different from each other. 
In contrast to this, our effective theory on domain walls 
does not become $U(N)$ gauge theory 
even when domain walls coincide.\footnote{
If we consider domain walls with degenerate masses for 
Higgs scalar fields, 
we have $U(N)$ Nambu-Goldstone modes 
for coincident domain walls
\cite{Shifman:2003uh,Eto:2005cc,Eto:2008dm}. 
Taking a duality has been achieved only for 
3+1 dimensional wall world-volume, 
where dual fields are non-Abelian 2-form fields 
rather than Yang-Mills fields \cite{Eto:2008dm}.
} 

This paper is organized as follows. 
In Sec.~\ref{sec:review_eq} 
we briefly explain the 1/4 BPS equations
in the $U(\NC)$ gauge theory with $\NF$ hypermultiplets.
In Sec.~\ref{sec:review_wv} we review 1/4 BPS wall-vortex systems
in the $U(1)$ gauge theory.
In Sec.~\ref{sec:lagrangian}
we first construct a general form of 
the effective Lagrangian of 1/4 BPS solitons
in the $U(\NC)$ gauge theory,
by applying the method to obtain 
a manifestly supersymmetric effective action on 
BPS solitons \cite{Eto:2006uw}. 
Next we use it to examine 
normalizability of zero modes of 1/4 BPS wall-vortex systems
in the $U(1)$ gauge theory.
Here we assume that each vacuum region between two adjacent domain walls
has the same number of vortices.
It is easy to see that zero modes related to domain walls or
 vortices with infinite lengths are 
non-normalizable. 
We find the center of mass of vortex-strings in each vacuum 
region is also non-normalizable, and
the other zero modes are normalizable.
In Sec.~\ref{sec:dynamics} we give examples of 
$(1,1,1)$, $(2,2,2)$, $(0,2,0)$ and $(n,0,n)$
in the $U(1)$ gauge theory with three flavors 
admitting two domain walls. 
Here $(n_1,n_2,n_3)$ represent configurations in which 
$n_1$ and $n_3$ strings end on the left (right) domain wall 
from outside and $n_2$ strings are stretched between the domain walls,
{see Fig.~\ref{fig:dbs} below}.
We obtain the effective Lagrangian explicitly in the strong coupling limit
as a nonlinear sigma model on the moduli space.
In the case of $(1,1,1)$, we find the position of the vortex living
in the middle vacuum is non-normalizable. 
We give the physical explanation of the 
divergence in the effective Lagrangian.
In the cases of $(2,2,2)$ and $(0,2,0)$,  
the relative positions of vortices in the middle region 
gives normalizable modes and, 
we find the 90 degree scattering for head-on collision of 
those vortices.
%
Metrics of both configurations can be expressed in terms of hypergeometric functions. 
The $(n,0,n)$ example is 
a bit strange in the sense that no vortices can move.
Since the region of the middle vacuum is finite in this case, there exists
a normalizable moduli parameter for the size of the middle region.
We obtain the effective Lagrangian for the modulus.
In Sec.~\ref{sec:particle}
we obtain vortex-string dynamics from a dual effective theory on 
domain walls.
A dual effective theory on $N$ well-separated domain walls 
is $U(1)^{N}$ gauge theory with $N$ real scalar fields 
parameterizing the wall positions, 
and endpoints of the vortex-strings can be viewed as
particles with scalar charges and electric charges.
We obtain a general effective Lagrangian which describes dynamics of 
charged particles.
We find a good agreement 
to the results obtained by the moduli space approximation 
of full 1/4 BPS configurations. 
Sec.~\ref{sec:conclusion} is devoted to conclusion and discussion.
In Appendix.\,\ref{appendix:A}, we evaluate the K\"ahler metrics for $(2,2,2)$ and $(0,2,0)$ configurations. In Appendix.\,\ref{appendix:B}, the asymptotic K\"ahler metrics are examined. In Appendix.\,\ref{appendix:C}, we show that the asymptotic metric obtained in Sec.\,\ref{sec:particle} is K\"ahler by writing down the K\"ahler potential explicitly. In Appendix.\,\ref{appendix:D}, we discuss the dual effective theory on multiple domain walls.
\newpage

\section{Composite solitons of walls, vortices, and monopoles}\label{sec:review}
\subsection{BPS equations and their solutions}\label{sec:review_eq}
Let us here briefly present our model admitting 
the 1/4 BPS composite solitons 
of domain walls, vortices, and monopoles 
(see \cite{Eto:2006pg} for a review). 
Our model is 3+1 dimensional $\mathcal{N}=2$ 
supersymmetric $U(\NC)$ gauge theory with 
$\NF\,(>\NC)$ massive hypermultiplets 
in the fundamental representation. 
The bosonic components 
in the vector multiplet are 
gauge fields $W_M~(M=0,1,2,3)$, 
the two real adjoint scalar fields 
$\Sigma_\alpha~(\alpha=1,2)$, 
and those in the hypermultiplet are 
the $SU(2)_R$ doublets of 
the complex scalar fields $H^i~(i=1,2)$, 
which we express as $\NC \times \NF$ matrices. 
The bosonic part of the Lagrangian is given by 
\beq
\mathcal{L} \hs{-2} &=& \hs{-2} \Tr \left[ - \frac{1}{2g^2} F_{MN}F^{MN} 
+ \frac{1}{g^2} \mathcal D_M \Sigma_\alpha \mathcal D^M \Sigma^\alpha 
+ \mathcal D_M H^i (\mathcal D^M H^i)^\dag \right] - V, \label{eq:lag}\\
V \hs{-2} &=& \hs{-2} \Tr \left[ \frac{1}{g^2} \sum_{a=1}^3 (Y^a)^2 
+ (H^iM-\Sigma_1 H^i)(H^iM-\Sigma_1 H^i)^\dag 
+ \Sigma_2 H^i(\Sigma_2 H^i)^\dag 
- \frac{1}{g^2} [ \Sigma_1, \Sigma_2]^2 \right] 
\label{eq:pot}
\eeq
where $g$ is a $U(N)$ gauge coupling constant, 
and we have defined 
$Y^a \equiv \frac{g^2}{2} \left(c^a \mathbf 1_{\NC} - {(\sigma^a)^j}_i H^i(H^j)^\dagger \right)$ 
with $c^a$ an $SU(2)_R$ triplet of 
the Fayet-Iliopoulos (FI) parameters. 
In the following, we choose the FI parameters as $c^a=(0,0,c>0)$ 
by using $SU(2)_R$ rotation without loss of generality. 
We use the space-time metric $\eta_{MN}=\text{diag\,}(+1,-1,-1,-1)$ 
and the covariant derivatives are defined as 
$\mathcal D_M \Sigma_\alpha =\partial_M \Sigma_\alpha + i[W_M,\Sigma_\alpha],~\mathcal D_M H^i=(\partial_M+iW_M)H^i$, and the
field strength is defined as 
$F_{MN}=-i[\mathcal D_M,\mathcal D_N]=\partial_M W_N-\partial_N W_M +i[W_M,W_N]$. 
$M$ is a real $\NF \times \NF$ diagonal mass matrix, 
$M=\text{diag\,} (m_1,m_2,\cdots,m_{\NF})$. 
In this paper we consider non-degenerate 
real masses, chosen as $m_1>m_2>\cdots>m_{\NF}$.

If we turn off all the mass parameters, 
the moduli space of vacua 
is the cotangent bundle over 
the complex Grassmannian $T^\ast Gr_{\NF,\NC}$ 
\cite{Lindstrom:1983rt}. 
Once the mass parameters $m_A~(A=1,\cdots,\NF)$ 
are turned on and chosen to be fully non-degenerate 
($m_A \neq m_B$ for $A\neq B$), 
the almost all points of the vacuum manifold 
are lifted and only $\NF!/\left[\NC!(\NF-\NC)!\right]$ discrete points 
on the base manifold $Gr_{\NF,\NC}$ 
are left to be the supersymmetric vacua \cite{Arai:2003tc}. 
Each vacuum is characterized by a set of $\NC$ different indices 
$\left< A_1\cdots A_{\NC}\right> $ such that 
$1\leq A_1 < \cdots < A_{\NC} \leq \NF$. 
In these discrete vacua, 
the vacuum expectation values are determined as 
\beq
&\left<H^{1rA}\right> = \sqrt{c} \, {\delta^{A_r}}_A,\hs{5} \left<H^{2rA}\right> = 0,& \non
& \left<\Sigma_1 \right> = {\rm diag} \, (m_{A_1},\cdots,m_{A_{\NC}}), \hs{5} \left<\Sigma_2\right> = 0, &
\label{vacuum}
\eeq
where 
the color index $r$ runs from 1 to $\NC$, and 
the flavor index $A$ runs from 1 to $\NF$. 

The 1/4 BPS equations for composite solitons of 
walls, vortices, and monopoles 
can be obtained by the usual 
Bogomol'ny completion 
of the energy density \cite{Isozumi:2004vg,Tong:2005un,Eto:2006pg,Shifman:2007ce} as 
\begin{eqnarray}
 {\cal D}_2 \Sigma - F_{31} =0,
\quad 
 {\cal D}_1 \Sigma - F_{23} =0, 
\label{eq:4eq:mvw-1}\\
 {\cal D}_3 \Sigma - F_{12} 
- \frac{g^2}{2}\left(c{\bf 1}_{N_{\rm C}} - HH^{\dagger}\right)  =0 ,
\label{eq:4eq:mvw-2}\\
 {\cal D}_1 H + i  {\cal D}_2 H =0, 
\quad 
 {\cal D}_3 H +  \Sigma H-H M=0,
\label{eq:4eq:mvw-3}
\end{eqnarray}
where $H \equiv H^1,~\Sigma \equiv \Sigma_1$ 
and $H^2,\,\Sigma_2$ have been suppressed since 
they do 
not contribute to soliton 
solutions for $c>0$. 
These equations describe composite solitons 
consisting of monopoles, 
vortices with codimensions in the $z \equiv x^1+i x^2$ plane, 
and 
walls perpendicular to the $x_3$ direction.\footnote{
When there exists a flux on 
a domain wall worldvolume, the domain wall 
is tilted \cite{Isozumi:2004vg} 
resembling with non-commutative monopoles. 
}
The Bogomol'ny bound for the energy density $\cal E$ is given as 
\beq
{\cal E} \geq t_{\rm w}+t_{\rm v}+t_{\rm m}+\p_m J_m. 
\label{eq:bound}
\eeq 
Here $t_{\rm w},\,t_{\rm v},\,t_{\rm m}$ are
the energy densities for walls, vortices and monopoles, respectively, 
given by 
\begin{eqnarray}
t_{\rm w}= c \, \partial _3 {\rm Tr}\Sigma,\quad 
t_{\rm v}= -c\,{\rm Tr} B_3,\quad
t_{\rm m}= \frac{2}{g^2} \partial_m {\rm Tr}(\Sigma B_m), 
\label{energy-den}
\end{eqnarray}
where $B_m = \frac{1}{2} \epsilon_{mnl} F_{nl}~(m,n,l=1,2,3)$. 
The monopole charge $t_{\rm m}$ can be 
either positive or negative, 
corresponding to monopoles and boojums, respectively. 
The last term in Eq.\,(\ref{eq:bound}) containing $J_m~(m=1,2,3)$, 
which are defined by 
\begin{eqnarray}
J_1&\equiv&{\rm Re}\left(-i{\rm Tr}(H^\dagger {\cal D}_2H)\right),\hs{5}
J_2\equiv{\rm Re}\left(i{\rm Tr}(H^\dagger {\cal D}_1H)\right),\nonumber\\
J_3&\equiv&-{\rm Tr}(H^\dagger (\Sigma -M)H), 
\end{eqnarray}
is a correction term which does not 
contribute to the total energy. 

Since Eq.(\ref{eq:4eq:mvw-1}), 
which are equivalent to $\left[\D_1+i\D_2,\D_3+\Sigma\right]=0$, 
provides the integrability condition 
for the operators $\D_1+i\D_2$ and $\D_3+\Sigma$, 
we can introduce an $N_{\rm C}\times N_{\rm C}$ 
invertible complex matrix function 
$S(z, \bar z, x_3)\in GL(N_{\rm C},\mathbf{C})$ 
defined by \cite{Isozumi:2004vg}
\begin{eqnarray}
\Sigma + iW_3 &\equiv& S^{-1}\partial_3 S,\hs{15}
\left(({\cal D}_3+\Sigma)S^{-1}=0\right),
\label{def-S-3} \\
W_1+iW_2&\equiv &-2iS^{-1}\bar \partial S, \hs{10}
\left(({\cal D}_1+i{\cal D}_2)S^{-1}=0\right) , 
\label{def-S-bar}
\end{eqnarray}
with $\bar \partial\equiv \partial/\partial \bar z$. 
With 
the form of Eq.(\ref{def-S-3}) and Eq.(\ref{def-S-bar}), 
Eq.(\ref{eq:4eq:mvw-1}) is satisfied, and 
Eq.(\ref{eq:4eq:mvw-3}) is solved by 
\begin{eqnarray}
 H = S^{-1}(z,\bar z, x_3) H_0(z) e^{M x_3}.
\label{sol-H}
\end{eqnarray}
Here $H_0(z)$ is an $N_{\rm C} \times N_{\rm F}$ 
matrix whose elements are arbitrary holomorphic functions of $z$. 
We call it the ``moduli matrix" 
since it contains all the moduli parameters of solutions 
as we will see shortly. 
Let us define an $N_{\rm C}\times N_{\rm C}$ 
Hermitian matrix 
\beq
 \Omega \equiv SS^\dagger ,
\eeq 
invariant under the $U(N_{\rm C})$ gauge transformations. 
The remaining BPS equation (\ref{eq:4eq:mvw-2}) 
can be rewritten in terms of $\Omega$ as \cite{Isozumi:2004vg}
\beq
&\displaystyle \frac{1}{g^2c} \left[ 4 \partial_z (\Omega^{-1} \bar{\partial}_z \Omega ) + {\partial}_3 (\Omega^{-1} \partial_3 \Omega ) \right]
= {\bf 1}_{N_{\rm C}}- \Omega^{-1} \Omega_0,& \label{eq:master_wvm} \\
&\displaystyle \Omega_0 \equiv \frac{1}{c}H_0 e^{2M x_3}H_0^\dagger.&
\eeq
This equation is called the master equation for 
the wall-vortex-monopole system. 
This reduces to the master equation 
for the 1/2 BPS domain walls if we omit 
the $z$-dependence $(\p_z = \p_{\bar z} = 0)$ 
while that for the 1/2 BPS vortices 
if we omit the $x_3$-dependence $(\p_3=0)$ and set $M=0$. 
It determines $S$ for a given moduli matrix $H_0$ 
up to the gauge symmetry $S \rightarrow SU^\dagger,~U \in U(\NC)$ 
and then the physical fields can be obtained 
through Eqs.~(\ref{def-S-3}), (\ref{def-S-bar}) and (\ref{sol-H}). 
The master equation Eq.\,(\ref{eq:master_wvm}) 
has a symmetry which we call ``$V$-transformations" 
\begin{align}
H_0(z) \rightarrow V(z)H_0(z),~~~~S(z, \bar z, x_3) \rightarrow V(z)S(z, \bar z, x_3).
\end{align}
where $V(z)\in GL(N_{\rm C}, {\bf C})$ has components 
holomorphic with respect to $z$. 
The moduli matrices related by 
this $V$-transformation are physically equivalent 
$H_0(z) \sim V(z) H_0(z)$ 
since they do not change the physical fields. 
Therefore the total moduli space of this system, 
defined by all topological sectors patched together, 
is given by a set of the whole 
holomorphic 
matrix $H_0(z)$ divided by 
the equivalence relation $H_0(z) \sim V(z) H_0(z)$. 
Therefore the parameters contained in the moduli matrix $H_0(z)$ 
after fixing the redundancy of the $V$-transformation 
can be interpreted as the moduli parameters, 
namely the coordinates of the moduli space of the BPS configurations. 

\subsection{Composite solitons of vortices and domain walls}\label{sec:review_wv}

The moduli matrix offers a powerful tool to study 
the moduli space of the 1/4 BPS composite solitons, 
because it exhausts all possible BPS configurations. 
%
%
In this paper we study the dynamics in 
the Abelian-Higgs model with $\NF\,(\ge2)$ flavors, 
in which the moduli matrix is an $\NF$-vector. 
%
%
To this end we summarize here how the moduli matrix 
represents 1) the SUSY vacua, 2) 1/2 BPS domain walls, 
3) 1/2 BPS vortices and 4) 1/4 BPS composite states. 

1) $\NF$ discrete SUSY vacua.~
In the $A$-th vacuum $\left<A\right>$ ($A=1,2,\cdots,\NF$), 
only $A$-th element is nonzero with the rests being zero, 
\beq
 \left<A\right>: ~ H_0 ~=~ (0,\cdots,0,1,0,\cdots,0).
\eeq

2) 
$\NF-1$ multiple 1/2 BPS domain walls. 
When the $A$-th and the $B$-th elements $(A>B)$
and the elements between them are 
nonzero constants in $H_0$, 
it represents $A-B$ multiple domain walls 
interpolating between two vacua 
$\left< A\right>$ and $\left< B\right>$. 
The most general configurations are obtained when 
all the elements are nonzero constants: 
\beq
 H_0 = (h_1,h_2,\cdots,h_{\NF}),\quad h_A \in {\bf C}.
 \label{eq:minimal_wall}
\eeq
{
If some of $\{h_A\}$ vanish, the corresponding vacua disappear.
Namely, the domain walls adjacent to the vacua collapse.
}
In order to estimate the position of the 
domain wall interpolating between 
$\left< A\right>$ and $\left< A+1\right>$, 
let us define the weight of the vacuum $\left<A\right>$ by 
\beq
 \exp({\cal W}^{\left<A\right>}) \equiv h_A e^{m_Ax_3}.
\eeq
In the region where the only one of the 
weights 
$\exp({\cal W}^{\left<A\right>})$ is large, 
the solution of the master equation 
Eq.\,(\ref{eq:master_wvm}) is approximately 
given by $\Omega \approx \exp({\cal W}^{\left<A\right>})$. 
In such regions, the energy density of the domain wall $t_{\rm w}$ 
vanishes since it is given by 
$t_{\rm w} = \frac{c}{2} \p_3^2 \log \Omega$. 
The $A$-th domain wall exists where 
the weights $\exp({\cal W}^{\left<A\right>})$ and 
$\exp({\cal W}^{\left<A+1\right>})$ 
of the two vacua $\left< A\right>$ and $\left< A+1\right>$ 
are balanced. 
Its position $x_3 = X^A$ can be estimated by 
\beq
 \Delta m_A X^{A} + i \sigma^A \simeq \log \left(
 \frac{h_{A+1}}{h_A}\right),
 \label{eq:weight_minomal}
\eeq
with $\Delta m_A \equiv m_A-m_{A+1} > 0$. 
Here the imaginary part $\sigma^A$ represents 
an associated phase modulus of the wall. 
The tension of this domain wall is $c \Delta m_A$. 

3) 1/2 BPS vortices in the vacuum $\left<A\right>$. 
When only the $A$-th element $h_A(z)$ of $H_0$ is 
a polynomial function of $z$ with the rests zero, 
\beq
H_0 &=& (0,\cdots,0,h_A(z),0,\cdots,0),
\label{eq:minimal_vortex}\\
h_A(z)&=&v_A(z-z_{\langle A \rangle 1})(z-z_{\langle A \rangle 2})
\cdots(z-z_{\langle A \rangle k_A}),  
  \label{eq:h_A}
\eeq
it represents multiple vortices 
in 
the vacuum $\left<A\right>$ 
extending to 
infinity ($x_3\rightarrow \pm \infty$). 
The degree of the polynomials $n_A = \deg [h_A(z)]$ 
is identical to the number of 
the vortices in the vacuum $\left<A\right>$, 
and the 
zeros 
$z_{\langle A \rangle 1}, \cdots, z_{\langle A \rangle k_A}$ 
of 
$h_A(z)$ represent the vortex positions. 
We call the infinitely long straight vortices, 
generated by the above moduli matrix,  
the ANO vortices. 
The tension of each vortex is $2\pi c$, and 
its transverse size is of the order $1/(g\sqrt c)$. 
The ANO vortex becomes singular in the
strong gauge coupling limit $g\to\infty$.

\begin{figure}[htb]
\begin{center}
\includegraphics[width=160mm]{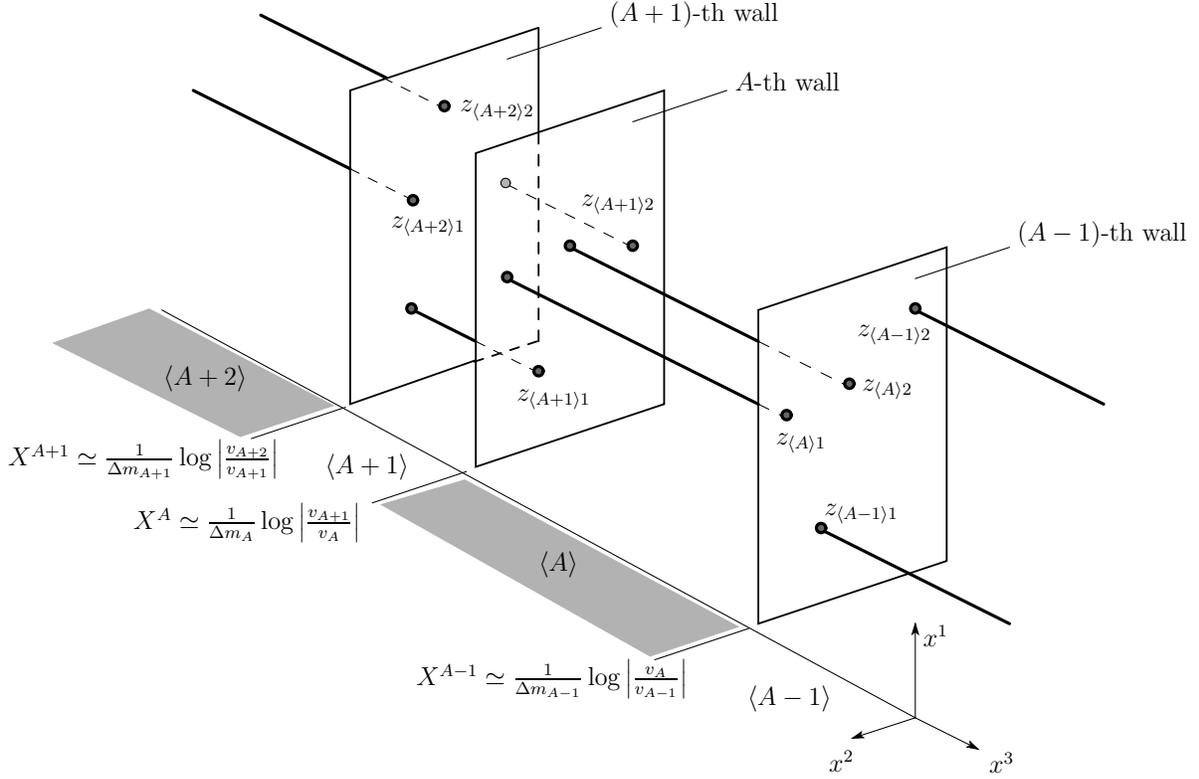}
\caption{
An example of the D-brane soliton 
$(\cdots,n_{A-1},n_A,n_{A+1},n_{A+2},\cdots)=(\cdots,2,2,2,2,\cdots)$ in the $U(1)$ gauge theory.
}
\label{fig:dbs}
\end{center}
\end{figure}

4) 1/4 BPS states {(D-brane solitons)}.~
The most general composite states of vortex-strings ending on 
(or stretched between) domain walls are given by the moduli matrix 
\beq
H_0 &=& (h_1(z), h_2(z),\cdots,h_{\NF}(z)),  
\label{eq:minimal_wall-vortex}
\eeq
where the $A$-th 
element $h_A(z)$ 
is 
of the form of Eq.\,(\ref{eq:h_A}) 
with the degree $n_A$. 
Here $n_A$ vortices exist in the vacuum $\left<A\right>$ 
and are 
suspended between $(A-1)$-th and the $A$-th domain walls for 
$A \neq 1,N_{\rm F}$, 
or ending on the first or the ($N_{\rm F}-1$)-th domain wall 
for $A = 1,N_{\rm F}$.  
We denote such D-brane soliton by $(n_1,n_2,\cdots,n_{\NF})$, see Fig.~\ref{fig:dbs}.

\begin{figure}[htb]
\begin{center}
\includegraphics[width=180mm]{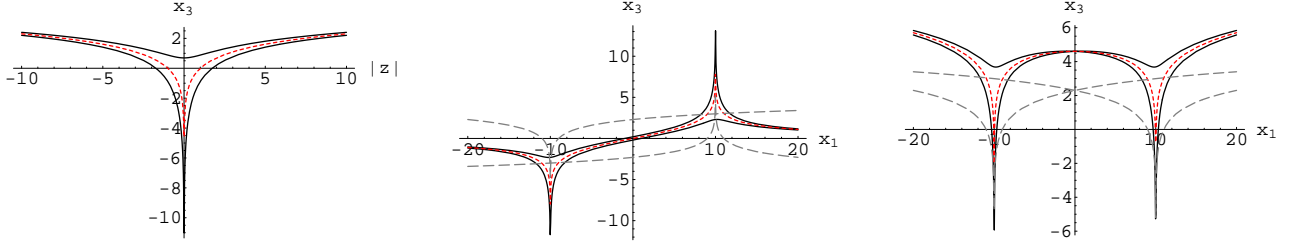}
\caption{The \{left, middle, right\} panel 
shows 
\{(0,1), (1,1), (0,2)\} configuration, respectively.
The solid line is  
the 
exact solution for the equal energy density contour of ${\cal E}=1/2$
while the broken line is 
an approximate curve given by Eq.\,(\ref{eq:log-bend}).
Gray long-dashed lines correspond to the configurations that one of two vortices is
removed away to infinity.
The parameters are chosen to be $c=1$ and $M={\rm diag}(1/2,-1/2)$.}
\label{fig:compare}
\end{center}
\end{figure}

Let us define 
a $z$-dependent generalization of the weight of 
the vacuum $\langle A \rangle$ by 
\beq
\exp \left({\cal W}^{\langle A \rangle}(z)\right) \equiv h_A(z)e^{m_Ax_3}.
\label{eq:weight}
\eeq
Domain walls are curved 
and their positions in $x_3$-direction depend on $z$ in general. 
The $z$-dependent 
position $X^A(z, \bar z)$ of 
the $A$-th domain wall and its associated ($z$-dependent) phase 
$\sigma(z, \bar z)$ can 
be estimated by equating two weights (\ref{eq:weight}), 
to yield
\beq
 \Delta m_A X^A(z, \bar z) + i \sigma^A(z,\bar z) \simeq \log \left(
 \frac{h_{A+1}(z)}{h_A(z)}\right).
\label{eq:log-bend}
\eeq
One can quickly see 
an approximate configurations from this rough estimation.
Exact solutions and the approximations (\ref{eq:log-bend}) 
are compared 
in Fig.~\ref{fig:compare}.
If a vortex ends on a domain wall,
it pulls the domain wall towards its direction to give 
the logarithmic bending of the domain wall (\ref{eq:log-bend}). 
When the same number of vortices end on the domain wall from  both sides, 
that is, $n_A=n_{A+1}$, then the domain wall is asymptotically flat 
\beq
\Delta m_A X^A(z, \bar z) + i\sigma^A
\to \log \left( \frac{v_{A+1}}{v_A} \right),
\qquad {\rm as}\quad |z| \to \infty.
\eeq
The correction terms of order $\log z$ correspond to deformation by
the vortices end on the domain walls. 
We can read the deformation near the $i$-th vortex at 
$z=z_{\left<A\right>i}$ in the vacuum $\left<A\right>$, 
which ends on the $A$-th domain wall from the right 
\beq
\!\!\! 
\Delta m_A X^A(z_{\left<A\right>i}, \bar z_{\left<A\right>i}) 
\simeq 
- \log \epsilon + \log \left| \frac{v_{A+1}}{v_A} \right|
+ \sum_j \log \left|z_{\left<A\right>i} - z_{\left<A+1\right>j}\right|
- \sum_{j\neq i} \log \left|z_{\left<A\right>i} - z_{\left<A\right>j}\right|.
\label{eq:vor_length}
\eeq
Here the first term with $0< \epsilon \ll 1$ (UV cut off) comes from the vortex at $z = z_{\left<A\right>i}$ and
the second term represents the host $A$-th domain wall. The third and the fourth terms correspond to the
deformation by the rest of vortices.
The deformation by the rest of vortices ending from the same side make the vortex longer while 
that by the other vortices ending from the opposite side shorten the vortex, 
see Fig.~\ref{fig:compare}.

Furthermore, one can estimate transverse size of vortices as follows.
If we look at region sufficiently away from domain walls, we can 
ignore $x_3$ dependence of the configurations. 
In such a region one can take a slice with fixed $x_3$ and 
can regard the 
configuration as semilocal vortices in 2+1 dimensions. The configuration is
determined by $\Omega_0$ typically taking the form of
\beq
\Omega_0 = |z-z_0(x_3)|^2 + |a(x_3)|^2,
\eeq 
up to an overall factor independent of $z$. 
Here $z_0$ stands for the position and $|a|$ for 
the transverse size of the semilocal vortex. 
Let us show two concrete examples of the D-brane soliton 
$(n_1, \cdots, n_{N_{\rm F}})$. 
\begin{itemize}
\item 
$(n_1, n_2)=(0,1)$ 
case:
For instance, we consider the mass matrix $M={\rm diag.}(m/2, -m/2)$ 
with the moduli matrix $H_0=(1, z-z_0)$ and obtain 
\beq
\Omega_0 = e^{m x_3} + |z-z_0|^2 e^{-m x_3} 
= e^{- m x_3} \left( |z-z_0|^2 + e^{2 m x_3} \right).
\eeq
The ($z$-dependent) transverse size can be read as 
$|a(x_3)| = e^{ m x_3}$, 
see 
Fig.~\ref{fig:compare} (left-most).
\item 
$(n_1, n_2, n_3)=(1,1,1)$ case: 
For instance, we consider the mass matrix $M={\rm diag.}(m/2, 0, -m/2)$ 
with the moduli matrix $H_0=(z-z_1, e^{ml/4}(z+z_1), z-z_1)$ 
for 
the coincident outer vortices 
\beq
\Omega_0 &=& |z-z_1|^2 e^{mx_3} + e^{ml/2}|z+z_1|^2 
+ |z-z_1|^2 e^{-mx_3} \nonumber\\
&=&  2 |z-z_1|^2 \cosh mx_3 + e^{ml/2} |z+z_1|^2 \nonumber\\
&=& (A+B) \left[
\left| z - \frac{A-B}{A+B} z_1 \right|^2 
+ \left(1 - \left(\frac{A-B}{A+B}\right)^2\right)|z_1|^2
\right],
\eeq
with $A\equiv 2 \cosh mx_3$ and $B\equiv e^{ml/2}$.
The physical meaning of $l$ is the distance 
between two domain walls.
The ($z$-dependent) position of the vortex is given by 
$z_0(x_3) = \frac{A-B}{A+B} z_1$ and their sizes are by
$|a(x_3)| = |z_1|\sqrt{1 - \left(\frac{A-B}{A+B}\right)^2}$.
Therefore, one can see that $z_0(x_3 \to \pm \infty) \to z_1$
and the position of the middle vortex is $z_0(x_3=0) = -z_1$ if $l \gg 1$.
The size of the outer vortices reduces 
$2|z_1|\sqrt{\frac{B}{A}} \sim 2|z_1|e^{{m \over 2}({l \over 2}- |x_3|)} 
\to 0$
as $|x_3| \to \infty$. On the other hand, if the separation of 
the two walls are sufficiently large 
so that $B \gg A$, the size is estimated by
$2|z_1|\sqrt{\frac{A}{B}} \sim 2|z_1|e^{-{m \over 2}({l \over 2}- |x_3|)}$.
Thus the center of the middle vortex ($x_3=0$) has the smallest size
$|a(x_3=0)| = 2 |z_1| e^{-ml/2}$. 
It 
is exponentially small with respect to the wall distance $l$,
but still finite.
However, the size becomes zero when all the vortices are 
coincident ($z_1=0$).
\end{itemize}
The vortices ending on the domain walls are not the usual ANO vortices. 
They are also deformed by domain walls and their transverse 
sizes are not constant along $x_3$ any more. 
The sizes depend on the positions the vortices. 
The ANO vortices appear 
{at $z=z_1,z_2,\cdots,z_{k'}$}
only when 
all elements of 
the moduli matrix have common zeros as 
\beq
H_0 (z) = (z-z_1)(z-z_2) \cdots (z-z_{k'}) \times H_0'(z), 
\label{eq:ANO}
\eeq
without poles in $H_0'(z)$. 
We call such moduli matrix as 
factorizable. 
In the strong gauge coupling limit, 
these ANO vortices become singular. 
On the other hand, 
the other general vortices ending on (stretched between) domain walls 
remain regular with finite transverse sizes in 
the strong coupling 
limit.



\section{Effective Lagrangian of 1/4 BPS wall-vortex systems}\label{sec:lagrangian}
\subsection{General form of the effective Lagrangian}
Now let us construct the effective Lagrangian of the
full 1/4 BPS
composite solitons.\footnote{
Our main interest in this paper is 
dynamics of vortices between domain walls 
in the Abelian gauge theory. 
However, the general formula obtained in this section 
can be applied to other 
composite solitons in non-Abelian gauge theory.} 
Zero modes on the background BPS solutions 
will play a main role in the effective theory 
while all the massive modes will be ignored in the following. 
As we will see shortly, 
the composite solitons have normalizable zero modes, 
and also non-normalizable zero modes. 
Only the normalizable zero modes can be promoted 
to dynamical degrees of freedom in the effective theory. 
In this subsection we construct a formal form of the effective Lagrangian 
without identifying which zero modes are normalizable. 
We will discuss the problem of the normalizability in the next subsection  
by extending our method to obtain 
a manifestly supersymmetric effective action 
on BPS solitons \cite{Eto:2006uw}.

If there are normalizable zero modes $\phi^i$, 
we can give them weak dependence on time 
(slow move approximation {\it \`a la} Manton 
\cite{Manton:1981mp,Manton:2004tk}) 
\beq
H_0\left(\phi^i \right) \rightarrow H_0 \left(\phi^i(t)\right) \label{eq:prom}.
\eeq
We introduce ``the slow-movement order parameter" $\lambda$, 
which is assumed to be much smaller than 
the other typical mass scales in the problem. 
There are two characteristic mass scales: 
one is mass difference $|\Delta m|$ of hypermultiplets, 
and the other is $g\sqrt{c}$ in front of the master equation. 
Therefore, we assume that
\beq
\lambda \ll \text{min}(\,|\Delta m|\,,\,g\sqrt{c}\,).
\eeq
The non-vanishing fields in the 1/4 BPS background have 
contributions independent of $\lambda$, namely we assume that 
\beq
H^1=\mathcal{O}(1),
\quad W_m=\mathcal O(1),
\quad \Sigma_1=\mathcal{O}(1).
\eeq
The derivatives of these fields with respect to time 
are assumed to be of order $\lambda$ 
expressing the weak dependence on time. 
The vanishing fields in the background 
can now have non-vanishing values, 
induced by the fluctuations of 
the moduli parameters of order $\lambda$. 
Therefore, we assume that 
\begin{align}
\partial_0=\mathcal{O}(\lambda),
\quad W_0=\mathcal{O}(\lambda),
\quad H^2=\mathcal{O}(\lambda),
\quad \Sigma_2 = \mathcal{O}(\lambda).
\end{align}
Then the covariant derivative $\D_0 = {\cal O}(\lambda)$ 
has consistent $\lambda$ dependence. 

If we expand the full equations of motion of the Lagrangian 
(\ref{eq:lag}) in powers of $\lambda$, 
we find that the 
$\mathcal{O}(1)$ equations 
are automatically satisfied due to 
the BPS equations (\ref{eq:4eq:mvw-1})-(\ref{eq:4eq:mvw-3}).
The next leading $\mathcal{O}(\lambda)$ equation 
is the equation for $W_0$, which is called the Gauss law 
\beq
0=-\frac{2}{g^2}\mathcal D_m F_{m 0}+\frac{2i}{g^2}
[\Sigma_1, \mathcal D_0 \Sigma_1]
+i(H^1\mathcal D_0 H^{1\dag}-\mathcal D_0 H^1 H^{1\dag}),
\label{eq:lambda}
\eeq
{with $m=1,2,3$.}
In order to obtain the effective Lagrangian of order 
$\lambda^2$ of the composite solitons, 
we have to solve this equation and determine the 
configuration of $W_0$.

As a consequence of Eq.(\ref{eq:prom}), 
the moduli matrix $H_0(\phi^i)$ depends on time through the 
time-dependent moduli parameter $\phi^i(t)$. 
Note that for the fields which depend on time only 
through $\phi^i(t)$, the derivatives with respect to 
time satisfy 
\beq
\partial_0=\delta_0+\delta_0^\dag,
\eeq
where we have defined the differential operators 
$\delta_0$ and $\delta^\dag_0$ by
\beq
\delta_0 \equiv 
\sum_i \partial_0 \phi^i \frac{\p}{\p \phi^i},
\quad
\delta_0^\dag 
\equiv \sum_i \partial_0 \bar \phi^i 
\frac{\p}{\p \bar \phi^i},
\eeq
in order to distinguish chiral $\phi^i$ and anti-chiral 
$\bar \phi^i$ multiplets of preserved supersymmetry. 
Using these operators, 
the Gauss law (\ref{eq:lambda}) 
can be solved to yield \cite{Eto:2006uw}
\beq
W_0=i(\delta_0 S^\dag S^{\dag-1}-S^{-1}\delta_0^\dag S).
\eeq

The effective Lagrangian is obtained 
by substituting these solutions into 
the fundamental Lagrangian (\ref{eq:lag}) and 
integrating over the codimensional coordinates $x^1$, $x^2$ and $x_3$. 
We retain the terms up to $\mathcal O(\lambda^2)$ 
since we are interested in the leading 
nontrivial part in powers of $\lambda$, 
and we ignore total derivative terms which do not 
contribute to the effective Lagrangian. 
Then the effective Lagrangian 
for the composite solitons can be obtained as 
\beq
L_{\rm eff} &=& K_{i \bar j} \frac{d \phi^i}{dt} \frac{d \bar \phi^j}{dt} \\
K_{i \bar j} &=& \int d^3 x \, \Tr \Biggl[
\p_i \p_{\bar j} \left(c \, \log \Omega + \frac{1}{2g^2} 
(\Omega^{-1} \partial_3 \Omega)^2 \right) \notag \\
&& \hs{10} + \frac{4}{g^2} \bigg( \p_{\bar z} ( \Omega \p_i \Omega^{-1})
\p_{\bar j} (\Omega \p_z \Omega^{-1})
- \p_{\bar z} ( \Omega \p_z \Omega^{-1})
\p_{\bar j} (\Omega \p_i \Omega^{-1}) \bigg)
\Biggr].
\label{eq:kahler-met}
\eeq
This is a nonlinear sigma model whose target space 
is the moduli space of the 1/4 BPS configurations. 
The metric on the moduli space is a K\"ahler metric, 
which can be obtained from 
the following K\"ahler potential\footnote{Although this K\"ahler {potential} is divergent, it can be made finite without changing the K\"ahler metric by the K\"ahler transformation, namely by adding terms $f(x,\phi)$ and $\overline{f(x,\phi)}$ which are (anti-)holomorphic with respect to the normalizable moduli parameters to the integrand of Eq.\,(\ref{eq:kahler-pot}).}
\begin{eqnarray}
K&=& \int d^3 x\, {\rm Tr}\,\left[c\,\psi+c\,e^{-\psi}\Omega_0+
\frac{1}{2g^2}(e^{-\psi}\p_3 e^{\psi})^2
+\frac{4}{g^2}\int ^1_0 dt \int ^t_0 ds\,
\bar\partial \psi e^{-sL_\psi}\partial \psi\right]
\label{eq:kahler-pot}
\eeq
where $\psi \equiv \log \Omega $
and the operation $L_\psi$ is defined by 
\begin{eqnarray}
L_\psi \times X=[\psi,\, X]. 
\label{eq:adjoint-operation}
\end{eqnarray} 
This general form reduces to the effective Lagrangian for either 
1/2 BPS domain walls or 1/2 BPS vortices 
if one considers the moduli matrix of the corresponding 1/2 BPS states.

\subsection{Normalizability of zero modes}\label{sec:normalizable}

The moduli matrix contains 
both normalizable and non-normalizable zero modes 
because it exhausts all possible configurations.  
There exist two kinds of zero modes 
appearing in the moduli matrix as we saw in the previous section. 
One is related to positions and phases of domain walls 
which form complex numbers. 
The other represents the positions of vortices. 
In general zero modes changing the boundary conditions at infinities 
are non-normalizable. 
For examples, zero modes related to domain walls 
or vortices with infinite lengths 
are apparently non-normalizable because 
the infinite extent of the solitons brings 
divergence in the integration. 
However the opposite is not true; 
zero modes fixing the boundary conditions 
are not always normalizable but sometimes are non-normalizable. 
Purpose of this section is to examine if zero modes 
for vortices with finite lengths 
stretched between domain walls are normalizable or not.

Now let us analyze the divergences of 
the K\"ahler potential (\ref{eq:kahler-pot}) 
in order to find out which modes are normalizable and which are not. 
Since the solutions of the master equation (\ref{eq:master_wvm}) 
have been assumed to be smooth, the divergences of the K\"ahler 
potential can appear only from the integration around the 
boundaries at infinity. 
The composite solitons have two kinds of boundaries: 
one is along $|z| \to \infty$ where we see no vortices, 
and the other is along $x_3 \to \pm \infty$ where we see no domain walls. 
We will discuss the behaviors of solutions 
near these two boundaries separately. 

From now on, we consider Abelian gauge theories ($N_{\rm C}=1$) 
for simplicity. 
First let us consider the boundary along $|z| \to \infty$. 
The master equation (\ref{eq:master_wvm}) 
can be rewritten in terms of $\psi=\log \Omega$ as 
\beq
\left( 4\p_z \p_{\bar z} + \p_{3}^2 \right) \psi
=g^2c \left(1-e^{-\psi}\Omega_0 \right)
\label{eq:master_wvm_psi}
\eeq
For simplicity, we will assume all domain walls 
are asymptotically flat, that is, each vacuum 
has the same number of vortices. 
Let us denote the number of vortices as $k$. 
Then $\Omega_0$ is given by 
\beq
\Omega_0 &\equiv& \frac{1}{c} H_0 e^{2Mx_3} H_0^\dag 
= \sum_{A=1}^{\NF} |h_A(z)|^2 e^{2m_Ax_3} \\
h_A(z) &=& v_A (z-z_{\left<A\right>1})(z-z_{\left<A\right>2})\cdots
(z-z_{\left<A\right>k}) = v_A \left( z^k - a_A z^{k-1} + \cdots \right).
\eeq
The moduli parameter $v_A$ controls 
the weight 
of the vacuum $\left<A\right>$ 
in Eq.\,(\ref{eq:weight}), 
and thus is related to positions of 
the domain walls 
separating the vacuum $\left<A\right>$ 
from the adjacent vacua. 
The moduli parameters $a_A$ are related to 
the center of 
mass $Z_A^c$ of 
vortices in the vacuum 
$\left<A\right>$ as 
\beq 
Z^c_A \equiv \frac{z_{\left<A\right>1} + \cdots + z_{\left<A\right>k}}{k} = \frac{a_A}{k}.
\eeq
Let us introduce new functions defined as 
\beq
\tilde{h}_A(z) &\equiv& \frac{h_A(z)}{z^k}, \\
\tilde{\Omega}_0(z, \bar z, x_3) &\equiv& \frac{\Omega_0(z, \bar z, x_3)}{|z|^{2k}}, \\
\tilde \psi(z, \bar z, x_3) &\equiv& \psi(z, \bar z, x_3)-\log |z|^{2k}.
\eeq
The master equation (\ref{eq:master_wvm_psi}) does not change 
in terms of these functions except for the appearance of 
the delta function\footnote{
This redefinition transform our master equation to 
the so-called Taubes's equation \cite{Taubes:1979tm} 
in the case of vortices without domain walls. 
}
in $z$. 
Since we are interested in the boundary along $|z| \to \infty$, 
let us ignore the delta function in the following discussion. 
\beq
\left( 4 \p_z \p_{\bar z} + \p_{3}^2 \right) \tilde\psi
=g^2c \left(1-e^{-\tilde\psi}\tilde\Omega_0 \right).
\label{eq:master_wvm_tilpsi}
\eeq
If we take the limit $|z| \to \infty$, $\tilde{\Omega}_0$ becomes 
\beq
\tilde{\Omega}_0 \to \Omega_{0{\rm w}} \equiv \sum_{A=1}^{\NF} |v_A|^2 e^{2m_Ax_3},
\eeq
which is nothing but the source for domain walls without vortices. 
Therefore, the solution $\tilde \psi(z, \bar z, x_3)$ 
approaches the domain wall solution in large $|z|$ region. 
We denote it as $\psi_{\rm w}(x_3)$ 
\beq
\tilde \psi(z, \bar z, x_3) \to \psi_{\rm w}(x_3) \qquad {\rm as} \quad |z| \to \infty. 
\eeq
Now let us analyze the effects of vortices on the asymptotic behavior. 
Note that $\tilde \Omega_0$ can be expanded as 
\beq
\tilde \Omega_0(z, \bar z, x_3) = \Omega_{0{\rm w}}(x_3) - \sum_{A=1}^{\NF} |v_A|^2 \left( \frac{a_A}{z} + \frac{\bar a_A}{\bar z} \right) e^{2m_A x_3} + {\cal O}\left(\frac{1}{z^2}\right).
\label{eq:exp_omega0}
\eeq
We will assume $\tilde \psi$ can be also expanded as 
\beq
\tilde \psi(z, \bar z, x_3)=\psi_{\rm w}(x_3)+\frac{\varphi(x_3)}{z}+\frac{\bar \varphi(x_3)}{\bar z}
+{\cal O}\left(\frac{1}{z^2}\right).
\label{eq:exp_psi}
\eeq
If we substitute the asymptotic forms (\ref{eq:exp_omega0}) and 
(\ref{eq:exp_psi}) into the master equation (\ref{eq:master_wvm_tilpsi}), 
and expand it in terms of $z$,
we find that $\O(1)$ equation gives the master equation for domain walls 
\beq
\p_3^2 \,\psi_{\rm w}=g^2c \left(1-e^{-\psi_{\rm w}} \Omega_{0{\rm w}} \right),
\label{eq:master_order0}
\eeq
and $\O(z^{-1})$ equation gives
\beq
\p_3^2\,\varphi = g^2c \left(\Omega_{0{\rm w}} \varphi
+ \sum_A^{\NF} |v_A|^2 a_A e^{2 m_A x_3} \right) e^{-\psi_{\rm w}}.
\label{eq:master_orderz}
\eeq
Using the equation (\ref{eq:master_order0}), 
we can find the solution of the equation (\ref{eq:master_orderz}) as 
\beq
\varphi(x_3) = - \sum_{A=1}^{\NF} v_A a_A \frac{\p \psi_{\rm w}(x_3)}{\p v_A}.
\label{eq:master_z}
\eeq
Let us now substitute these asymptotic behaviors 
(\ref{eq:exp_psi}) and (\ref{eq:master_z}) 
into the K\"ahler potential (\ref{eq:kahler-pot}). 
Using the fact that the solution of the master equation 
is the extremum of the K\"ahler potential, we obtain 
\beq
K &=& \int d^2z \,\left(K_{\rm w} + \frac{1}{|z|^2} 
\sum_{A,B=1}^{\NF} (a_A v_A)(\bar a_B \bar v_B) 
\frac{\p^2 K_{\rm w}}{\p v_A \p \bar v_B}
+\O \left( \frac{1}{z^4} \right) \right) \notag \\
&\simeq& \pi L^2 K_{\rm w} + 2 \pi \log L 
\sum_{A,B=1}^{\NF} (a_A v_A)(\bar a_B \bar v_B) 
\frac{\p^2 K_{\rm w}}{\p v_A \p \bar v_B}+{\rm const.}+\O(L^{-2})
\label{eq:KP}
\eeq
where $L$ is the infrared cutoff $|z| < L$, and 
$K_{\rm w}$ is the K\"ahler potential for domain walls 
\beq
K_{\rm w}(v_A, \bar v_A) &=& \int d x_3 \, {\rm Tr} \, \left[ c \, \psi_{\rm w} + c \, e^{-\psi_{\rm w}} \Omega_{0{\rm w}} + \frac{1}{2g^2} ( e^{-\psi_{\rm w}} \p_3 e^{\psi_{\rm w}} )^2 \right] \notag \\
&\approx& \sum_{A=1}^{\NF-1} \frac{c}{\Delta m_A} \left( \log \left| \frac{v_{A+1}}{v_A} \right| \right)^2, 
\eeq
where the last line is valid for well-separated walls. 
From the K\"ahler potential Eq.\,(\ref{eq:KP}) 
we find that the leading terms in the K\"ahler metric 
for the moduli parameters $v_A$ are proportional to $L^2$ 
and diverge in the limit $L \rightarrow \infty$. 
Therefore moduli parameters $v_A$, 
which are contained in $\Omega_{0{\rm w}}$, 
correspond to non-normalizable zero modes. 
This is because the infinitely extended domain walls 
move with infinite kinetic energy 
when the parameters $v_A$ vary. 
However, the above result says that 
the center of mass positions of vortices, 
$a_A$, in each vacuum are also non-normalizable 
even if the vortices have finite lengths. 
The divergent part of the effective Lagrangian 
associated with the motion of the parameters $a_A$ is 
\beq
2 \pi \log L \sum_{A,B=1}^{\NF} \frac{da_A}{dt} 
\frac{d\bar a_B}{dt} v_A \bar v_B 
\frac{\p^2 K_{\rm w}}{\p v_A \p \bar v_B} 
~\approx~ \pi c \log L \sum_{A=1}^{\NF-1} 
\frac{1}{\Delta m_A} \left| \frac{d a_A}{dt} 
- \frac{d a_{A+1}}{dt} \right|^2.
\label{eq:com_vor}
\eeq
The intuitive explanation is given as follows. 
In the presence of vortices, the positions of domain walls 
actually depend on the positions of vortices. 
For example, the position and the corresponding 
phase of the $A$-th domain wall 
interpolating the two vacua 
$\left<A\right>$ and $\left<A+1\right>$ is given by
\beq
\Delta m_A X^A(z, \bar z) + i \sigma^A(z,\bar z) = \log \left( \frac{v_{A+1}}{v_A} \right) + \log \left( \frac{z^k- a_{A+1} z^{k-1} + \cdots}{z^k - a_A z^{k-1} + \cdots} \right).
\eeq
Let us perturb the center of mass positions of vortices 
$a_A$ and $a_{A+1}$ with $v_A$ and $v_{A+1}$ fixed, 
to yield  
\beq
\delta (\Delta m_A X^A + i \sigma^A) =
\delta a_A \frac{z^{k-1}}{z^k-a_A z^{k-1} + \cdots}
- \delta a_{A+1} \frac{z^{k-1}}{z^k - a_{A+1} z^{k-1} + \cdots} .
\eeq
Therefore, if $a_A$ is promoted to 
a dynamical degrees of freedom 
to have the weak dependence on time, 
there appears kinetic energy of domain wall 
with the tension $T_A \equiv c \Delta m_A$ 
\beq
\int d^2 z \, \frac{T_A}{2} \left(\frac{d X^A}{dt}\right)^2
\approx \frac{\pi c}{2 \Delta m_A} \log L 
\left| \frac{d a_A}{dt} - \frac{d a_{A+1}}{dt} \right|^2.
\label{eq:kinetic-wall}
\eeq
The same amount of kinetic energy appears 
from the phase $\sigma^A$ of the domain wall. 
Thus the non-normalizability of $a_A$ 
{given in Eq.\,(\ref{eq:com_vor})}
can be understood as the divergent 
kinetic energy of domain walls. 

Now let us consider the boundaries 
along $x_3 \to \pm \infty$ directions. 
As in the previous case, we define the following new functions 
{for the limit $x_3 \to + \infty$}
\beq
\check{\Omega}_0(z, \bar z, x_3) &\equiv& \Omega_0(z, \bar z, x_3)
e^{-2m_1x_3}=\sum_{A=1}^{\NF}|h_A(z)|^2e^{-2(m_1-m_A)x_3}, \\
\check{\psi}(z, \bar z, x_3)&\equiv&\psi(z, \bar z, x_3)-2m_1x_3.
\eeq
Note that we have chosen the mass parameters such that 
$m_1>m_2>\cdots>m_{\NF}$. 
The master equation (\ref{eq:master_wvm_psi}) does not change 
in terms of these functions as before. 
If we take the limit $x_3 \to \infty$, $\check{\Omega}_0$ becomes 
\beq
\check{\Omega}_0 \to \Omega_0^{\rm v} \equiv |h_1(z)|^2,
\eeq
which is nothing but the source for vortices 
in vacuum $\langle 1 \rangle$. 
Therefore, the solution $\check{\psi}(z, \bar z, x_3)$ 
approaches the vortex solution in large $x_3$ region, 
which we denote as $\psi_{\rm v}(z, \bar z)$ 
\beq
&&\check{\psi}(z, \bar z, x_3) \to \psi_{\rm v}(z,\bar z) \qquad {\rm as}\quad x_3 \to \infty,\\
&&\qquad 4\p_z\p_{\bar z} \psi_{\rm v} = g^2c\left(1 - e^{-\psi_{\rm v}} \Omega_0^{\rm v} \right).
\eeq
We are interested in the effects of domain walls on the asymptotic behavior.
Note that $\check{\Omega}_0$ behaves 
\beq
\check{\Omega}_0 = |h_1(z)|^2+|h_2(z)|^2e^{-2(m_1-m_2)x_3}+\cdots
\eeq
in large $x_3$ region.
The second term is strongly suppressed by the exponential factor
in contrast to the previous case.
The solution $\check{\psi}(z, \bar z, x_3)$
should also behave as 
\beq
\check{\psi}(z, \bar z, x_3)=\psi_{\rm v}(z, \bar z)+\phi(z, \bar z)e^{-m_v x_3}+ {\cal O}(e^{-2m_v x_3})
\eeq
where $m_v$ stands for the lowest mass scale 
of the bulk modes in the right most vacuum $\left<1\right>$. 
Since the second term is exponentially 
suppressed and does not 
give a 
divergence, 
the parameters contained in the function $\phi(z, \bar z)$ 
correspond to normalizable zero modes. 
Therefore, only the function $\psi_{\rm v}(z, \bar z)$ 
has non-normalizable zero modes, 
which are positions of vortices living in 
vacuum $\langle 1 \rangle$. 
The same argument holds for $x_3 \to -\infty$ direction. 

In summary, in the case of flat domain walls, 
non-normalizable zero modes are 
positions of domain walls $v_A$, 
the center of mass of vortices in each vacuum $Z^c_A=a_A/n_A$, 
and positions of infinitely long vortices 
$z_{\left<1\right>i}$ and $z_{\left<\NF\right>i}$ 
in vacuum $\langle 1 \rangle$ 
and vacuum $\langle \NF \rangle$, respectively.

\section{Dynamics of 1/4 BPS wall-vortex systems}\label{sec:dynamics}

Now let us construct the effective Lagrangian of 
vortices between domain walls. 
We will discuss the Abelian gauge theory with three flavors. 
The mass parameters are taken as 
$M={\rm diag}\,(\frac{m}{2},0,-\frac{m}{2})$, and 
we denote the numbers of vortices in three vacua by 
$(n_1,n_2,n_3)$. 
In what follows, we will take the strong gauge coupling limit $(g\to\infty)$ 
in order to calculate the effective action analytically. 
In the strong coupling limit, the master equation Eq.\,(\ref{eq:master_wvm}) 
becomes an algebraic equation and analytically solved as 
\beq
\Omega ~=~ \Omega_0 ~\equiv~ \frac{1}{c} H_0 e^{2 M x_3} H_0^\dagger. 
\eeq
The K\"ahler metric Eq.\,(\ref{eq:kahler-met}) 
also takes a simple form in the strong coupling limit 
\beq
K_{i \bar j} = c \int d^3 x \, \p_i \p_{\bar j} \log \det \Omega_0.
\label{eq:kahler-met-strong}
\eeq
Although the ANO vortices linearly extending to infinity
like Eq.(\ref{eq:ANO}) shrink to singular configurations 
since their sizes $1/(g\sqrt{c})$ tend to zero in this limit, 
vortex strings with finite length between domain walls do not
(its size behaves as $e^{-m L}$ where $L$ is separation between
walls).
Therefore we can construct
low energy effective theories for 
vortex strings between domain walls 
in the strong coupling limit. 

\subsection{Numbers of vortices: (1,1,1)}\label{sec:111}
First let us consider the case in which each vacuum has a single vortex. 
This configuration admits no normalizable zero modes. 
However, it will give us an explicit example of 
the non-normalizable modes which we have discussed 
in Sec.\ref{sec:normalizable}. 
The general form of the moduli matrix is given by 
\beq
H_0 = \sqrt{c} \, \left(
v_1 (z-z_{\left<1\right>1}),\,
v_2 (z-z_{\left<2\right>1}),\,
v_3 (z-z_{\left<3\right>1}) \right).
\eeq
Since we are interested in the vortex in vacuum $\langle 2 \rangle$, 
we set $z_{\left<1\right>1} = z_{\left<3\right>1}=0$ and 
$v_1=v_3=1$, and define $v_2\equiv v$ and $z_{\left<2\right>1} \equiv z_0$ 
\beq
H_0 = \sqrt{c} \, \left(z,\, v(z-z_0),\, z \right).
\label{eq:moduli-111}
\eeq
The positions of domain walls can be 
estimated by weights of vacua (\ref{eq:weight}). 
Both domain walls are asymptotically flat, 
and the asymptotic distance between these domain walls, 
that is, the length of the vortex in vacuum $\langle 2 \rangle$, 
is given by 
\beq
l_{\langle 2 \rangle}=\frac{4}{m} \log |v|.
\label{eq:111-length}
\eeq
Energy densities of configurations 
in a plane containing vortices are shown for several 
values of moduli parameters in Fig.\ref{fig:111}. 
\begin{figure}[htb]
\begin{center}
\begin{tabular}{cccc}
\includegraphics[height=4cm]{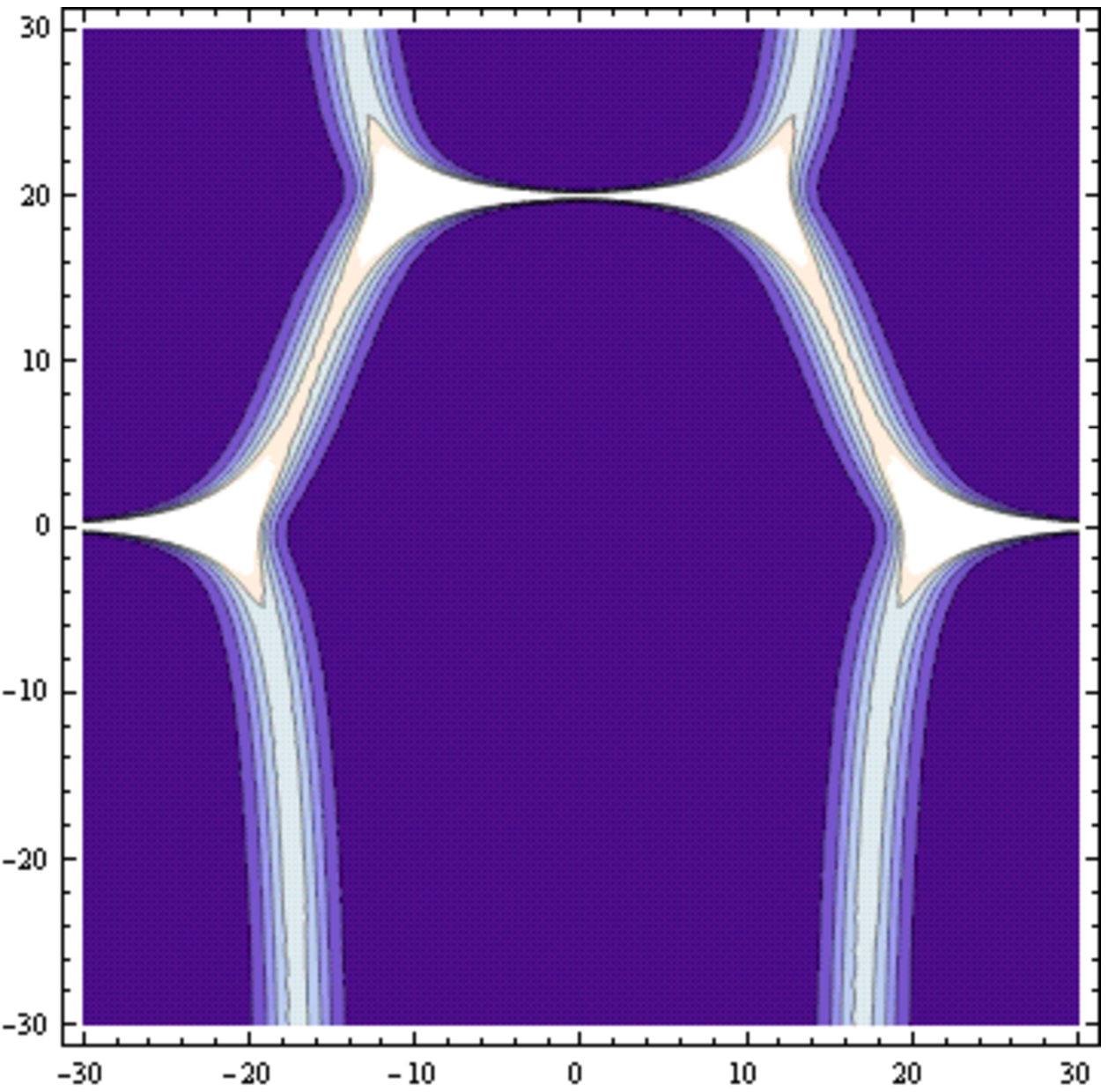} &
\includegraphics[height=4cm]{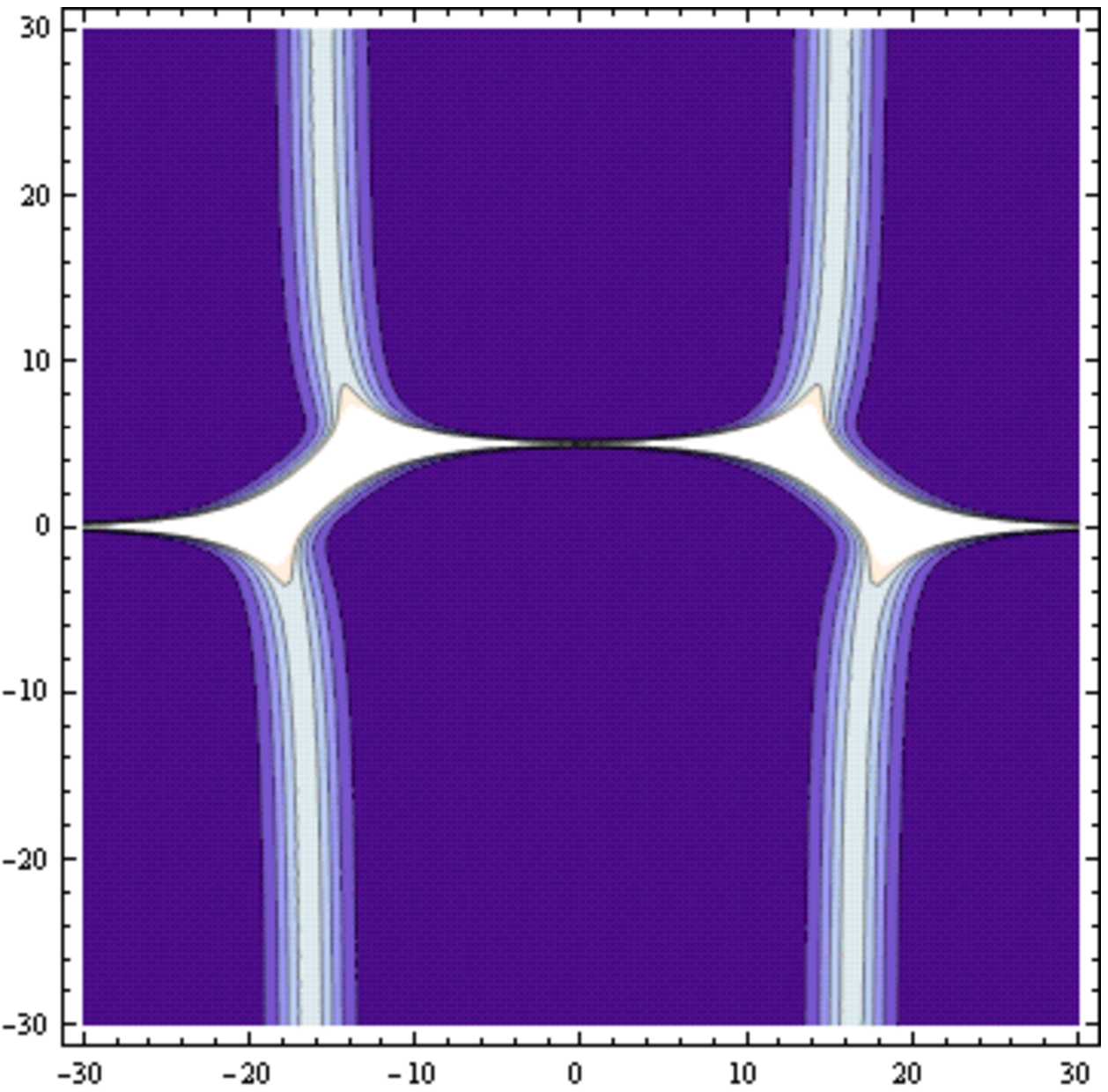} &
\includegraphics[height=4cm]{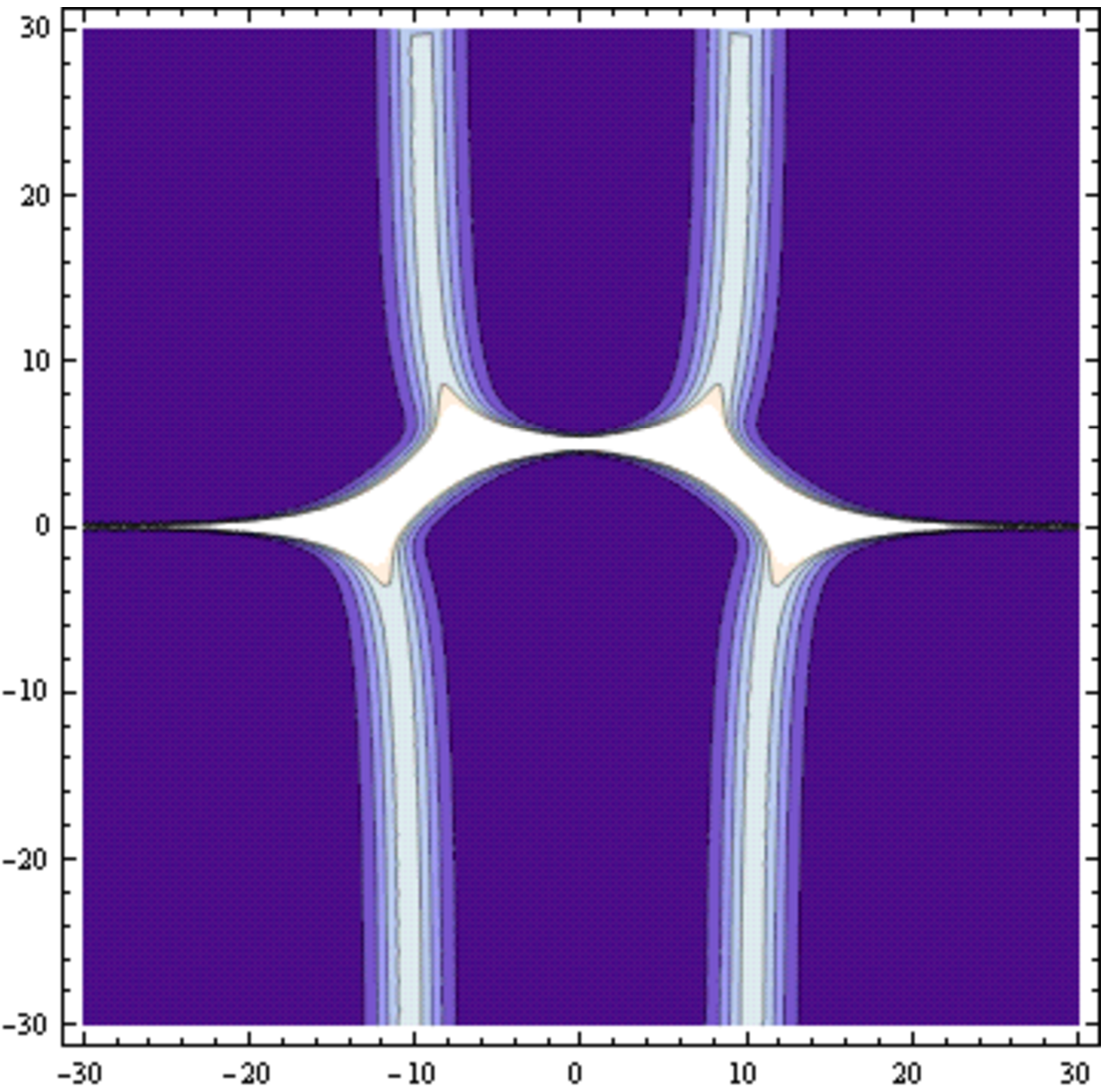} &
\includegraphics[height=4cm]{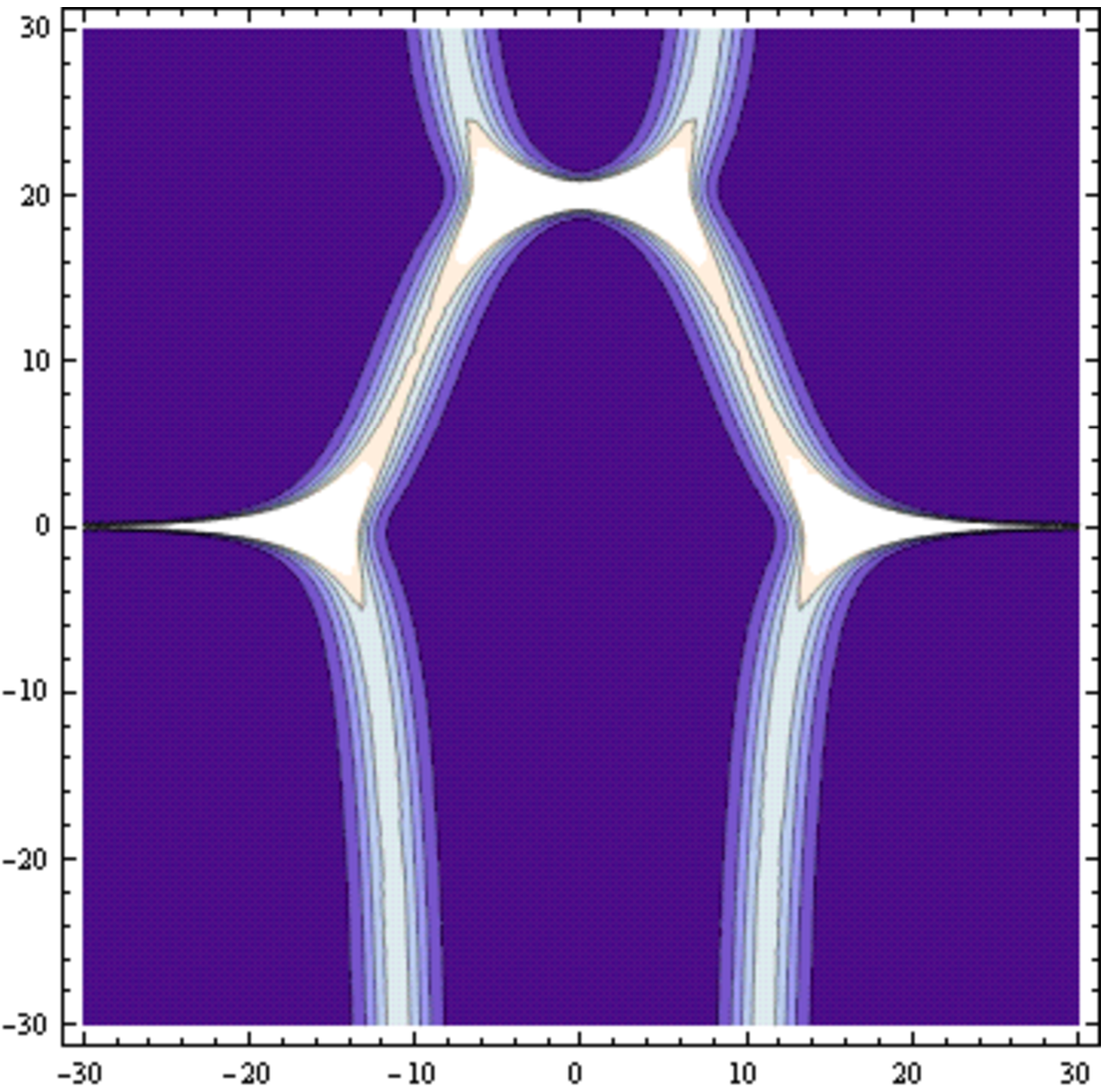} \\
$v=e^8,\, z_0=20$&$v=e^8,\,z_0=5$ & $v=e^5,\,z_0=5$&$v=e^5,\,z_0=20$
\end{tabular}
\caption{
The energy densities in a plane containing vortices in the 
strong coupling limit $g\to \infty$
with $c=1$ and $m=1$. 
Vertical lines are walls and horizontal lines are vortices. 
} 
\label{fig:111}
\end{center}
\end{figure}

Now we give the weak time dependence to $z_0$, 
and investigate the dynamics of the middle vortex. 
The explicit solution of the master equation (\ref{eq:master_wvm}) 
can be obtained in the strong coupling limit $g\to \infty$ as 
\beq
\Omega ~\to~ \Omega_0 ~=~ |z|^2 e^{mx_3} + |v|^2 |z-z_0|^2 + |z|^2 e^{-mx_3}.
\eeq
Let us substitute this solution into the K\"ahler metric 
(\ref{eq:kahler-met-strong}). 
Leaving the integration along the $x_3$-coordinate, 
the K\"ahler metric can be calculated as 
\beq
K_{z_0 \bar z_0} &\equiv& c \int d^3 x \frac{2|vz|^2 
\cosh(mx_3)}{(|v|^2 |z-z_0|^2 + 2 |z|^2 \cosh mx_3)^2}\notag \\
&=& \pi c \int dx_3 \, 
\Biggl[ \frac{2 |v|^2 \cosh mx_3}{(2\cosh mx_3 + |v|^2)^2} 
\log \frac{L^2}{|z_0|^2} 
+ \frac{|v|^2(|v|^2-2\cosh mx_3)}{(2\cosh mx_3 + |v|^2)^2}\notag \\
&& \hs{30} - \frac{2|v|^2 \cosh mx_3}{(2\cosh mx_3 + |v|^2)^2}
\log \left(\frac{2|v|^2 \cosh mx_3}{(2\cosh mx_3 + |v|^2)^2} \right) \Biggr],
\label{eq:metric_111}
\eeq
where $L$ is the infrared cutoff in the $z$-plane $|z|<L$. 
As we saw in the previous section, the 
K\"ahler metric 
contains 
the logarithmic divergence. 
The complicated metric (\ref{eq:metric_111}) reduces to 
a simple form 
when we take the limit of $|v| \to \infty$ 
\beq
K_{z_0 \bar z_0} \approx \pi c \left(
\frac{4}{m} \log |v|
-\frac{4}{m} \log |z_0|
+\frac{4}{m} \log L
+{\rm const.}
\right).
\label{eq:large_metric_111}
\eeq
The physical meaning of the metric is clear in this form.
Since the tension of the vortex is $2 \pi c$ 
and its length is given in Eq.\,(\ref{eq:111-length}), 
the first term 
in equation
(\ref{eq:large_metric_111}) corresponds to kinetic energy of the vortex.
According to equation (\ref{eq:kinetic-wall}) and the following comments, 
the kinetic energy of two domain walls can be calculated as 
\beq
T_{\rm wall} = \frac{4 \pi c}{m} \log L \, \frac{dz_0}{dt} \frac{d\bar z_0}{dt},
\eeq
{
where we have identified $z_0 = a_2$ and $\Delta m_A = m/2$ in Eq.\,(\ref{eq:kinetic-wall}).}
This coincides with the third term in equation (\ref{eq:large_metric_111}).
{This is the origin of the non-normalizability of $z_0$.}

Note that the moduli space has the singularity at $z_0=0$. 
This is because we have fixed the vortices 
in vacuum $\langle 1 \rangle$ and vacuum $\langle 3 \rangle$ 
at the same position. 
When the vortex in $\langle 2 \rangle$ also comes to the same position, 
they result in a single ANO vortex which is infinitely long 
and becomes singular in the strong coupling limit 
with its shrinking size $1/(g \sqrt{c}) \to 0$. 
The singularity can be removed by dislocating the outer vortices. 
We will discuss this issue in the next example. 

\subsection{Numbers of vortices: (2,2,2)}\label{sec:222}
Let us next consider the case 
in which each vacuum has a pair of vortices. 
The general form of the moduli matrix is given by 
\begin{align}
H_0 = \sqrt{c} \, \Big(
v_1 (z-z_{\left<1\right>1})(z-z_{\left<1\right>2}),\, 
v_2 (z-z_{\left<2\right>1})(z-z_{\left<2\right>2}),\, 
v_3 (z-z_{\left<3\right>1})(z-z_{\left<3\right>2}) \Big).
\end{align}
Since we are interested in the relative motion of vortices 
in vacuum $\langle 2 \rangle$, we set 
$z_{\left<1\right>i}=0=z_{\left<3\right>j}$ and 
$v_1=v_3=1$, and 
define $v_2\equiv v$ and 
$z_{\left<2\right>1}=-z_{\left<2\right>2} \equiv z_0$
\beq
H_0 = \sqrt{c} \, \left(z^2,\, v(z^2-z_0^2),\, z^2 \right).
\label{eq:moduli-222}
\eeq
The distance between two domain walls 
is also given by equation (\ref{eq:111-length}) 
in the present case. 
Energy densities of configurations 
in a plane containing vortices are shown for several 
values of moduli parameters in Fig.\ref{fig:222}. 
\begin{figure}[htb]
\begin{center}
\begin{tabular}{cccc}
\includegraphics[height=4cm]{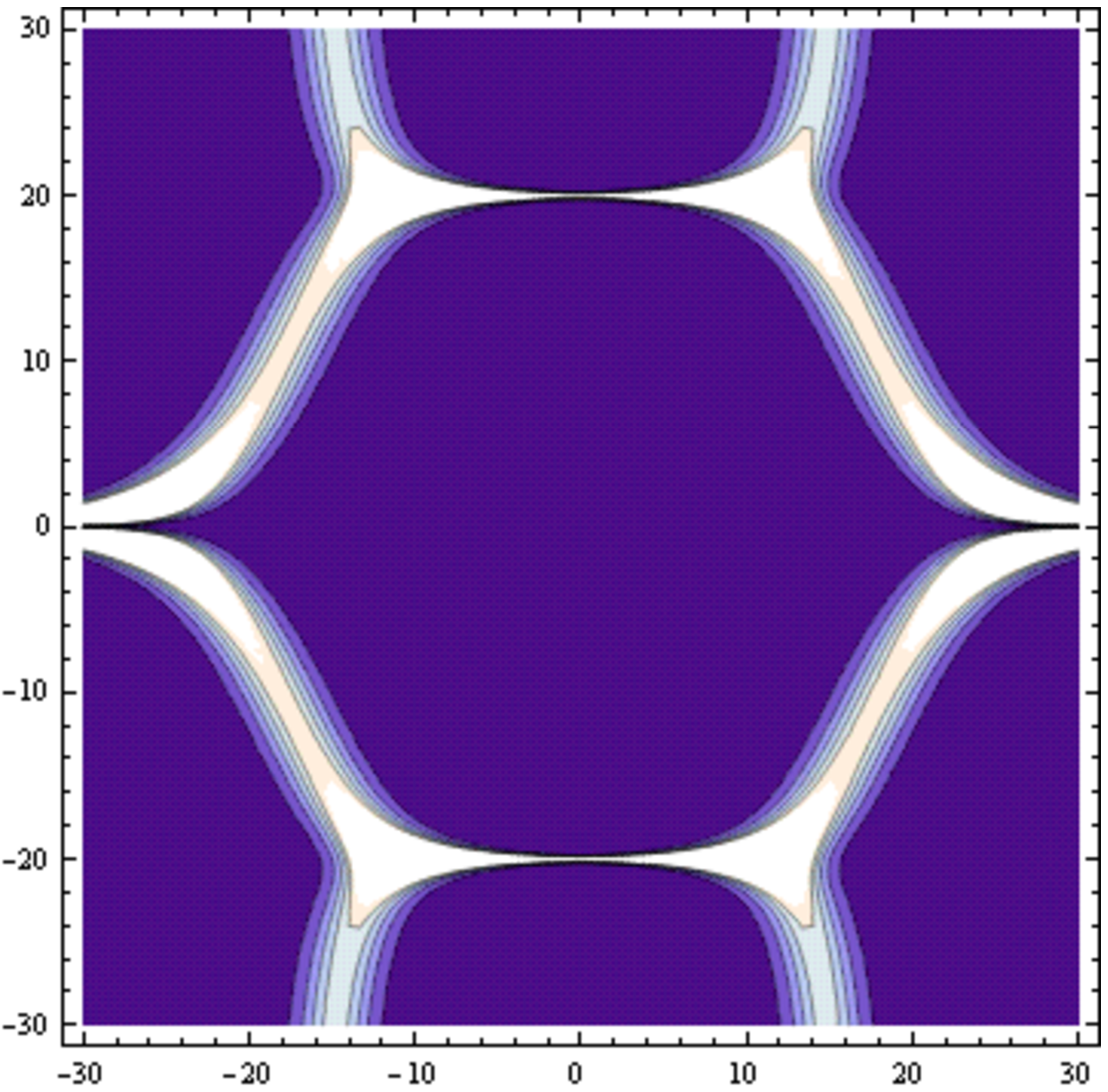}&
\includegraphics[height=4cm]{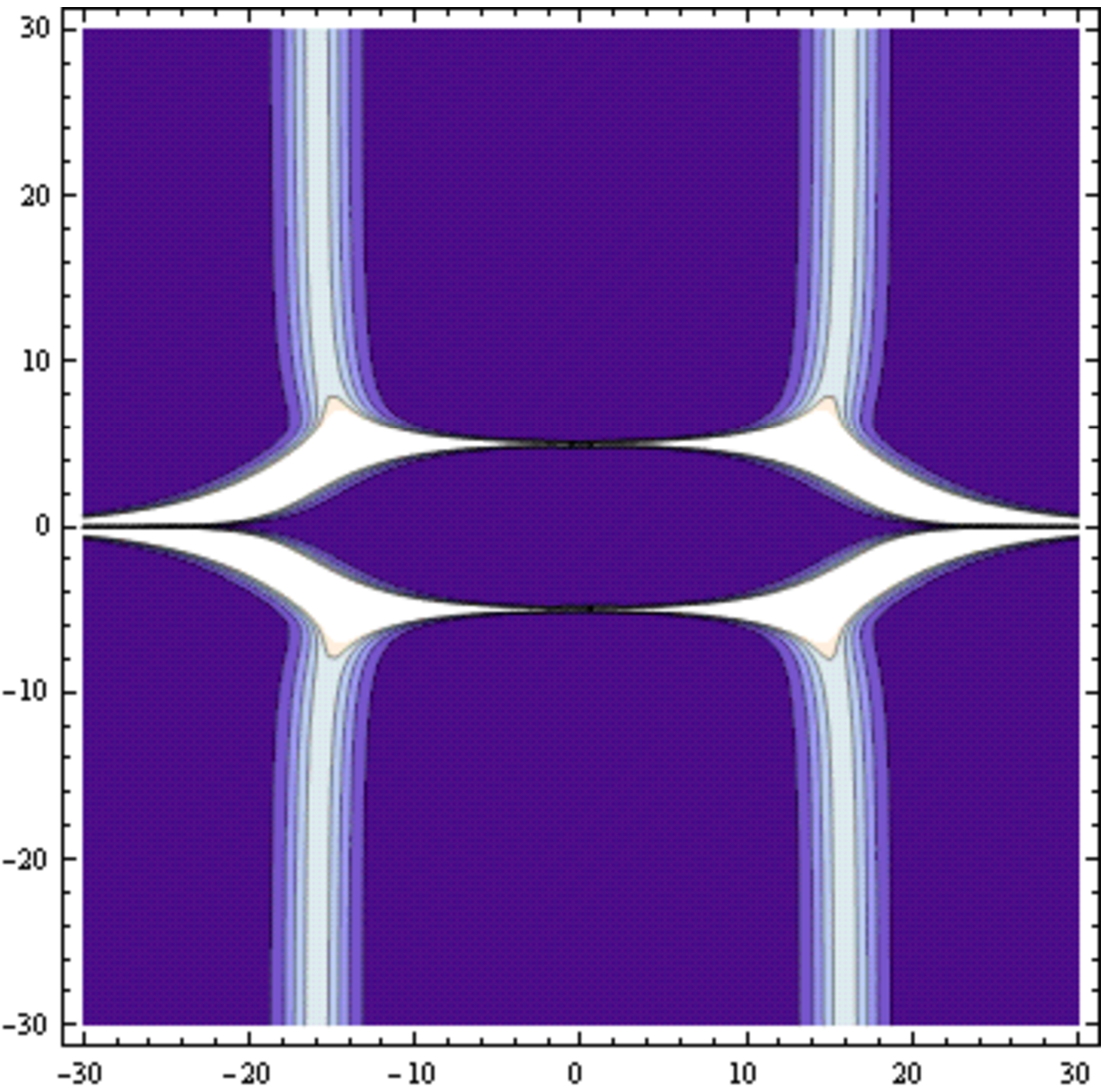}&
\includegraphics[height=4cm]{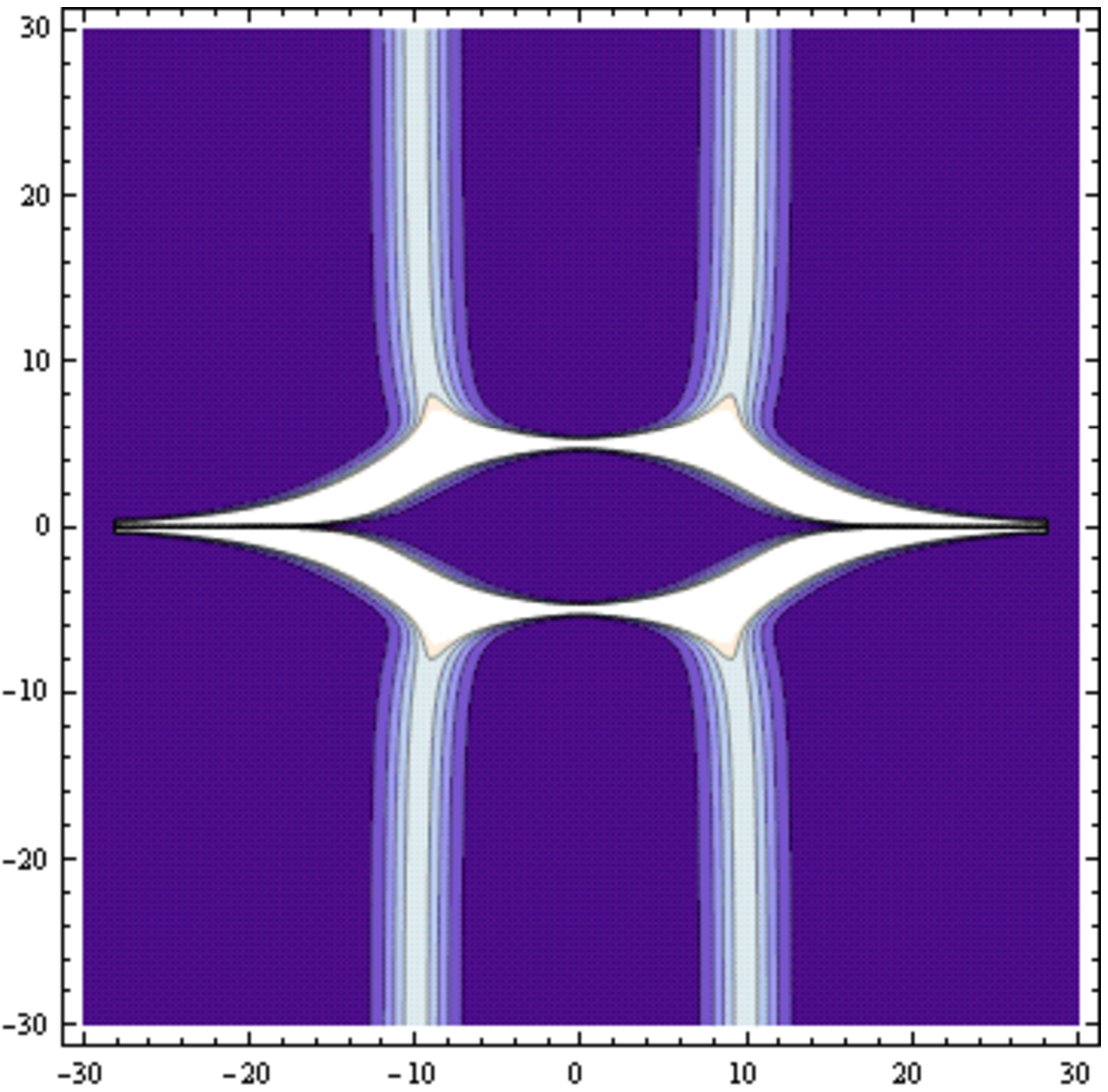}&
\includegraphics[height=4cm]{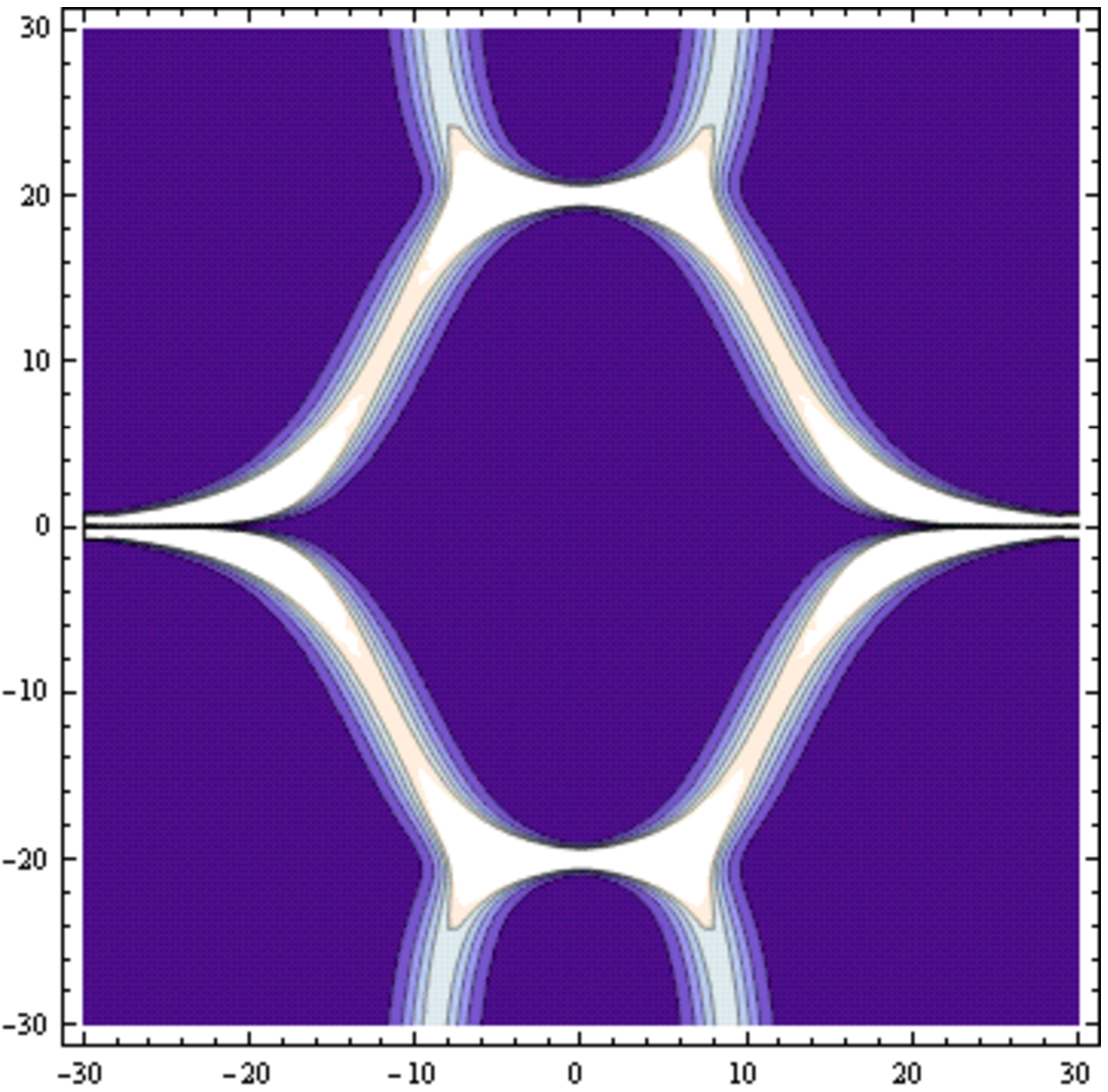}\\
$v=e^8,\,z_0=20$ & $v=e^8,\,z_0=5$ 
& $v=e^5,\,z_0=5$ & $v=e^5\,\,z_0=20$
\end{tabular}
\caption{
The energy densities in a plane containing vortices in the 
strong coupling limit $g\to \infty$
with $c=1$ and $m=1$. 
Vertical lines are walls and horizontal lines are vortices. 
}
\label{fig:222}
\end{center}
\end{figure}

Now let us give the weak time dependence to $z_0$, 
and investigate the dynamics of the middle vortices. 
The explicit solution of the master equation (\ref{eq:master_wvm}) 
can be obtained in the strong coupling limit $g\to \infty$ 
\beq
\Omega ~\to~ \Omega_0 ~=~ 
|z|^4 e^{mx_3} + |v|^2|z^2-z_0^2|^2 + |z|^4 e^{-mx_3}.
\eeq
Let us substitute this solution 
into the K\"ahler metric (\ref{eq:kahler-met-strong}). 
After integrating the $z$-coordinates, we obtain 
the K\"ahler metric 
\begin{eqnarray}
K_{z_0 \bar z_0} = 2 \pi c \int dx_3 \, k E(k) 
\label{eq:metric_222}
\end{eqnarray}
where $E(k)$ is the complete elliptic integral of the second kind 
\beq
E(k) \equiv \int_0^{\frac{\pi}{2}} d\theta \sqrt{1 - k^2 \sin^2 \theta}, 
\label{eq:EllipticE}
\eeq 
with the $x_3$ dependent parameter $k$ 
\begin{equation}
k = \left(\frac{|v|^2}{2\cosh mx_3 + |v|^2} \right)^{1/2}. 
\label{eq:parameter_222}
\end{equation}
The metric does not depend on $z_0$ and 
its value can be written as 
a sum of the hypergeometric functions (see Appendix \ref{appendix:A}). 
The asymptotic value of this metric for large $|v|$ is given by (see Appendix \ref{appendix:B})
\beq
K_{z_0 \bar z_0} ~\approx~ \frac{8\pi c}{m} \log |4v|.
\eeq
The leading term in the effective Lagrangian coincides with 
the kinetic energy of two vortices with length $l_{\left<2\right>} = \frac{2}{m} \log |v|^2$ 
and tension $T_{\rm v} =  2 \pi c$
\beq
L_{\rm eff} &\approx& \left( \frac{8\pi c}{m} \log |v| + \frac{16\pi c}{m} \log 2 
\right) |\dot z_0|^2 = \left( T_{\rm v} l_{\left<2\right>} + \frac{16\pi c}{m} \log 2 \right)|\dot z_0|^2,
\label{eq:large_metric_222} 
\eeq
The independence of 
$L_{\rm eff}$ on the IR cutoff $L$ shows the normalizability 
of the moduli $z_0$. 
Therefore it makes sense to consider its dynamics. 
Since we cannot distinguish two vortices, 
the geometry of the moduli space is a cone, ${\bf C}/{\bf Z}_2$. 
Here $\mathbf Z_2$ denotes the exchange of the vortices and 
acts on the coordinate as $z_0 \rightarrow -z_0$. 
In fact, 
a good coordinate of the moduli space 
is not $z_0$ but $z_0^2$, which appears naturally 
in the moduli matrix (\ref{eq:moduli-222}). 
The moduli space has the singularity at $z_0=0$. 
As we explained in the previous section, 
this is because we have fixed 
the outer vortices at the same position. 
When the vortices in vacuum $\langle 2 \rangle$ 
also come to the same position, 
they result in two ANO vortices which are singular 
in the strong coupling limit $g \rightarrow \infty$.
The singularity can be removed by dislocating the outer 
vortices. 
For instance, let 
us consider the moduli matrix given in the form 
\beq
H_0 = \sqrt{c} \, \left((z-z_1)^2,\, v(z^2-z_0^2),\, (z+z_1)^2 \right).
\eeq
The vortices in vacuum $\langle 1 \rangle$ are located at $z=z_1$, 
and the vortices in vacuum $\langle 3 \rangle$ are at $z=-z_1$. 
The K\"ahler metric in strong coupling limit is given as 
\beq
K_{z_0 \bar z_0} 
= c \, \int d^3x \, 
\frac{4|v|^2 |z_0|^2 \left(|z-z_1|^4 e^{mx_3} 
+ |z+z_1|^4 e^{-mx_3} \right)}
{(|z-z_1|^4 e^{mx_3} + |v|^2 |z^2-z_0^2|^2 + |z+z_1|^4 e^{-mx_3})^2}.
\eeq
The metric starts at $\O \left(|z_0|^2 \right)$, 
and can be expanded around $z_0=0$ as 
\beq
ds^2 &\simeq& \left|z_0\right|^2 
\Big( A + B z_0^2 + \bar B \bar z_0^2 + 
\mathcal O(|z_0|^4) \Big) d z_0 d \bar z_0 \notag \\
&=& \Big( A + B Z + \bar B Z + \mathcal O(|Z|^2) \Big) dZ d\bar Z,
\eeq
where $Z \equiv z^2_0$ is a good complex coordinate on the moduli space. 
Since the constant $A \equiv (K_{z_0 \bar z_0}/2|z_0|^2)|_{z_0=0}$ 
is non-zero, the scalar curvature does not diverge at $Z=0$. 
Therefore, the moduli space is non-singular at the origin, and 
the vortices scatter with right-angle in head-on collisions. 
On the other hand, the asymptotic metric for 
$|v|^2 \gg 1 \gg |z_1/z_0|$ is given by (see Appendix \ref{appendix:B})
\begin{eqnarray}
K_{z_0 \bar z_0} \approx  
\frac{8\pi c}{m} \log \left| \frac{4vz_0^2}{z_0^2-z_1^2}\right|. 
\label{eq:asym_metric_222}
\end{eqnarray}
This coincides with Eq.\,(\ref{eq:large_metric_222}) when 
$z_1=0$.
The leading term is again identified with the kinetic term 
of the vortex of length
$l_{\left<2\right>} = \frac{2}{m}\log|v|^2$.

\subsection{Numbers of vortices: (0,2,0)}\label{sec:020}
Let us next consider the case 
in which only the middle vacuum has a pair of vortices.  
This is the case where walls are not asymptotically flat, but 
may be useful as a building block for more complicated configurations. 
The general form of the moduli matrix is given by 
\begin{align}
H_0 = \sqrt{c} \, \left(v_1,\, v_2 (z-z_{\left<2\right>1})
(z-z_{\left<2\right>2}),\, v_3 \right).
\end{align}
We are interested in the relative motion of vortices in 
vacuum $\langle 2 \rangle$. 
Although we have not discussed the cases 
in which domain walls are logarithmically 
bending in Sec.\,\ref{sec:normalizable}, 
it turns out that the relative motion of the two vortices 
is normalizable zero mode even in such cases. 
Let us set $v_1=v_3=1$, 
$z_{\left<2\right>1}=-z_{\left<2\right>2} 
\equiv z_0$ and $v_2 \equiv v$
\beq
H_0 = \sqrt{c} \, \left(\,1,\, v (z-z_0)(z+z_0),\, 1\, \right).
\label{eq:moduli-020}
\eeq
Energy densities of configurations 
in a plane containing vortices are shown for several 
values of moduli parameters in Fig.\ref{fig:020}. 
\begin{figure}[htb]
\begin{center}
\begin{tabular}{cccc}
\includegraphics[height=4cm]{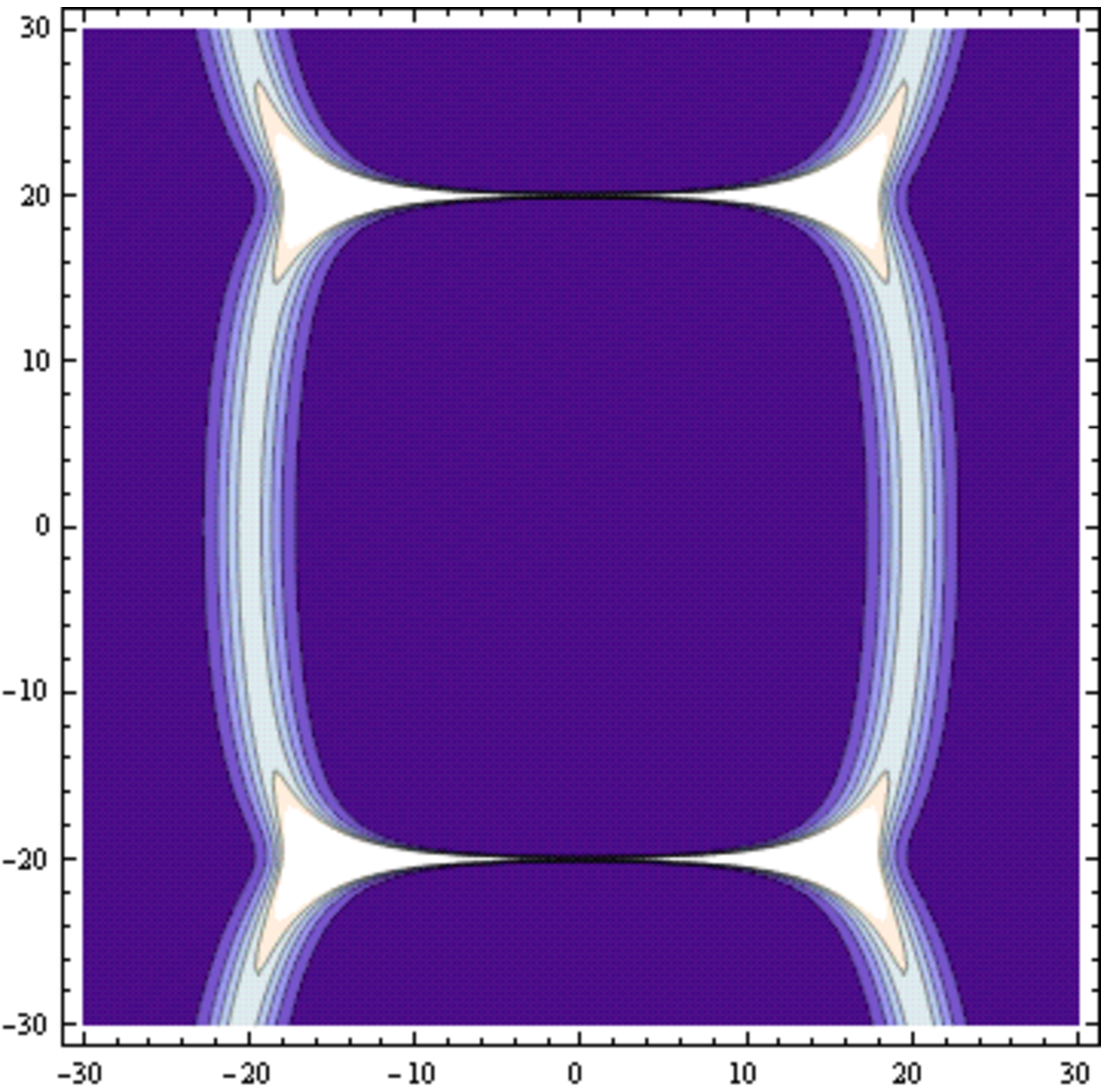} &
\includegraphics[height=4cm]{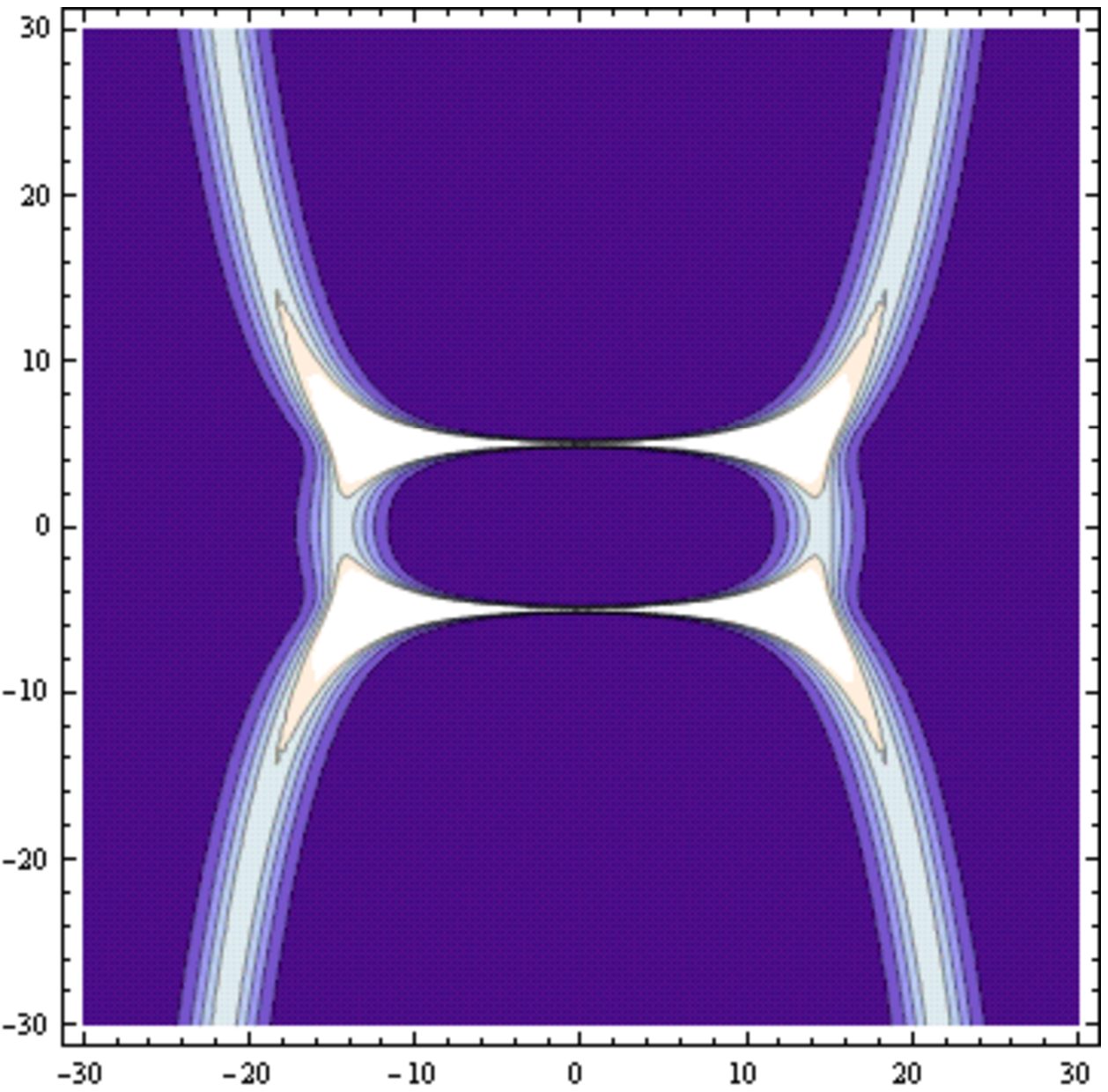} &
\includegraphics[height=4cm]{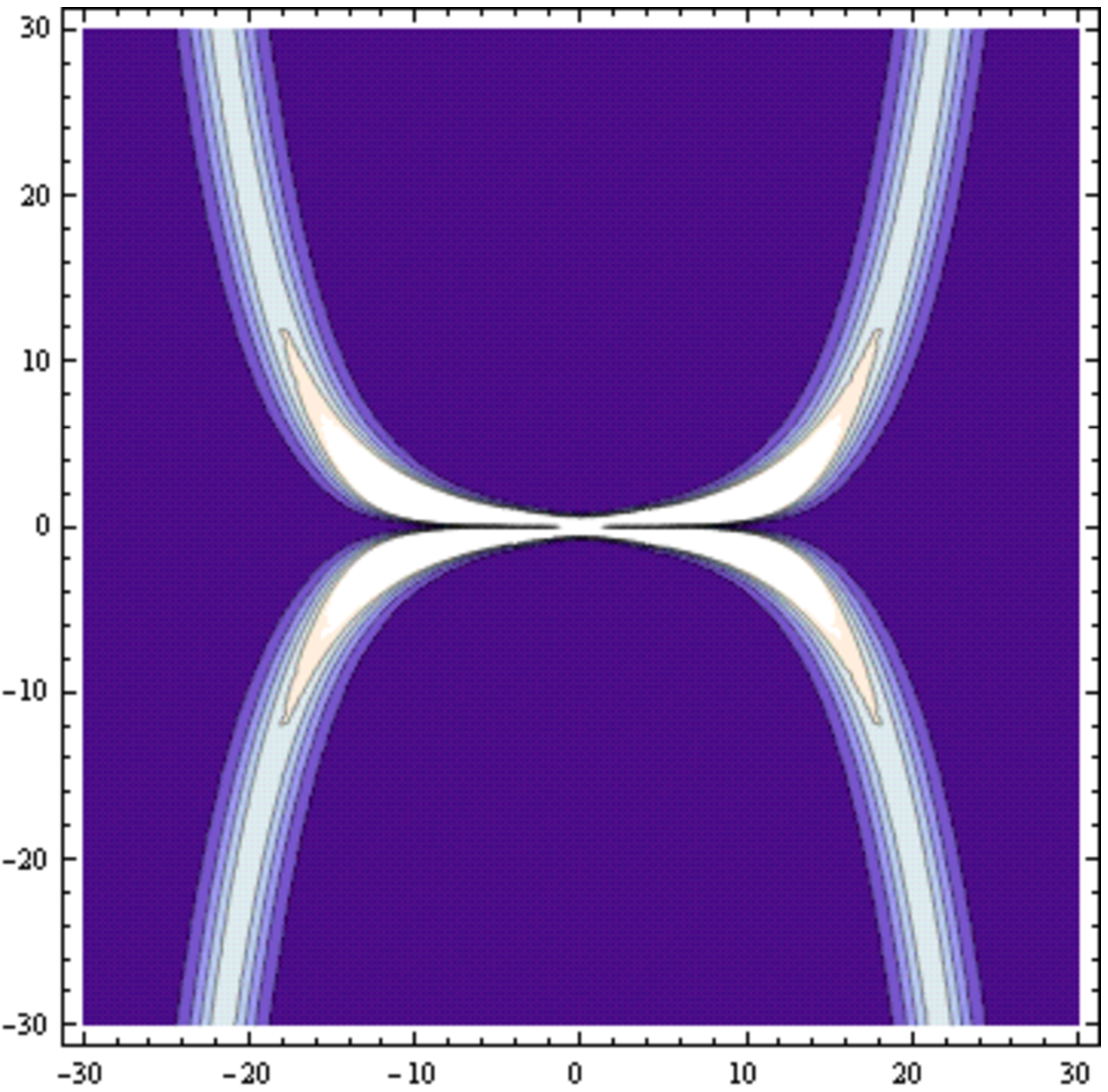} &
\includegraphics[height=4cm]{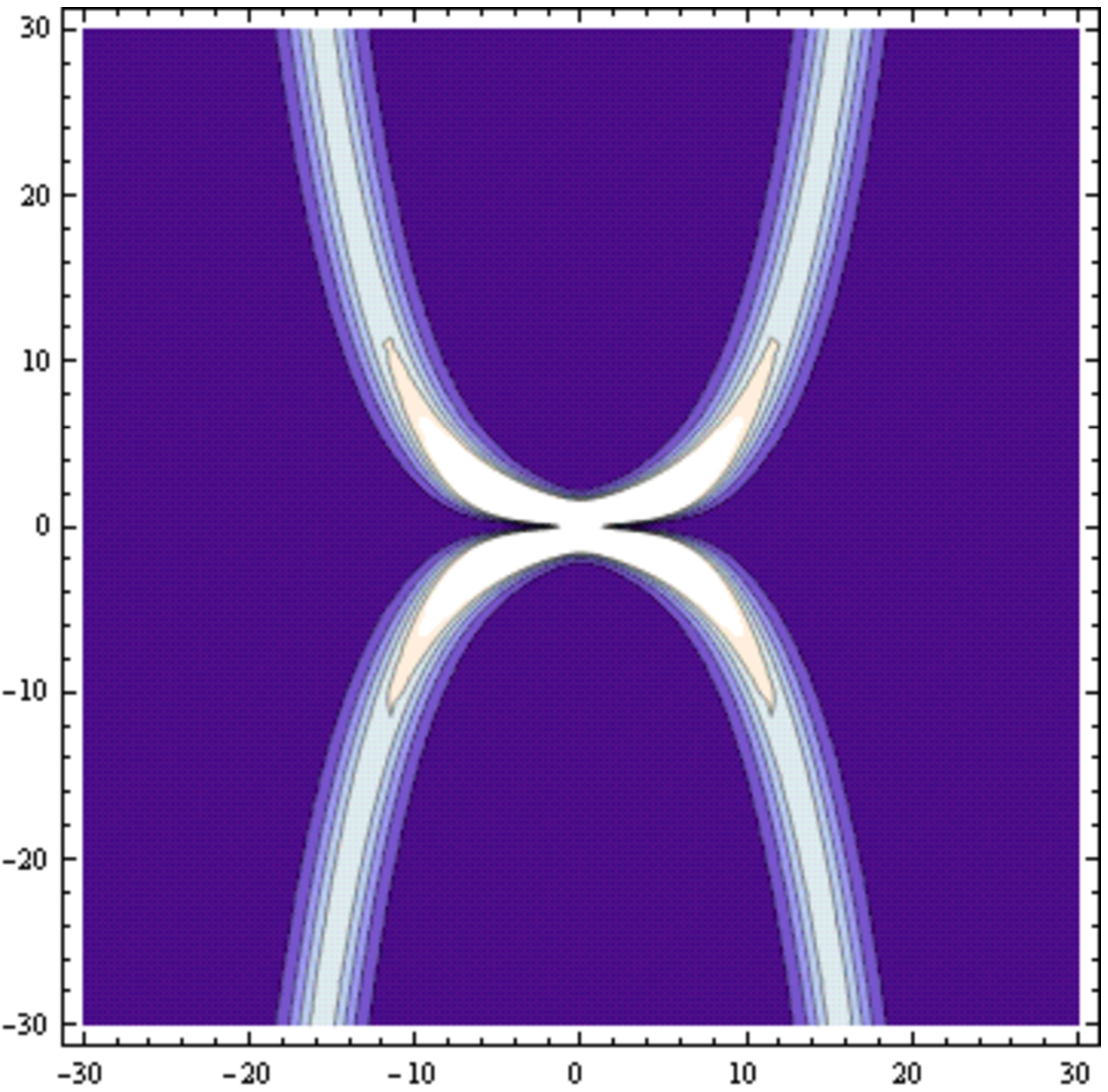} \\
$v=e^4,\,z_0=20$ & $v=e^4,\,z_0=5$ 
& $v=e^4,\,z_0=0$ & $v=e^2,\,z_0=0$
\end{tabular}
\caption{
The energy densities in a plane containing vortices in the 
strong coupling limit $g\to \infty$
with $c=1$ and $m=1$. 
Vertical lines are walls and horizontal lines are vortices. 
}
\label{fig:020}
\end{center}
\end{figure}

Now let us give the weak time dependence to $z_0$, 
and investigate the dynamics of the middle vortices. 
The explicit solution of the master equation (\ref{eq:master_wvm}) 
can be obtained in the strong coupling limit $g\to \infty$ 
\beq
\Omega ~\to~ \Omega_0 ~=~ e^{mx_3} + |v|^2 |z^2-z_0^2|^2 + e^{-mx_3}.
\eeq
Let us substitute this solution 
into the K\"ahler metric (\ref{eq:kahler-met-strong}). 
After integrating the $z$-coordinates 
similarly to the case of 
the number of vortices $(2, 2, 2)$ in the previous subsection, 
we obtain the K\"ahler metric as 
an integral over the complete elliptic integral 
of the second kind $E(k)$ defined in 
Eq.\,(\ref{eq:EllipticE}) 
\begin{eqnarray}
K_{z_0 \bar z_0}=2 \pi c \int dx_3 \, k E(k),
\quad {\rm with} \quad 
k=\left(\frac{|v z_0^2|^2}{2 \cosh mx_3 + |v z_0^2|^2} \right)^{1/2}. 
\label{eq:metric_020}
\end{eqnarray}
The metric has the same form as 
that of the previous example (\ref{eq:metric_222}). However,
the variable $k$ in $E(k)$ is now 
defined differently from the case of $(2, 2, 2)$, $v$ is 
now replaced by $vz_0^2$. 
Integrating over $x_3$, we obtain the K\"ahler metric 
as a sum of the hypergeometric functions 
(see Appendix 
\ref{appendix:A}). 
If we expand the K\"ahler metric around $|v z_0^2|^2 = 0$, we obtain 
\beq
dx^2 = 2 K_{z_0 \bar z_0} d z_0 d \bar z_0 
&\rightarrow& \frac{\pi^{\frac{3}{2}}c}{m} 
\left( \Gamma(1/4)^2 - \frac{3}{2} \Gamma(3/4)^2 |v z_0^2|^2 
+ \mathcal O(|v z_0^2|^4) \right) \left| v z_0^2 \right| 
d z_0 d \bar z_0 \notag \\
&=& \frac{\pi^{\frac{3}{2}}|v|c}{4m} 
\left( \Gamma(1/4)^2 - \frac{3}{2} \Gamma(3/4)^2 |v Z|^2 
+ \mathcal O(|v Z|^4) \right) d Z d \bar Z.
\eeq
Since the coordinate $Z \equiv z_0^2$ is a good coordinate 
even at the origin, 
it shows that the moduli space is non-singular at the origin and 
the vortices scatter with right-angle in head-on collisions. 
If we take the opposite limit $|v z_0^2|^2 \to \infty$, 
the metric can be calculated as 
(see Appendix \ref{appendix:B}) 
\beq
K_{z_0 \bar z_0} \approx \frac{8 \pi c}{m} \log |4v z_0^2|. 
\label{eq:asym_metric_020}
\eeq
Since the domain walls are logarithmically bending in the present case, 
the definition of the distance between domain walls 
is not clear. However, at the center of mass of two vortices, 
the distance between domain walls is given by 
\beq
l_{\langle 2 \rangle} = \frac{4}{m} \log |v z_0^2|.
\label{eq:020-length}
\eeq
It can be considered as 
the typical lengths of the vortices (see Fig.\ref{fig:020}). 
Therefore, the above asymptotic metric (\ref{eq:asym_metric_020}) 
can be understood as the kinetic energy of two vortices. 
Fig.\,\ref{fig:metric} shows a numerically calculated 
metric and the moduli space embedded into $\mathbf R^3$. 
\begin{figure}[htp]
\begin{center}
\begin{tabular}{cc}
\includegraphics[height=45mm]{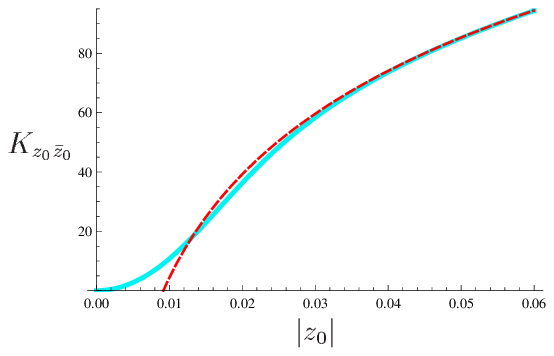} & \hs{20}
\includegraphics[height=50mm]{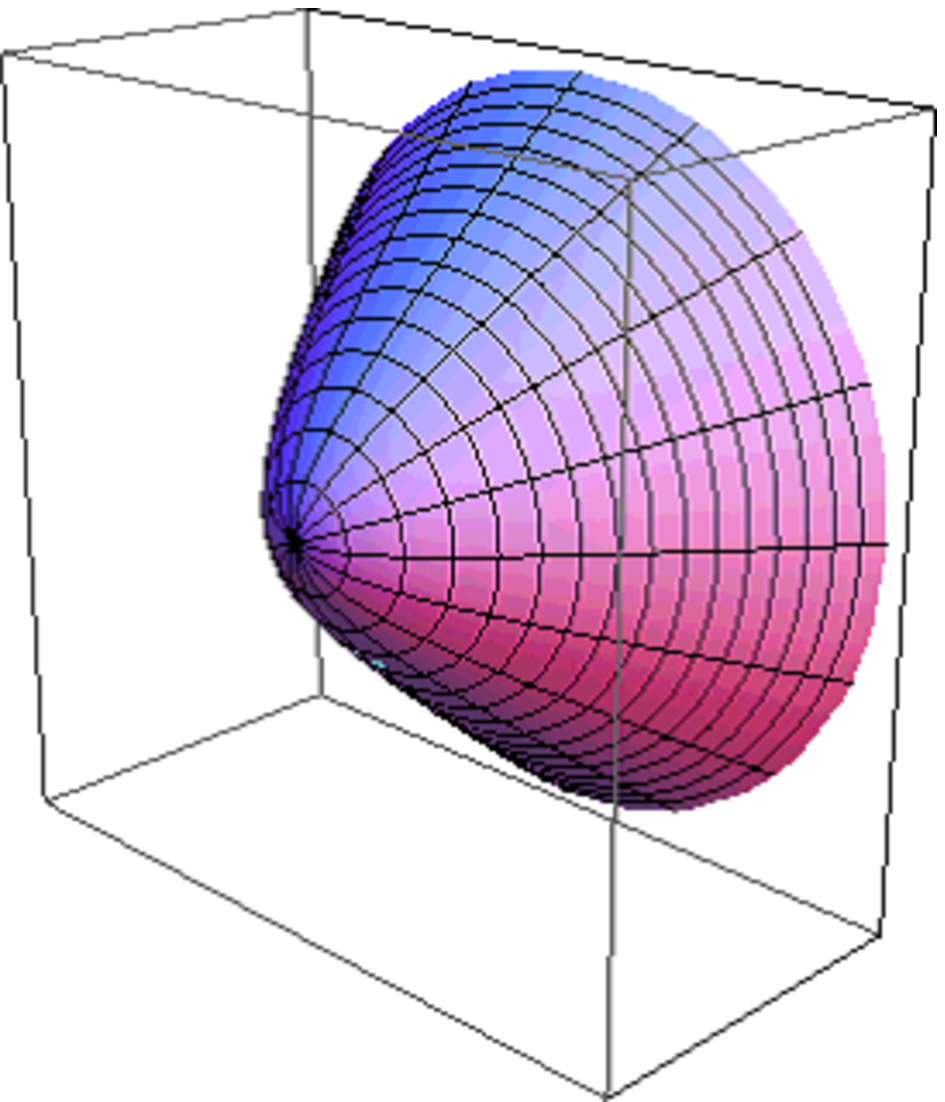} \\
(a) & (b) 
\end{tabular}
\caption{A metric of the moduli space for $c=1,~m=1,~v=e^8$. (a) Numerically calculated metric (solid line) and the asymptotic metric $K_{z_0 \bar z_0} \approx \frac{8 \pi c}{m} \log |4vz_0^2|~~(|vz_0^2|\rightarrow \infty)$ (dashed line). (b) The moduli space isometrically embedded into three dimensional Euclidean space $\mathbf R^3$.}
\label{fig:metric}
\end{center}
\end{figure}

\subsection{Numbers of vortices: (\mbox{\boldmath $n$},\,0,\,\mbox{\boldmath $n$})}\label{sec:202}
In Sec.\,\ref{sec:normalizable} we have seen that the moduli parameters $v_A~(A=1,\cdots,\NF)$ correspond to non-normalizable zero modes if there exist the same number of vortices in each vacuum region. In fact, this is not necessarily the case if there exist different numbers of vortices in each vacuum region. 
The simplest such example is the configuration described by the following moduli matrix 
\beq
H_0 = (z^n,\,v,\,z^n). 
\eeq
The K\"ahler metric for the moduli parameter $v$ is finite for $n \geq 2$. 
\begin{figure}
\begin{center}
\begin{tabular}{cccc}
\includegraphics[width=40mm]{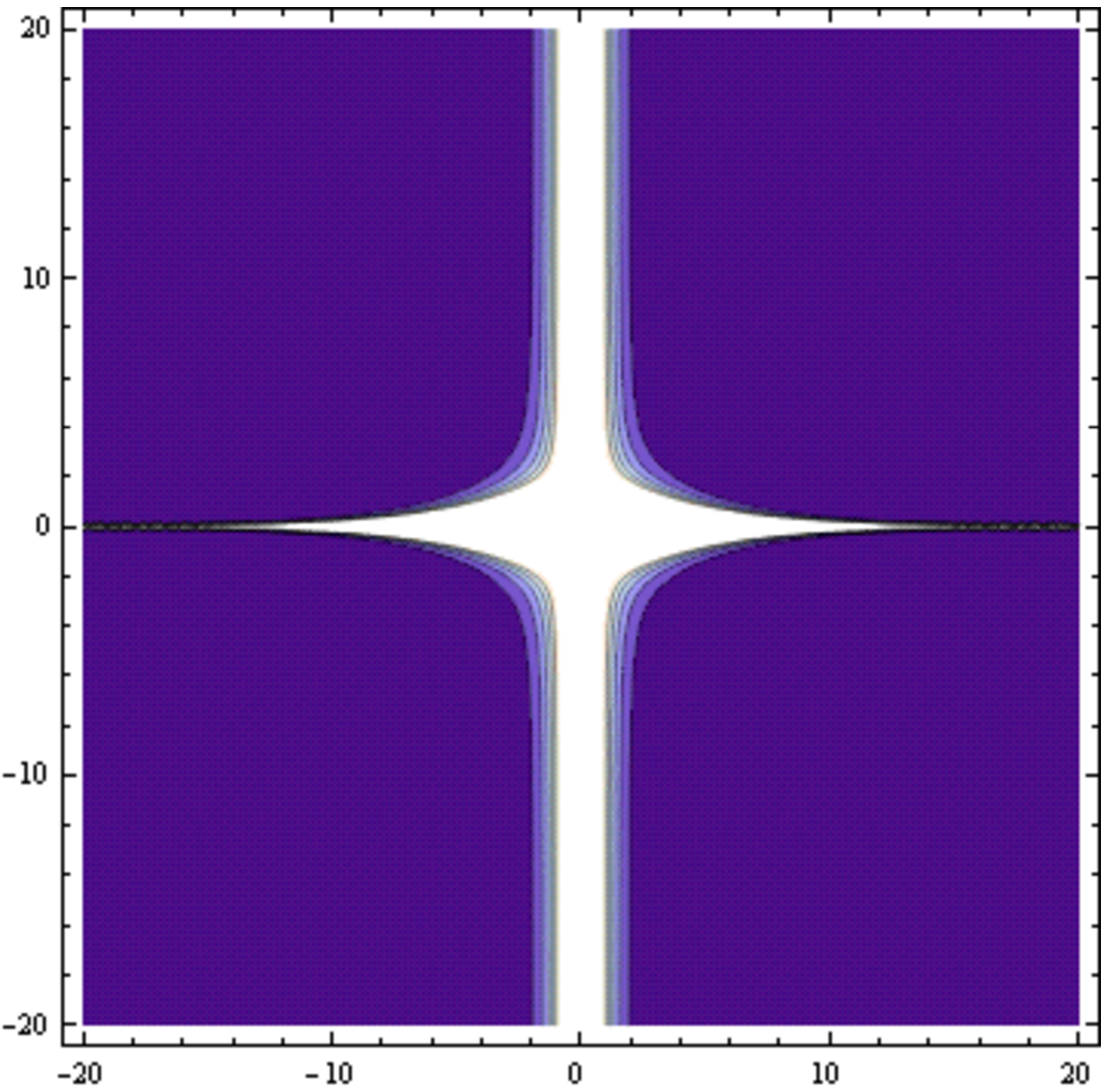} & 
\includegraphics[width=40mm]{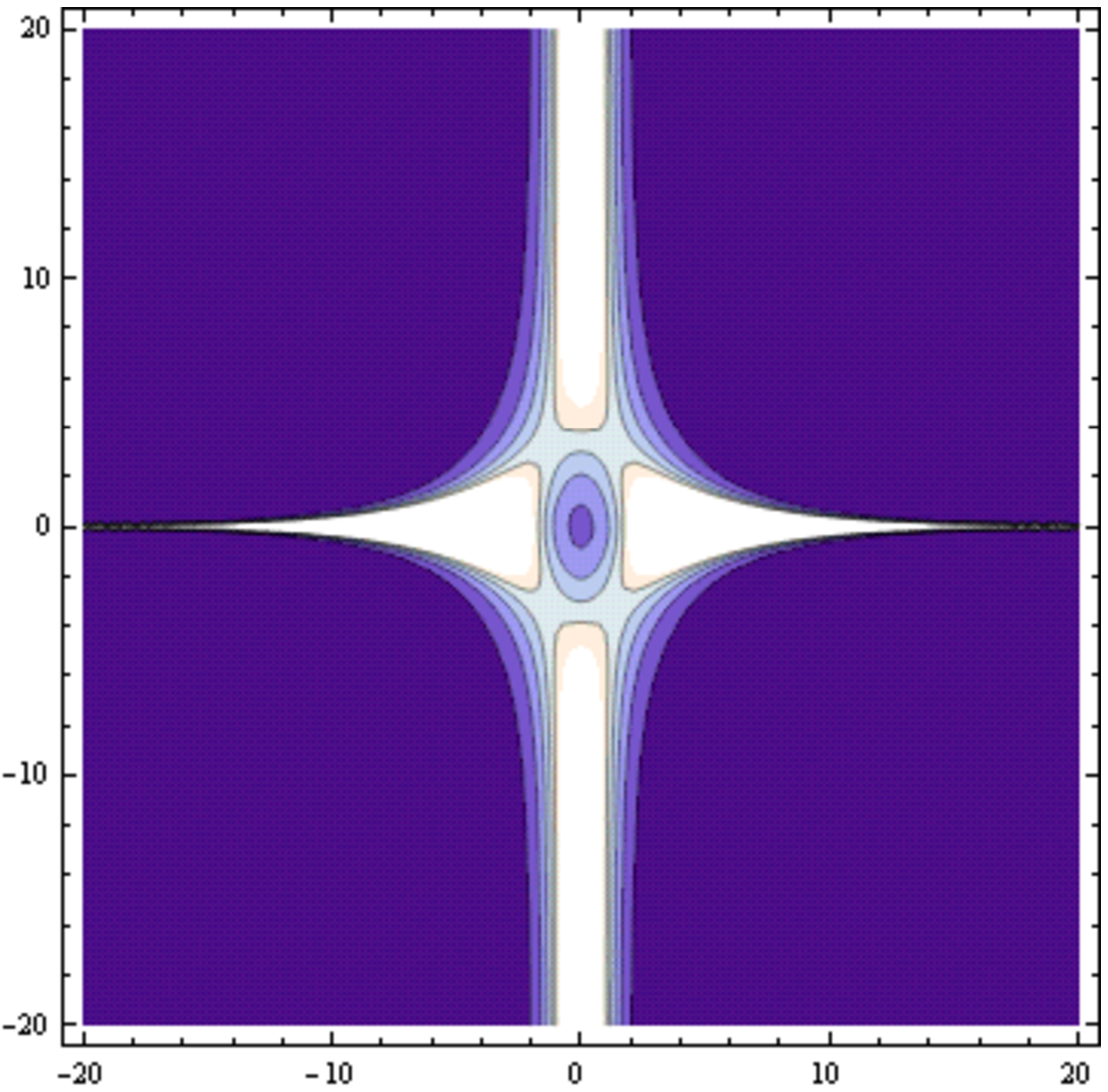} &
\includegraphics[width=40mm]{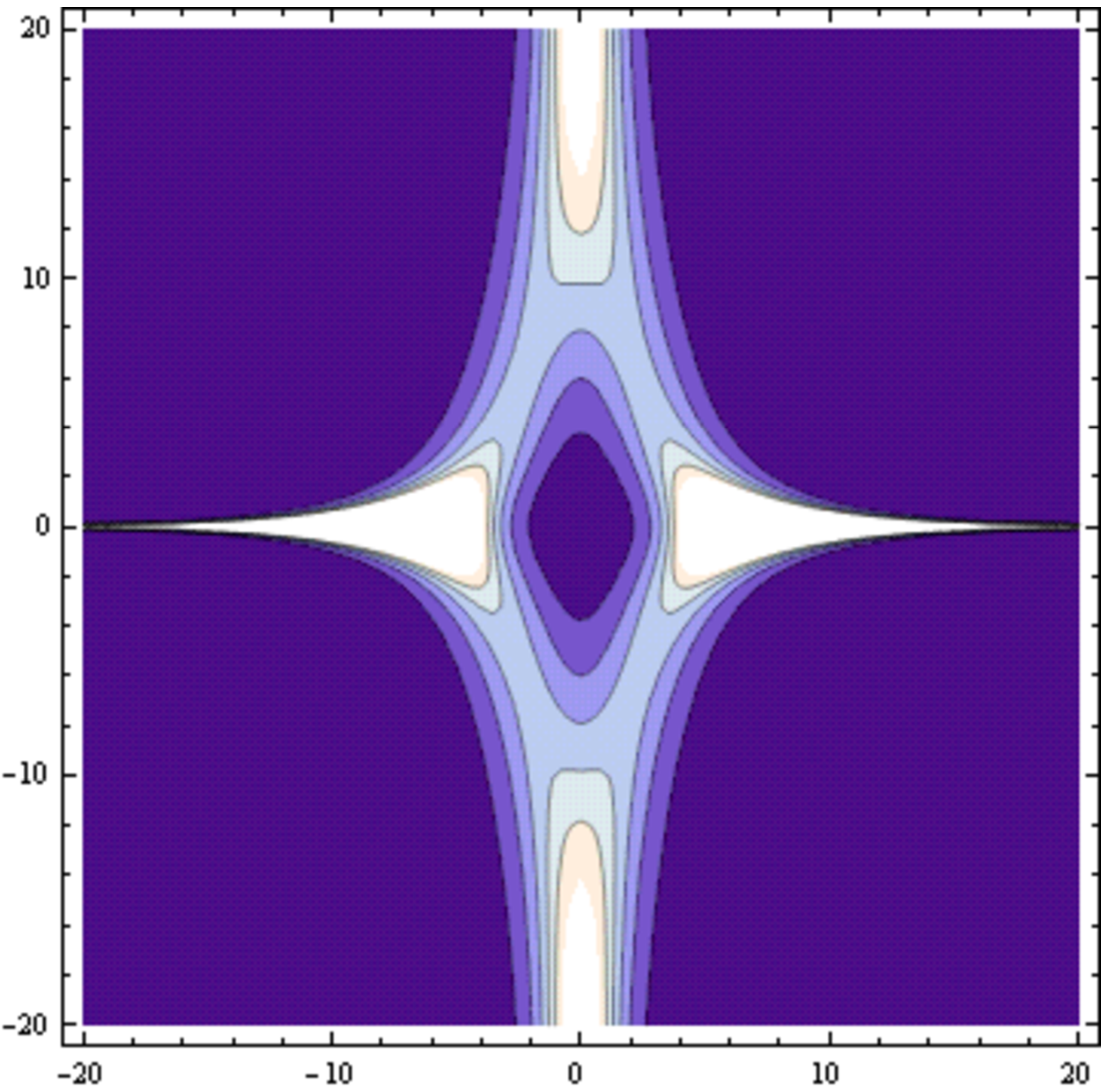} &
\includegraphics[width=40mm]{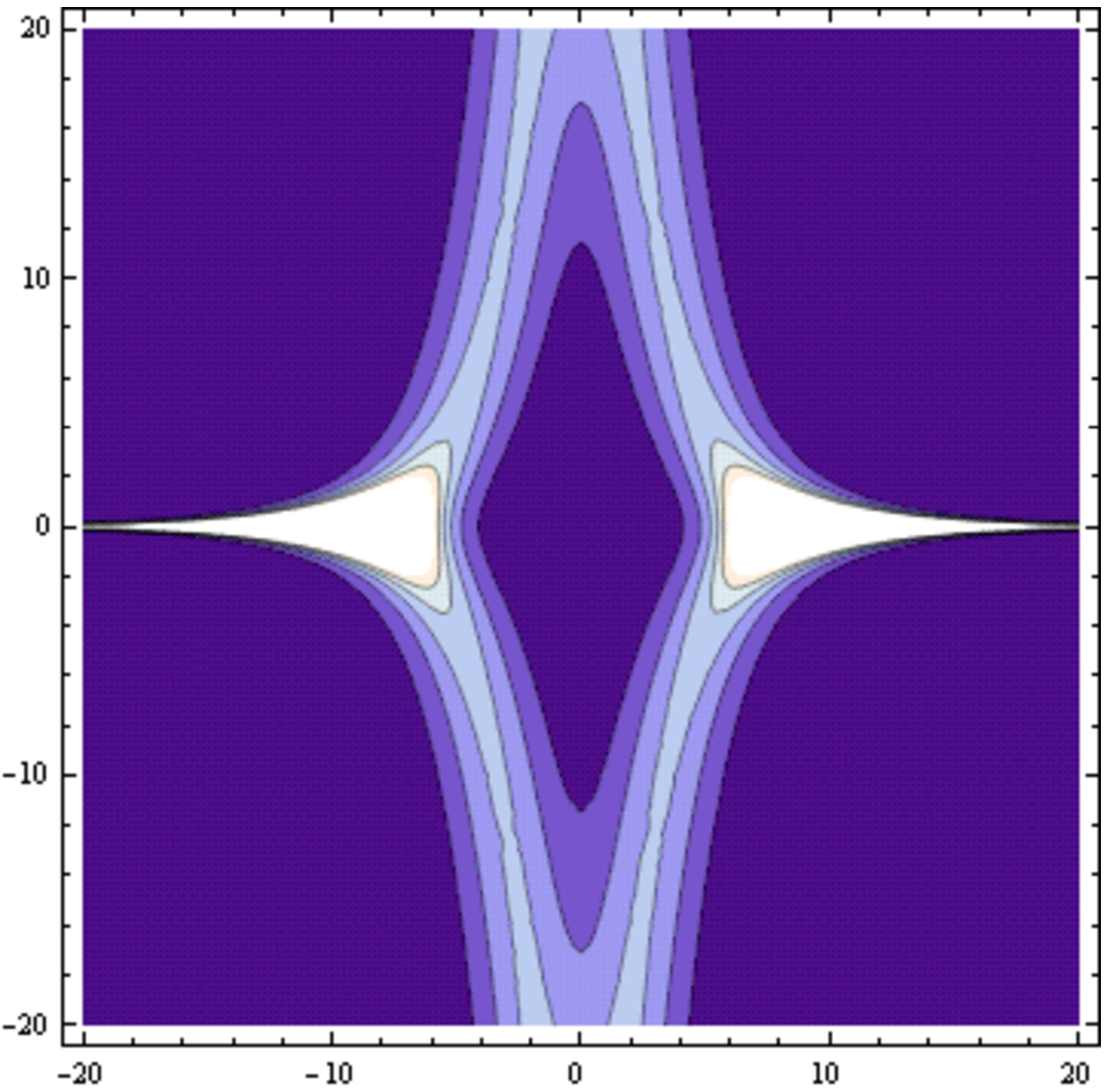} \\
(a) $v = 50$ & (b) $v = 200$ & (c) $v = 800$ & (d) $v = 3200$
\end{tabular}
\caption{The energy densities in a plane containing vortices in the 
strong coupling limit $g\to \infty$ with $m=1,~c=1,~n=2$. 
Vertical lines are walls and horizontal lines are vortices. }
\label{fig:202}
\end{center}
\end{figure}
The relative distance of two walls are determined from 
Eq.\,(\ref{eq:log-bend}) as 
$\Delta X(z,\bar z) = \frac{4}{m} \log |v/z^n|$. 
Therefore, in the region $|z| > |v|^{\frac{1}{n}}$, 
two walls are compressed into one wall located at $x_3=0$ 
and its position is unchanged under the variation of the 
moduli parameter $v\rightarrow v+\delta v$. 
Several plots of the energy densities are shown in Fig.\,\ref{fig:202}. 
The K\"ahler metric for the moduli parameter $v$ 
is given in the strong coupling limit by 
\beq
K_{v \bar v} ~=~ c \int d^3 x \frac{2 |z|^{2n} 
\cosh(m x_3)}{(2 |z|^{2n} \cosh(mx_3) + |v|^2)^2} 
~=~ \frac{\pi^2 c}{2 n^2 m} \frac{|v|^{\frac{2}{n}-2}}{ \sin(\pi/n)} 
\frac{\Gamma(\frac{1}{2n})^2}{\Gamma(\frac{1}{n})}.
\label{eq:metric202}
\eeq
In terms of the coordinate $u \equiv v^{\frac{1}{n}}$, 
the metric can be written as
\beq
ds^2 ~=~ 2 K_{v \bar v} dv d \bar v 
~=~ \frac{\pi^2 c}{m\sin(\pi/n)} 
\frac{\Gamma(\frac{1}{2n})^2}{\Gamma(\frac{1}{n})} du d \bar u.
\eeq
The moduli space is a cone $\mathbf C / \mathbf Z_n$ and 
has a singularity at $v=0$. 
In the limit $v \rightarrow 0$, the moduli matrix can be factorized as
\beq
H_0 = (z^{2n},\,v,\,z^{2n}) \rightarrow z^{2n} (1,\,0,\,1).
\eeq
This indicates the appearance of ANO vortices in the limit 
$v \rightarrow 0$. The existence of the singularity in the 
moduli space reflects the fact that the vortices become ANO 
vortices which are singular in the strong coupling limit 
$g \rightarrow \infty$.

\section{Vortex dynamics in a dual effective theory on walls}\label{sec:particle}

So far, we have calculated 
the metric on the 1/4 BPS moduli space and 
investigated the dynamics of vortices suspended 
between the domain walls, 
using the moduli space approximation. 
Now let us obtain the vortex dynamics from another point of view. 

\subsection{General formalism}\label{sec:general}
Let us first consider the single vortex 
ending on the single domain wall 
in the minimal Abelian-Higgs model with $\NF=2$ and 
see how the vortex ending on the wall 
appear in the effective theory on the domain wall worldvolume. 
The 1/2 BPS domain wall is described by 
a single complex parameter 
$\phi = e^{\Delta m X + i \sigma} \in {\bf C}^* 
\simeq {\bf C} -\{0\} \simeq {\bf R}\times S^1$ 
in 
the moduli matrix $H_0 = (h_1,h_2) \simeq (1,\phi)$, 
see Eq.\,(\ref{eq:minimal_wall}). 
The real part $X$ corresponds to the position of the domain wall, 
see Eq.\,(\ref{eq:weight_minomal}), 
and $\sigma$ is its phase. 
The effective theory on the 
wall 
turned 
out to be a free theory 
\beq
\mathcal L_{\rm w} = \frac{c \Delta m}{2} 
\left[ (\p_\mu X)^2 + \frac{1}{\Delta m^2} (\p_\mu \sigma)^2 \right]
\eeq
via the generic expression Eq.\,(\ref{eq:kahler-pot}) 
\cite{Eto:2006uw,Eto:2008dm} 
\beq
K_{\rm w} = \frac{c}{4\Delta m} \left(\log |\phi|^2 \right)^2
\label{eq:kahler_minimal}
\eeq
The moduli matrix given in Eq.\,(\ref{eq:minimal_wall-vortex}) 
tells us that we should identify 
the vortex ending on the wall at $z=z_0$ 
as the following configuration 
\beq
\phi(z,\bar z) = e^{\Delta m X(z,\bar z) +i\sigma(z,\bar z)}
= e^{\Delta m X_0 + i \sigma_0} \frac{(z-z_0)}{L},
\eeq
where 
we have introduced a ``boundary" at $|z| = L \gg |z_0|$ 
in the $z$-plane for later convenience. 
The parameter $L$ plays the role of the cutoff 
for the IR divergence of the non-normalizable modes. 
The constants $X_0$ and $\sigma_0$ respectively 
represent the position and the phase of 
the background domain wall at $z = L + z_0$. 
Notice that under the identification we have added two points 
$\phi = 0,\,\infty~(X=\pm \infty)$ to ${\bf C}^*$ 
resulting in the target space ${\bf C}P^1$. 
In this sense, the above realization of the vortex 
is thought of as the 1/2 BPS lump 
on the domain wall effective action.\footnote{
The BPS equation is $\bar \p_z \phi = 0$. 
{The solution of this BPS equation satisfies the equation 
of motion with a source term 
$\p_{\bar z} \p_z (X + i \sigma/\Delta m) = 
\pm \frac{\pi}{2\Delta m} \delta^2(z-z_0)$ 
corresponding to 
the addition of the points $X = \pm \infty$.}}
Let us set $X_0=\frac{1}{\Delta m} \log L,\,\sigma_0 = 0$ and $z_0=0$
 in the following for simplicity. 
The vortex causes the logarithmic bending of the domain wall 
\beq
X = \frac{1}{\Delta m} \log |\phi| = \frac{1}{\Delta m} \log |z|.
\label{eq:log_bending}
\eeq
This is consistent with the bulk point of view. 
We also find that if we walk around the vortex in the $z$-plane, 
the phase of the domain wall also winds once 
\beq
\sigma = \theta,\qquad (z = x_1 + ix_2 = r e^{i\theta}).
\label{eq:phase_winding}
\eeq

Eq.\,(\ref{eq:kahler_minimal}) is the free theory 
of the real scalar field $X$ and 
the periodic field $\sigma \in S^1$ in 2+1 dimensions. 
The phase degree of freedom of the domain wall 
$\sigma(x^\mu) \in S^1$ in 2+1 dimensional worldvolume 
can be dualized into an Abelian gauge field as \cite{Shifman:2002jm} 
\beq
F_{\mu \nu}=\frac{e^2}{2 \pi}\epsilon_{\mu \nu \rho}\partial^{\rho}\sigma,
\qquad
e^2 \equiv \frac{4 \pi^2 c}{\Delta m}.
\eeq
If we also rescale the scalar field $X(x^\mu)$ as 
\beq
\log \phi = \Delta m X + i\sigma = \frac{\Delta m}{2\pi c} \Phi + i \sigma
= \frac{2 \pi}{e^2} \Phi + i \sigma,
\label{eq:scale_scalar}
\eeq
the effective Lagrangian has the simple form
\beq
{\cal L}_{\rm w} = \left( \frac{1}{2e^2}\partial_\mu \Phi \partial^\mu \Phi
-\frac{1}{4e^2}F_{\mu \nu}F^{\mu \nu} \right).
\label{eq:dual_wall_1}
\eeq
In terms of the dual gauge field, 
the phase winding (\ref{eq:phase_winding}) 
corresponds to the electric field 
for a static source with unit electric charge 
\beq
F_{0r}=E_r=\frac{e^2}{2 \pi}\,\frac{1}{r},
\eeq
and the electrostatic potential is given by 
\beq
A_0=-\frac{e^2}{2 \pi} \log |z|.
\label{eq:lump_gauge}
\eeq
Furthermore, the logarithmic bending (\ref{eq:log_bending}) 
yields the scalar potential  
\beq
\Phi=\frac{e^2}{2 \pi} \log |z|.
\label{eq:lump_scalar}
\eeq
Therefore, the vortex at rest can be viewed 
as a charged particle in the effective theory, 
which gives the scalar field (\ref{eq:lump_scalar}) 
and the electric field (\ref{eq:lump_gauge}). 
When the electric charge (vortex) 
moves at a constant velocity $u=v^1+iv^2$,
($u \equiv \dot z_0$), the potentials 
are Lorentz boosted as, 
\beq
\Phi &=& \phantom{-} \frac{e^2}{2\pi}
\log |L_u(z-z_0)|, \label{eq:scalar_v} \\
A_\mu&=&
-\frac{e^2}{2\pi}\frac{v_\mu}{\sqrt{1-|u|^2}} 
\log|L_u(z-z_0)|, \label{eq:temporal_v}
\eeq
with $v^\mu=(1,v^1,v^2)$ and 
\begin{eqnarray}
 L_u(z-z_0)\equiv
\frac{1}{2}\left[ \left(\frac{1}{\sqrt{1-|u|^2}}+1\right)
(z-z_0) + \frac{u}{\bar u} \left(\frac{1}{\sqrt{1-|u|^2}}-1\right)
 (\bar z-\bar z_0) \right].
\end{eqnarray}
We can confirm that these configurations satisfy equations of motion 
with the 
moving
charged particle,
\begin{eqnarray}
 \partial_\mu\partial^\mu\Phi&=&-e^2\delta^2(z-z_0)\sqrt{1-|u|^2}\nn
\partial_\nu F^{\mu\nu}&=&-e^2\delta^2(z-z_0)v^\mu.
\end{eqnarray}
Notice that the vortex ending on the wall 
from the other side 
corresponds to 
the moduli matrix 
$H_0 = (z,1) \sim (1,1/z)$, namely 
$X = - \frac{1}{\Delta m} \log |z|$, $\sigma = - \theta$. 
This implies that it generates the potentials with 
the sign opposite to that 
in Eq.\,(\ref{eq:scalar_v}) and Eq.\,(\ref{eq:temporal_v}). 
Furthermore, if the BPS vortex is replaced 
by an anti-BPS vortex, we find that only the 
sign 
of the phase $\sigma$ flips without any change to $X$. 
We consider only BPS vortices in the following discussion.

\medskip

We can extend this analysis to the case of multiple domain walls.
When all the domain walls are well separated,
we can assume that the dual theory\footnote{
In fact, we can obtain the dual $U(1)^N$ gauge theory by
dualizing $N$ compact scalar fields $\sigma^A$, see Appendix \ref{appendix:D}
}
is a $U(1)^N$ gauge theory 
with $N$ neutral Higgs fields $\Phi^A$ ($A=1,\cdots,N$) 
\begin{eqnarray}
 {\cal L}_{\rm w}=
\sum_{A=1}^N\left[\frac{1}{2e^2_A}\partial_\mu \Phi^A \partial^\mu \Phi^A
-\frac{1}{4e^2_A}F_{\mu \nu}^AF^{A\,\mu \nu}\right]
\label{eq:multi_dual}
\end{eqnarray}
Here 
the scalar fields $\Phi^A$ ($A=1,\cdots,N$) are identified with the position of the domain wall between the vacua 
$\left<A\right>$ and $\left<A+1\right>$ 
as $X^A = \frac{1}{2\pi c}\Phi^A$. 
The constants $e_A~(A=1,\cdots,N)$ are the gauge coupling constants on the worldvolume of the domain wall between vacua $\left<A\right>$ and $\left<A+1\right>$,
\beq
e_A^2 = \frac{4 \pi^2 c}{\Delta m_A}.
\eeq
The $i$-th vortex living in vacuum $\langle B \rangle$ 
positioned at $z_{\left<B\right>i}$ with a velocity 
$u_{\left<B\right>i} \equiv \dot{z}_{\left<B\right>i}$ yields 
the scalar field and the electric field on the worldvolume of 
neighboring 
domain walls\footnote{
Note that $N$ domain walls divide the 3-dimensional space into 
$N+1$ different vacuum regions. 
We use indices $A$ and $B$ to label both the domain walls and vacuum regions: 
the indices $A$ and $B$ run from 1 to $N$ for domain walls, and the label $\langle A \rangle$ and $\langle B \rangle$ run from 1 to $N+1$ for vacuum regions, see Fig.~\ref{fig:dbs}.
}
\beq
(\Phi^A)_{(B,i)} &=& 
\phantom{-} (\delta_{A+1,B} - \delta_{AB}) \, \frac{e^2_A}{2\pi} \,
G(u_{\left<B\right>i} \,;\, z-z_{\left<B\right>i}), \label{eq:Phi} \\
(A_\mu^A)_{(B,i)} &=&
- (\delta_{A+1,B} - \delta_{AB}) \, \frac{e^2_A}{2\pi} \, 
\frac{v_{\mu\left<B\right>i}}{\sqrt{1-|u_{\left<B\right>i}|^2}}
G(u_{\left<B\right>i} \,;\, z-z_{\left<B\right>i}). \label{eq:A}
\eeq
where $v^\mu_{\left<B\right>i}=(1 ,\, {\rm Re}[u_{\left<B\right>i}],\, {\rm Im}[u_{\left<B\right>i}])$ 
and $G$ is the Green's function given by 
\begin{eqnarray}
G(u_{\left<B\right>i} \,;\, z-z_{\left<B\right>i})=\log|L_{u_{\left<B\right>i}}(z-z_{\left<B\right>i})| 
-\log L+f(u_{\left<B\right>i}), \hs{10} f(u) \approx \mathcal O(u^2).
\end{eqnarray}
Here we have added the last two terms 
so that the Green's function vanishes at the boundary $|z|=L$. 

Now let us assume that 
the dynamics of the $i$-th vortex living in vacuum 
$\langle A \rangle$ 
can be regarded as that of
an electric charge moving in the background potential 
produced by the other vortices. 
We shall suppose that the effect of the Lorentz scalar 
potential $\Phi^B$ ($B \neq A$) is to change the rest mass 
of the vortex ending 
on
the domain wall. 
This is consistent with the fact that the vortices 
cause the logarithmic bending of 
the domain wall and it 
leads to 
the change of the length of the 
other vortices ending on the domain wall, 
and thus the masses of the vortices.
With these assumptions, 
the Lagrangian for the $i$-th 
vortex 
in vacuum $\langle A \rangle$ 
is given by 
that for the charged particle 
\cite{Manton:1985hs,Manton:2004tk} 
\beq
L_i^{\langle A \rangle} &=& \sum_B (\delta_{A,B} - \delta_{A-1,B}) 
\left( - \widetilde \Phi^B \sqrt{1-|u_{\left<A\right>i}|^2} - \widetilde A_\mu^B v^\mu_{\left<A\right>i} \right) \nonumber \\
&\approx& \sum_B (\delta_{A,B} - \delta_{A-1,B}) 
\left( - \widetilde \Phi^B + \frac{\widetilde \Phi^B}{2} 
|u_{\left<A\right>i}|^2 - \widetilde A_\mu^B v^\mu_{\left<A\right>i} \right),
\label{eq:charge_lag}
\eeq
where $\widetilde \Phi^B,\, \widetilde A_0^B,\, \widetilde {\mathbf A}^B$ are the values of the fields produced by the other particles at the location of the particle $z=z_{\left<A\right>i}$
\beq
\widetilde \Phi^B \equiv \langle \Phi^B \rangle + \sum_{(C,j)} (\Phi^B)_{(C,j)} \big|_{z=z_{\left<A\right>i}}, \hs{5} 
\widetilde A_\mu^B = \sum_{(C,j)} (A_\mu^B)_{(C,j)} \big|_{z=z_{\left<A\right>i}}.
\eeq
Here we implicitly assume that the fields 
due to the particle in problem 
($i$-th vortex in vacuum $\langle A \rangle$) 
is excluded in the sum
and $\langle \Phi\rangle$ is VEV of the scalar field at 
the boundary $|z|=L$. 
Let us note that Eq.\,(\ref{eq:charge_lag}) gives the Lagrangian 
for the particle A under the background potential produced 
by all the other particles. 
To obtain the total Lagrangian for all particles including 
mutual interactions, we need to sum over the interaction 
terms only once for each pair of particles. 
Substituting Eqs.\,(\ref{eq:Phi}) and  (\ref{eq:A}) 
into Eq.\,(\ref{eq:charge_lag}) 
and summing up the kinetic terms and the interaction terms 
from all pairs of particles, 
we obtain the effective Lagrangian as 
\beq
L_{\rm eff} = \sum_{(A,i)} \frac{\langle \Phi^{A-1} \rangle - \langle \Phi^A
\rangle}{2} |u_{\left<A\right>i}|^2 + 
\underset{(A,i),\,(B,j)}{{\sum}'} 
\left( \frac{C_{AB}}{2} 
 \log \frac{|z_{\left<A\right>i} -
z_{\left<B\right>j}|}{L}\right) |u_{\left<A\right>i} - u_{\left<B\right>j}|^2, 
\label{eq:charge_lag2}
\eeq
\vspace{-0.5cm}
\beq
C_{AB} &\equiv& 2 \pi c \left[ \left( \frac{1}{\Delta m_A} + \frac{1}{\Delta m_{A-1}} \right) \delta_{AB} - \frac{1}{\Delta m_A} \delta_{A,B-1} - \frac{1}{\Delta m_B} \delta_{A-1,B} \right]. 
\eeq
where $\sum'$ means that the sum is taken 
only once for 
each 
pair of the index sets 
$(A,i)$ and $(B,j)$ such that $(A,i) \not = (B,j)$. 

The general Lagrangian Eq.(\ref{eq:charge_lag2})
can be 
interpreted as an asymptotic 
effective Lagrangian for the vortices between the domain walls. 
The dynamics of the vortices are well described 
by the Lagrangian Eq.\,(\ref{eq:charge_lag2}) 
when the domain walls are well-separated in $x_3$-direction 
and the vortices are well-separated in $z$-plane. 
We will compare it with the results obtained
in Sec.\ref{sec:dynamics} by taking several examples in the following.
The general form itself also has some good properties. 
One is that the sigma model metric is K\"ahler as shown in
Appendix \ref{appendix:C}.
The other is that the IR divergences $\log L$
in Eq.(\ref{eq:charge_lag2}) are completely 
canceled out when $\sum_{i} u_{\left<A\right>i} = 0$ and 
$u_{\left<1\right>i}=u_{\left<N+1\right>i}=0$,
that is, center of mass of vortices in each vacuum and
semi-infinite vortices in vacuum $\langle 1 \rangle$
and $\langle N+1 \rangle$ do not move.
This is consistent with the argument of normalizability
in Sec.~\ref{sec:normalizable}.
Furthermore, it
correctly reproduces the IR divergence in Eq.(\ref{eq:com_vor}) 
for the center of mass of vortices
in each vacuum.

Before concluding this subsection, let us 
comment on the effect of
bulk coupling constant $g$ 
which we have ignored in the discussion above. 
If the coupling constant is finite, 
we should take into account the boojum charges, 
which have negative contributions to the energy 
corresponding to the binding energy between vortices 
and domain walls. 
Since vortices become lighter by the amount of boojum charges,
the kinetic terms in 
the effective Lagrangian (\ref{eq:charge_lag2}) 
should be replaced as 
\beq
\frac{M_{{\rm v}\langle A \rangle}}{2} |u_{\left<A\right>i}|^2 = \frac{\langle \Phi^{A-1} \rangle - \langle \Phi^A
\rangle}{2} |u_{\left<A\right>i}|^2 ~\to~
\frac{\langle \Phi^{A-1} \rangle - \langle \Phi^A
\rangle+B_g^{A-1}+B_g^{A}}{2} |u_{\left<A\right>i}|^2.
\label{eq:shift}
\eeq
Here $B_g^A$ is the boojum charge between $A$-th domain wall and
a vortex living in vacuum $\langle A \rangle$
\beq
B_g^A=-\frac{2 \pi \Delta m_A}{g^2}<0.
\label{eq:boojum}
\eeq
Another interpretation of the shifts of the vortex masses Eq.\,(\ref{eq:shift}) is given as follows. 
If the gauge coupling constant $g$ is finite, $A$-th domain 
wall has its typical width 
\cite{Shifman:2002jm}, \cite{Eto:2006pg} 
\beq
d_A \equiv \frac{2 \Delta m_A}{g^2 c} = - \frac{B_g^A}{\pi c}.
\eeq
Since the length of the vortices $l_{\langle A \rangle}$ is measured as the distance between the surfaces of $A$-th and $(A-1)$-th domain walls, the mass of the vortices $M_{{\rm v}\langle A \rangle} \equiv 2\pi c \, l_{\langle A \rangle}$ is given by 
\beq
M_{{\rm v}\langle A \rangle} ~=~ 2 \pi c \, l_{\langle A \rangle} &=& 2 \pi c \left( \frac{\langle \Phi^{A-1} \rangle - \langle \Phi^A \rangle}{2\pi c} - \frac{d_{A-1}+d_A}{2} \right) \notag \\
&=& \langle \Phi^{A-1} \rangle - \langle \Phi^A \rangle+B_g^{A-1}+B_g^{A}.
\eeq
Here $(\langle \Phi^{A-1} \rangle - \langle \Phi^A \rangle)/2 \pi c$ is the distance between the middle points of $A$-th and $(A-1)$-th domain walls. 
For more details, see Appendix \ref{appendix:D}.

\subsection{Numbers of vortices : (1,1,1)}
Let us consider the Abelian gauge theory with three flavors 
and assume that each vacuum has a single vortex. 
We have already obtained the effective Lagrangian 
for the middle vortex in Sec.\,\ref{sec:111}. 
We use the same mass parameters given as 
$M={\rm diag\,}(\frac{m}{2},0,-\frac{m}{2})$, 
and set the outer vortices at 
$z_{\left<1\right>1}=z_{\left<3\right>1}=0$ as before. 
The gauge couplings in the effective theory 
on the domain walls are given by 
\beq
e^2\equiv e_1^2=e_2^2=\frac{8\pi^2 c}{m}.
\eeq
Since the first domain wall is positioned at $X^1 = \frac{2}{m} \log |v|$
and the second domain wall is at $X^2 = - \frac{2}{m} \log |v|$,
the vacuum expectation value of the adjoint scalar field is
\beq
\langle \Phi^1 \rangle = \frac{e^2}{2 \pi} \log |v|, \hs{5} \langle \Phi^2 \rangle = -\frac{e^2}{2 \pi} \log |v|.
\eeq
If we substitute these to Eq.(\ref{eq:charge_lag2}),
we obtain the effective Lagrangian for the middle vortex
\beq
L_{\rm eff} = \pi c \left(
\frac{4}{m} \log |v| - \frac{4}{m} \log \frac{|z_0|}{L} \right) |\dot{z}_0|^2.
\eeq
This result coincides with the asymptotic metric in Eq.(\ref{eq:large_metric_111}).

\subsection{Numbers of vortices : (2,2,2)}
Next let us consider the case 
in which each vacuum has a pair of vortices. 
We have already obtained the effective Lagrangian 
for the relative motion of the middle vortices in Sec.\,\ref{sec:222}. 
The gauge couplings and the vacuum expectation value of the scalar field 
are the same as in the previous example. 
Let us set the vortices in vacuum $\langle 1 \rangle$ at 
$z_{\left<1\right>1}=z_{\left<1\right>2}=z_1$ 
and the vortices in vacuum $\langle 3 \rangle$ at 
$z_{\left<3\right>1}=z_{\left<3\right>2}=-z_1$ as in Sec.\,\ref{sec:222}. 
Since we are interested in the relative motion 
of the vortices in vacuum $\langle 2 \rangle$, 
we take $z_{\left<2\right>1}=-z_{\left<2\right>2}=z_0$. 
The effective Lagrangian is given by 
\beq
L_{\rm eff} = \pi c \left( \frac{8}{m} \log |v| - \frac{8}{m} \log |z_0-z_1| - \frac{8}{m} \log |z_0+z_1| + \frac{16}{m} \log |2z_0| \right) |\dot{z}_0|^2,
\eeq
{Note that the divergence terms are exactly canceled out.}
The second term comes from the interactions with the vortices 
in vacuum $\langle 1 \rangle$, and the third from the vortices 
in vacuum $\langle 3 \rangle$. 
The last term represents the interactions of 
two vortices in vacuum $\langle 2 \rangle$. 
The effective Lagrangian for the relative motion 
of two vortices can be obtained as 
\beq
L_{\rm eff} =
\left[ \frac{8 \pi c}{m} \log |v| + \frac{16 \pi c}{m} \log 2 + 
\O \left( \left(\frac{z_1}{z_0}\right)^2, \left(\frac{\bar z_1}{\bar z_0}\right)^2 \right) \right] |\dot{z}_0|^2.
\eeq
This coincides with the previous result Eq.(\ref{eq:large_metric_222}). 

\subsection{Numbers of vortices : (0,2,0)}
Next let us consider the case 
in which only the middle vacuum has a pair of vortices. 
We have already obtained the effective Lagrangian 
for the relative motion of the vortices in Sec.\,\ref{sec:020}. 
The gauge couplings are 
the 
same as in the previous examples. 
In this case, walls logarithmically bend even at the boundary and 
the VEV of the scalar fields 
depend on 
the cutoff $L$ as,
\beq
\langle \Phi^1 \rangle = \frac{e^2}{2 \pi}\log (|v| L^2), \hs{5} 
\langle \Phi^2 \rangle = - \frac{e^2}{2 \pi}\log (|v| L^2).
\eeq
We will find that this vacuum expectation value 
gives the correct answer in the following. 
Since we are interested in the relative motion of the vortices, 
we take $z_{\left<2\right>1}=-z_{\left<2\right>2}=z_0$. 
The effective Lagrangian for the first vortex 
in vacuum $\langle 2 \rangle$ is given by 
\beq
L_{\rm eff} = \pi c \left( \frac{8}{m} \log (|v|L^2) + \frac{16}{m} 
\log \frac{|2z_0|}{L} \right) |\dot{z}_0|^2,
\eeq
The second term comes from the interaction of two vortices
and the cutoff dependence vanishes again. 
The effective Lagrangian for the relative motion 
of two vortices can be obtained as 
\beq
L_{\rm eff} = \left( \frac{8 \pi c}{m} \log
|v z_0^2| + \frac{16 \pi c}{m} \log 2 \right) |\dot{z}_0|^2.
\eeq
This coincides with the previous result 
in 
Eq.(\ref{eq:asym_metric_020}).

In summary, this method correctly reproduces 
the asymptotic metric on the moduli space. 
If the domain walls are well-separated in $x_3$-direction, 
and the 
vortices 
are 
well-separated 
from other vortices in $z$-plane, 
we can trust the Lagrangian in Eq.(\ref{eq:charge_lag2}). 

\section{Conclusion and Discussion}\label{sec:conclusion}

{
We have investigated dynamics of the 1/4 BPS solitons in ${\cal N}=2$ supersymmetric
$U(\NC)$ gauge theory with $\NF$ hypermultiplets in 3+1 dimensions.
The 1/4 BPS solitons are 
composite 
of different solitons: monopoles, boojums, 
vortex strings and parallel domain walls.
Neither the vortex strings of infinite length 
nor the domain walls can move because of 
their 
infinite masses.
On the other hand, the monopoles 
pieced by 
the vortices and the vortices of finite length
suspended  
between 
the domain walls may move.
We have considered 
two different methods 
to study this interesting dynamics of solitons 
;
the one is the 
so-called moduli approximation {\it \`a la} 
Manton and the other is 
the charged particle approximation for 
string endpoints in the wall effective action. 
After reviewing the moduli space of the 1/4 BPS states 
in Sec.~\ref{sec:review},
we have derived the formal low energy effective action 
which describes 
slow-move soliton dynamics and have specified which 
moduli parameters are
normalizable and which are not in Sec.~\ref{sec:lagrangian}.
Since we are primarily interested in 
the 1/4 BPS dynamics in the $U(1)$ gauge theory, 
we have no monopoles. 
Clearly only the vortices with 
finite length can have finite masses and have a chance 
to give a normalizable mode. 
In spite of the finite length and mass, 
we have found that the center of masses of the vortices 
in each vacuum are actually non-normalizable. 
In Sec.~\ref{sec:dynamics} we have dealt with several 
examples of (1,1,1), (2,2,2), (0,2,0) and ($n$,0,$n$).
In order to study it analytically, we have taken the strong 
gauge coupling limit where the 
gauge theory reduces to the massive ${\bf C}P^{\NF-1}$ 
nonlinear sigma model.
With the first example, we have seen that the low energy 
effective action can be intuitively understood 
as the normal kinetic energies of domain walls and the vortices. 
We have also found out that the 
origin of the non-normalizability of the middle vortex 
even though its mass is finite.
The (2,2,2) provides us with 
a simple 
example of the 
vortex dynamics. 
The dynamical degree of freedom is only the relative 
position of the vortices in the middle vacuum.
We have studied two situations. 
The first setup is tuned in such a way that all the 
outer semi-infinite vortices are
positioned at the origin of the $z$-plane and the 
center of mass of the middle vortices is put on the origin
as well. 
It turned 
out that the moduli space is
${\bf C}/{\bf Z}_2$ and we fall into its conical singularity 
as the middle vortices goes to the origin.
The next setup is taken so that the outer vortices are 
dislocated from the origin and
are put separately.
This removes the singularity and we have found
the 90 degree scattering for head-on collisions.
The (0,2,0) is the example where the domain walls are not 
asymptotically flat. We have seen
the 90 degree scattering for head-on collision also here.
The ($n$,0,$n$) is completely different from the others. 
There are no dynamical vortices but
there exists one complex parameter associated with the 
middle vacuum which is 
enclosed 
by the adjacent walls.
The metric of the moduli space have been found 
${\bf C}/{\bf Z}_n$, and the conical singularity
reflects that $n$ ANO vortices appear when the middle 
vacuum disappears.
Our last attempt to reveal the dynamics of the vortices 
ending on the domain walls have
been done from the view point of the effective action on 
the host domain walls.
As 
first 
shown in Ref.~\cite{Shifman:2002jm}, 
the effective action on the single domain wall
can be dualized to the $U(1)$ gauge theory with a free 
real scalar in the 2+1 dimensions.
The vortex ending on the wall can be, then, identified 
with an electrically charged particle.
We have applied the idea for the well separated 
$\NF-1$ domain wall system and
the vortices suspended between them. 
To this end, we have considered 
the $U(1)^{\NF-1}$ gauge theory with $\NF-1$ real scalar 
fields as the dual theory.
Vortices ending on a domain wall from right hand side 
have the opposite $U(1)$ charges
to those ending from the left hand side. 
Our effective action is given in 
Eq.\,(\ref{eq:charge_lag2}). 
It is worth while to mention that our Lagrangian is the 
2+1 dimensional analogue of the Lagrangian given by Manton 
who calculated the velocity dependent interactions 
between well separated BPS monopoles in the 3+1 dimensions 
\cite{Manton:1985hs}.
We have tested our second approach to the case of 
(1,1,1), (2,2,2) and (0,2,0).
It is gratifying that two different methods give us the same 
asymptotic interactions.

It is not easy to construct a string stretched between 
D-branes as a soliton of the Born-Infeld theory. 
On the contrary, our method of the moduli matrix allows us to 
construct easily the configurations of vortex-string stretched 
between walls. 
It is worth emphasizing that 
the dynamics of these composite solitons can be 
analyzed without any logical or practical difficulty 
in our method of the moduli matrix. 

It is an interesting future work to generalize our analysis 
to $U(N_{\rm C})$ gauge theory. 
For instance a characteristic feature of the non-Abelian gauge 
theories is that we can have monopoles (with positive energy contribution) 
rather than boojums (with negative energy contribution). 
It has been found in the case of webs of domain walls 
that the non-Abelian and Abelian junction can be interchanged 
during the course of geodesic motion \cite{Eto:2005cp}. 
A similar dynamical metamorphosis may also be expected in 
the case of wall-vortex-monopole system. 
It is also interesting to further generalize to arbitrary 
gauge groups \cite{Eto:2008yi} 
such as $SO$ gauge group \cite{Ferretti:2007rp}. 

In this paper we have assumed that 
the masses of the Higgs fields (hypermultiplets) 
are non-degenerate and real.
When some masses are degenerate the model enjoys 
non-Abelian flavor symmetry 
and a part of them broken by the wall configurations 
appear in the domain wall effective action 
\cite{Shifman:2003uh,Eto:2005cc,Eto:2008dm}, 
where some modes (called non-Abelian clouds) 
appear between domain walls \cite{Eto:2008dm}  
whereas the rests are localized around each wall as usual.
Accordingly 
non-Abelian vortex-strings \cite{Hanany:2003hp}
can end on 
these non-Abelian domain walls \cite{Shifman:2003uh}.
Classification of possible configurations is still 
lacking in this case. 
In particular non-Abelian semi-local vortex-strings 
\cite{Shifman:2006kd,Eto:2007yv} have not been studied 
in the presence of domain walls. 
For instance (non-)normalizability of orientational zero modes 
is quite non-trivial even in the absence of domain walls; 
they are non-normalizable for a single vortex 
with non-zero size moduli \cite{Shifman:2006kd} 
but become normalizable with vanishing size moduli \cite{Eto:2007yv}. 
Moreover relative orientational moduli of multiple vortices 
are normalizable \cite{Eto:2007yv}. 
Classification of possible configurations and 
dynamics of these configurations should be done 
as a natural extension of this paper.

Stationary time-dependent configurations 
carry a conserved Noether charge.
Such configurations are called dyonic (Q-)solitons.
Dyonic instantons were found  
as an extension of dyons and have been studied by many authors 
\cite{Lambert:1999ua}. 
Dyonic domain walls \cite{Lee:2005sv,Eto:2005sw} and 
dyonic network of domain walls \cite{Eto:2007uc} 
have been studied so far.
Dyonic non-Abelian vortices are also studied recently 
\cite{Collie:2008za}.
Dyonic extension of the wall-vortex system 
is still 1/4 BPS in four space-time dimensions \cite{Eto:2005sw}, 
which can be realized if we introduce 
imaginary masses for the Higgs fields (hypermultiplets) 
and a linear time dependence on corresponding phases
.

Our considerations 
of dynamics are classical so far. 
Quantization of solitons is one of 
interesting future directions. 
First, (semi-classical) first quantization of monopoles
was done by using the moduli space,  
to obtain quantum mechanics on the moduli space
\cite{Gibbons:1986df}. 
One 
should be able to 
generalize this 
even to a composite system. 
For instance by quantizing the sector $(0,2,0)$ of two strings, 
we will obtain quantum scattering of strings. 
It is interesting to compare this with 
a scattering of W-bosons 
since our configurations resemble with fundamental strings 
between D2-branes.\footnote{
We thank Koji Hashimoto for suggesting this problem.
}
Second, the second quantization of solitons 
is more interesting. 
It has been suggested that it is crucial to take account of 
quantum dynamics of solitons in order to see the precise 
parallel of our field theory solitons with the D-branes 
\cite{Tong:2005nf,Shifman:2006ea}. 
Second quantization of 
solitons is an intriguing delicate problem 
which is worth examining.

We have studied the moduli space and dynamics 
of 1/4 BPS composite systems such as domain wall webs 
(networks) \cite{Eto:2005cp}  
and vortex-strings between domain walls 
as in this paper. 
There exists another interesting 1/4 BPS composite system.
In 4+1 dimensions instantons are particle-like solitons,  
and they can lie inside vortex-sheets in the Higgs phase. 
So far instantons inside a straight vortex-plane as a host soliton 
were studied \cite{Hanany:2004ea,Eto:2004rz}. 
Their dynamics is identical to that of sigma model lumps, 
because the instantons can be regarded as 
lumps in the effective theory on the vortex-plane 
which is typically the ${\bf C}P^1$ model. 
Recently more interesting configurations 
of instantons living inside a vortex-network as a host soliton 
has been found \cite{Fujimori:2008ee}. 
In this case the host soliton has a geometry 
of 
Riemann surface 
so the instanton dynamics is more ample and interesting to explore. 
Solitons in different dimensions are connected by duality 
such as T-duality between domain walls and vortices \cite{Eto:2006mz,
Eto:2007aw}. In \cite{Eto:2007aw} it has been used  
to study 
statistical mechanics of vortices. 
This method should be extendible to the present case of 
vortex strings between domain walls. 

\section*{Acknowledgments}
T.F., T.N.~and M.N.~thank Koji Hashimoto for fruitful discussions. 
This work is supported in part by Grant-in-Aid for 
Scientific Research from the Ministry of Education, 
Culture, Sports, Science and Technology, Japan No.17540237
and No.18204024 (N.S.). 
The work of M.E.~and K.O.~is also supported by the Research Fellowships of the Japan Society for
the Promotion of Science for Research Abroad.
The work of T.F.~and T.N.~is 
supported by the Research Fellowships of the Japan Society for
the Promotion of Science for Young Scientists. 
The work of M.N.~is supported in part by Grant-in-Aid for Scientific
Research (No.~20740141) from the Ministry
of Education, Culture, Sports, Science and Technology-Japan.
T.N.~gratefully acknowledges 
support from a 21st Century COE Program at 
Tokyo Tech ``Nanometer-Scale Quantum Physics" by the 
Ministry of Education, Culture, Sports, Science 
and Technology, 
and support from the Iwanami Fujukai Foundation.

\appendix
\section{Evaluation of K\"ahler metric}\label{appendix:A}
In Eq.\,(\ref{eq:metric_222}) and Eq.\,(\ref{eq:metric_020}), we have seen that the K\"ahler metrics take the form
\beq
K_{z_0 \bar z_0} = 2 \pi c \int dx_3 \, k E(k), \hs{10} k \equiv \sqrt{\frac{|a|^2}{|a|^2 + 2 \cosh(mx_3)}},
\eeq
with $a = v$ for Eq.\,(\ref{eq:metric_222}) and $a = vz_0^2$ 
for Eq.\,(\ref{eq:metric_020}). This integral can be evaluated by expanding the integrand in terms of $a$ as 
\beq
k E(k) ~=~ \frac{\pi}{2} \sum_{n=0}^\infty (-1)^n \frac{1}{2n+1} \left( \frac{(2n+1)!!}{(2n)!!} \right)^2 \left( \frac{|a|}{\sqrt{2 \cosh(mx_3)}} \right)^{2n+1}.
\eeq
Then integrating term by term, we obtain the K\"ahler metric as 
\beq
K_{z_0 \bar z_0} &=& \pi^2 c \int_{-\infty}^\infty dx_3 \sum_{n=0}^\infty (-1)^n \frac{1}{2n+1} \left( \frac{(2n+1)!!}{(2n)!!} \right)^2 \left( \frac{|a|}{\sqrt{2 \cosh(mx_3)}} \right)^{2n+1} \notag \\
&=& \frac{\pi^{\frac{5}{2}} c}{\sqrt{2}m} \sum_{n=0}^\infty \left( - \frac{1}{2} \right)^n \frac{1}{2n+1} \left( \frac{(2n+1)!!}{(2n)!!} \right)^2 \frac{\Gamma \left( \frac{1}{4} + \frac{n}{2} \right)}{\Gamma \left( \frac{3}{4} + \frac{n}{2} \right)} |a|^{2n+1} \notag \\
&=& \frac{\pi^{\frac{3}{2}} c}{2m} |a| \bigg[ \Gamma\left(1/4\right)^2 {}_4 F_3 \left( \textstyle \frac{1}{4},\frac{1}{4},\frac{3}{4},\frac{5}{4};\frac{1}{2},\frac{1}{2},1; \frac{|a|^4}{4} \right) \notag \\
&{}& \hs{20} - \frac{3}{2} |a|^2 \Gamma\left(3/4\right)^2 {}_4 F_3 \left(\textstyle \frac{3}{4},\frac{3}{4},\frac{5}{4},\frac{7}{4};1,\frac{3}{2},\frac{3}{2}; \frac{|a|^4}{4} \right) \bigg], 
\eeq
where ${}_4 F_3(a_1,a_2,a_3,a_4;b_1,b_2,b_3;z)$ is the generalized hypergeometric function defined by 
\beq
{}_4 F_3(a_1,a_2,a_3,a_4;b_1,b_2,b_3;z) = \sum_{n=0}^{\infty} \frac{(a_1)_n(a_2)_n(a_3)_n(a_4)_n}{(b_1)_n(b_2)_n(b_3)_n} \frac{z^n}{n!},
\eeq
with $(a)_n \equiv a(a+1)(a+2) \cdots (a+n-1)$. 

\section{Asymptotic K\"ahler metric}\label{appendix:B}
In this section we derive the asymptotic K\"ahler metrics 
Eq.\,(\ref{eq:asym_metric_222}) and Eq.\,(\ref{eq:asym_metric_020}). 
In both cases the moduli matrix takes the form
\beq
H_0 = \left( \varphi_1(z),\, v \, \varphi_2(z,z_0),\, \varphi_3(z) \right), 
\eeq
with 
\beq
\begin{array}{ccc}
\varphi_1=(z-z_1)^2,~~\varphi_2=(z^2-z_0^2),~~\varphi_3=(z+z_1)^2 & ~~\mbox{for}~~ & (k_1,k_2,k_3)=(2,2,2), \\
\varphi_1=1,~~\varphi_2=(z^2-z_0^2),~~\varphi_3=1 & ~~\mbox{for}~~ & (k_1,k_2,k_3)=(0,2,0).
\end{array}
\eeq
the K\"ahler potential Eq.\,(\ref{eq:kahler-pot}) in the strong coupling limit is given by
\beq
K ~=~ c \int d^3 x \, \mathcal K ~=~ c \int d^3 x \, \log \left( |\varphi_1|^2 e^{mx_3} + |v|^2 |\varphi_2|^2 + |\varphi_3|^2 e^{-mx_3} \right).
\eeq
For $x_3 > x_0 \equiv \frac{1}{2m} \log \left|\varphi_3/\varphi_1\right|^2$, the integrand can be expanded as
\beq
\mathcal K ~=~ \log \left( |\varphi_1|^2 e^{mx_3} + |v|^2 |\varphi_2|^2 \right) - \sum_{n=1}^\infty \frac{1}{n} \left( - \frac{|\varphi_3|^2 e^{-mx_3}}{|\varphi_1|^2 e^{mx_3} + |v|^2 |\varphi_2|^2} \right)^n,
\eeq
and for $x_3 < x_0$ it can be expanded as 
\beq
\mathcal K ~=~ \log \left( |v|^2 |\varphi_2|^2 + |\varphi_3|^2 e^{-mx_3}\right) - \sum_{n=1}^\infty \frac{1}{n} \left( - \frac{|\varphi_1|^2 e^{mx_3}}{|v|^2 |\varphi_2|^2 + |\varphi_3|^2 e^{-mx_3}} \right)^n,
\eeq
We can show that the contributions to the metric from the terms in the infinite series vanish in the limit $v \rightarrow \infty$. 
Therefore the asymptotic K\"ahler metric is given by 
\beq
K \approx c \int d^2 z \left[ \int_{-\infty}^{x_0} dx_3 
\log \left(|v|^2 |\varphi_2 |^2 + |\varphi_3|^2 e^{-mx_3} \right) 
+ \int_{x_0}^\infty dx_3 
\log \left(|\varphi_1|^2 e^{mx_3} + |v|^2 |\varphi_2|^2 \right) \right], 
\eeq
and the asymptotic K\"ahler metric can be written as 
\beq
K_{z_0 \bar z_0} &=& \p_{z_0} \p_{\bar z_0} K \phantom{\bigg[} \notag \\
&\approx& c \int d^2 z \, |v|^2 |\p_{z_0} \varphi_2|^2 
\left[ \int_{-\infty}^{x_0} dx_3 
\frac{|\varphi_3|^2 e^{mx_3}}
{(|\varphi_3|^2 + |v|^2 |\varphi_2|^2 e^{mx_3})^2} 
+ \int_{x_0}^\infty dx_3 \frac{|\varphi_1|^2 e^{-mx_3}}
{(|\varphi_1|^2 + |v|^2 |\varphi_2|^2 e^{-mx_3})^2} \right] \notag \\
&=& \frac{2c}{m} \int d^2 z \, 
\frac{|v|^2 |\p_{z_0} \varphi_2|^2}
{|v|^2 |\varphi_2|^2 + |\varphi_3||\varphi_1|}. 
\label{eq:asymptotic}
\eeq
First, let us consider the case of $(k_1,k_2,k_3) = (2,2,2)$. By using Eq.\,(\ref{eq:asymptotic}) the asymptotic metric can be calculated as 
\beq
K_{z_0 \bar z_0} &=& \frac{2c}{m} \int d^2 z 
\frac{4|v|^2 |z_0|^2}{|v|^2|z^2-z_0^2|^2+|z^2-z_1^2|^2} \notag \\
&=& \frac{2\pi c}{m} \left| \frac{vz_0^2}{z_0^2-z_1^2} \right| 
\int_0^{2\pi} d\theta 
\frac{1}
{\sqrt{1+\frac{||v|^2z_0^2+z_1^2|^2}{|v|^2|z_0^2-z_1^2|^2}\sin^2(2\theta)}}.
\eeq
If we assume that $|v|\gg|z_1|/|z_0|$, the K\"ahler metric becomes 
\beq
K_{z_0 \bar z_0} &\approx& \frac{2\pi c}{m} 
\left| \frac{vz_0^2}{z_0^2-z_1^2} \right| 
\int_0^{2\pi} d\theta 
\frac{1}{\sqrt{ 1 + 
\left| \frac{vz_0^2}{z_0^2-z_1^2} \right|^2 \sin^2(2\theta)}} \notag \\
&=& \frac{8\pi c}{m} \left| \frac{vz_0^2}{z_0^2-z_1^2} \right| 
K \left( i \left| \frac{vz_0^2}{z_0^2-z_1^2} \right| \right), 
\eeq
where the complete elliptic integral of the first kind $K(k)$ 
is defined by 
\beq
K(k) \equiv \int_0^{\frac{\pi}{2}} d \theta \frac{1}{\sqrt{1-k^2 \sin^2 \theta}}.
\eeq
By using the asymptotic form of the complete elliptic integral 
\beq
k K(ik) ~\rightarrow~ \log(4k), \hs{10} k \gg 1,~k \in \mathbf R,
\label{eq:asy-K}
\eeq
we obtain the asymptotic K\"ahler metric as
\beq
K_{z_0 \bar z_0} ~\approx~ \frac{8\pi c}{m} \log \left| \frac{4vz_0^2}{z_0^2-z_1^2} \right|.
\eeq
Next, let us consider the case of $(k_1,k_2,k_3) = (0,2,0)$. By using Eq.\,(\ref{eq:asymptotic}) the asymptotic metric can be calculated as 
\beq
K_{z_0 \bar z_0} ~\approx~ \frac{2c}{m} \int d^2 x \frac{4|v|^2 |z_0|^2}{|v|^2 |z^2-z_0^2|^2 + 1} ~=~ \frac{8 \pi c}{m} |vz_0^2| K(i|vz_0^2|) ~\approx~ \frac{8 \pi c}{m} \log |4 v z_0^2|.
\eeq
Here we have used the asymptotic relation Eq.\,(\ref{eq:asy-K}). 

\section{K\"ahler potential for the asymptotic metric}\label{appendix:C}
In Sec.\,3 we showed that the moduli space of 1/4 BPS configurations is a K\"ahler manifold. In this section, we check that 
the K\"ahler condition is satisfied for the asymptotic metric obtained from the charged particle analysis in Sec.\,5. From Eq.\,(\ref{eq:charge_lag}), we can read the asymptotic metric as
\beq
g_{(A,i) \overline{(A,i)}} &=& \frac{\langle \Phi^{A-1} \rangle - \langle \Phi^A \rangle}{2} + \frac{1}{2} \sum_{(B,j) \not = (A,i)} C_{AB} \log \left|\frac{z_{\left<A\right>i}-z_{\left<B\right>j}}{L}\right|, \label{eq:metric1} \\
g_{(A,i) \overline{(B,j)}} &=& - \frac{1}{2} C_{AB} 
\log \left|\frac{z_{\left<A\right>i}-z_{\left<B\right>j}}{L}\right|, \hs{10} (A,i) \not = (B,j). 
\label{eq:metric2}
\eeq
This metric can be obtained from the following K\"ahler potential
\beq
K = \frac{\langle \Phi^{A-1} \rangle - \langle \Phi^A \rangle}{2} |z_{\langle A \rangle i}|^2 + \underset{(A,i),\,(B,j)}{{\sum}'} \frac{C_{AB}}{2} \left(\log \left|\frac{z_{\left<A\right>i}-z_{\left<B\right>j}}{L} \right| - 1 \right) \left|z_{\left<A\right>i}-z_{\left<B\right>j} \right|^2, 
\eeq
where $\sum'$ means that the sum is taken 
only once for a pair of the index sets 
$(A,i)$ and $(B,j)$ such that $(A,i) \not = (B,j)$. 
The existence of the K\"ahler potential means that the asymptotic metric Eq.\,(\ref{eq:metric2}) obtained in the charged particle analysis is a K\"ahler metric. 
The normalizable part of the moduli space, which is free from the divergence in $L \rightarrow \infty$ limit, is a subspace defined by 
\beq
&z_{\left<1\right>i} = {\rm const}.~~(i=1,\cdots,k_1), \hs{5} z_{\left<N\right>i} = {\rm const}.~~(i=1,\cdots,k_N),& \notag \\ 
&\displaystyle \sum_{i=1}^{k_A} z_{\left<A\right>i} = {\rm const}.~~(A=2,\cdots,N-1).&
\label{eq:constraint}
\eeq
The metric on this complex submanifold is given by the induced metric of (\ref{eq:metric1}) and (\ref{eq:metric2}).
The pullback of the K\"ahler form
\beq
\omega ~\equiv~ i \p \bar \p K ~=~ i \sum_{(A,i),(B,j)} g_{(A,i) \overline{(B,j)}}~ d z_{\left<A\right>i} \wedge d \bar z_{\left<B\right>j}.
\label{eq:kahlerform}
\eeq 
onto the subspace Eq.\,(\ref{eq:constraint}) gives a closed two form $\omega^\ast$. This is because the K\"ahler form Eq.\,(\ref{eq:kahlerform}) is a closed two form and the exterior derivative commutes with pullback. 
Therefore the submanifold Eq.\,(\ref{eq:constraint}) is also a K\"ahler manifold endowed with the K\"ahler form $\omega^\ast$, which is finite in the infinite cutoff limit $L \rightarrow \infty$.
\section{Dual effective theory on multiple domain walls}\label{appendix:D}
Effective theory on $N (\ge 2)$ domain walls is described by the positions of $N$ domain walls $X^A=\tilde X^A/\Delta m_A$ and the associated phases $\sigma^A$ $(A=1,2,\cdots,N)$ as
\beq
{\cal L}_{\rm w}=\frac{1}{2} {\cal G}_{AB} 
\left(\p_\mu \tilde X^A \p^\mu \tilde X^B +
\p_\mu \sigma^A \p^\mu \sigma^B \right).
\label{eq:multi_dw}
\eeq
Here ${\cal G}_{AB}$ is the K\"ahler metric on the domain wall moduli space, 
and depends only on $\tilde X^A$
\beq
{\cal G}_{AB}(\tilde X) = \frac{1}{2} \frac{\p^2 K_{\rm w}}{\p\tilde X^A \p \tilde X^B}.
\eeq
When all the domain walls are well-separated $X^1 \gg X^2 \gg \cdots \gg X^N$, the metric of the domain wall moduli space is a flat metric
\beq
{\cal G}_{AB} \simeq \frac{c}{\Delta m_A} \delta_{AB}.
\label{eq:defPhi}
\eeq
We would like to obtain the dual Lagrangian by dualizing
the periodic
 scalar fields $\sigma^A$.
First let us define scalar fields $\Phi^A~(A=1,2,\cdots,N)$ by
\beq
\Phi_A \equiv \pi \frac{\p K_{\rm w}}{\p \tilde X^A}.
\eeq
The derivative of $\Phi_A$ with respect to $\tilde X^B$
gives the metric on the domain wall moduli space as
\beq
{\cal G}_{AB} = \frac{1}{2\pi} \frac{\p \Phi_A}{\p \tilde X^B} = \frac{1}{2\pi} \frac{\p \Phi_B}{\p \tilde X^A}.
\eeq
Since we can assume the existence of an inverse of the metric
\beq
{\cal G}^{AC}{\cal G}_{CB} = \delta^A_{~B},\qquad
{\cal G}^{AB} = 2\pi \frac{\p \tilde X^A}{\p \Phi_B} = 2\pi \frac{\p \tilde X^B}{\p \Phi_A},
\eeq
the set of $\Phi_A (A=1,2,\cdots,N)$ can be interpreted as
a new coordinates on the domain wall moduli space.
Note that the definition of $\Phi_A$ is not invariant 
under the K\"ahler transformation
\beq
K_{\rm w}(\phi, \bar \phi) \rightarrow K_{\rm w}(\phi,\bar \phi) + f(\phi) + \overline{f(\phi)}.
\eeq
However, we can always fix the definition of $\Phi_A$ so that the asymptotic values of $\Phi_A$ take the form
\beq
\Phi_A \simeq 2 \pi c X^A,
\label{eq:coord_equiv}
\eeq
when all the domain walls are well-separated. 
Next let us define one form fields $\tilde F^A_{\mu}~(A=1,2,\cdots,N)$ by
\beq
\tilde F^A_{\mu} \equiv \p_\mu \sigma^A,
\eeq
and interpret them as new dynamical fields 
obeying the Bianchi identity $ \epsilon^{\mu \nu \rho} \p_\nu \tilde F^A_{\rho} = 0$.
In order to rewrite the Lagrangian in terms of
$\tilde F^A_{\mu}$, we have to add a term with Lagrange multipliers $A_{A\mu}$
\beq
{\cal L}_F \propto \epsilon^{\mu \nu \rho}
A_{A\mu} \p_\nu \tilde F^A_{\rho}.
\eeq
Then, if we eliminate $\tilde F^A_{\mu}$ using the equation of motion, 
we obtain $U(1)^N$ gauge theory with 
$N$ neutral scalar fields
\beq
\tilde{\cal L}_{\rm w} = \frac{1}{4\pi^2} {\cal G}^{AB} \left(
\frac{1}{2}\p_\mu \Phi_A \p^\mu \Phi_B-\frac{1}{4}
F_{A\mu \nu}F_B^{\mu \nu} \right),
\eeq
where $F_{A \mu \nu}=\p_\mu A_{A\nu}-\p_\nu A_{A\mu}$.
When all the domain walls are well-separated, the effective Lagrangian is
simply given by
\beq
\tilde{\cal L}_{\rm w} \simeq \sum_{A=1}^N\left(
\frac{1}{2e_A^2}\p_\mu \Phi_A \p^\mu \Phi_A-\frac{1}{4e_A^2}
F_{A\mu \nu}F_A^{\mu \nu} \right), \qquad
e_A^2 \equiv \frac{4 \pi^2 c}{\Delta m_A}.
\eeq

The new scalar fields $\Phi_A$ have also an interesting physical meaning.
We have assumed that scalar fields $X^A$ represent positions of domain walls.
However, it is not 
precisely 
correct when 
a domain 
wall approaches to another domain 
wall.
Let us focus on $(A-1)$-th and $A$-th domain walls, and define
their center of mass $X_0$ and the relative distance $R_A$ by
\beq
\Delta m_{A-1} X^{A-1} = \Delta m_{A-1} X_0 + \frac{\mu_A}{2} R_A, \hs{10}
\Delta m_A X^A = \Delta m_A X_0 - \frac{\mu_A}{2} R_A,
\eeq
where $\mu_A$ is defined by
\beq
\mu_A \equiv \frac{2 \Delta m_{A-1} \Delta m_A}{\Delta m_{A-1} + \Delta m_A}.
\eeq
The relative distance 
$R_A$ can be negative, which does not mean the interchange of domain walls but the compression of two walls, namely they become a single wall in the limit of $R_A \to -\infty$. 
Therefore the parameter $R_A$ loses its meaning as relative distance when the distance between the walls becomes small.
An interesting property of the new coordinates $\Phi_A$
is that their differences are bounded from below by boojum charges
defined in Eq.(\ref{eq:boojum})
\beq
\frac{2 \pi}{\mu_A}\frac{\p K}{\p R_A}~=~\Phi_{A-1}-\Phi_A
~=~ |B_g^{A-1}|+|B_g^A| + \O\left(e^{\mu_A R_A} \right), \qquad R_A \ll - \mu_A.
\label{eq:position_bound}
\eeq
Since $|B^A_g|/\pi c = 2 \Delta m_A / g^2 c$ is equal to the width of $A$-th domain wall,
$(|B^{A-1}_g|+|B^A_g|)/2 \pi c$ can be interpreted as the lower bound
of distance between 
the middle points of 
$(A-1)$-th and $A$-th domain walls.
Therefore, $\Phi_A/2 \pi c$ instead of $X^A$
represents the correct position of $A$-th domain wall
since it has the correct lower bound Eq.(\ref{eq:position_bound}) and asymptotically coincides with $X^A$ (see Fig.\,\ref{fig:Phi}). 
\begin{figure}[t]
\begin{center}
\includegraphics[width=80mm]{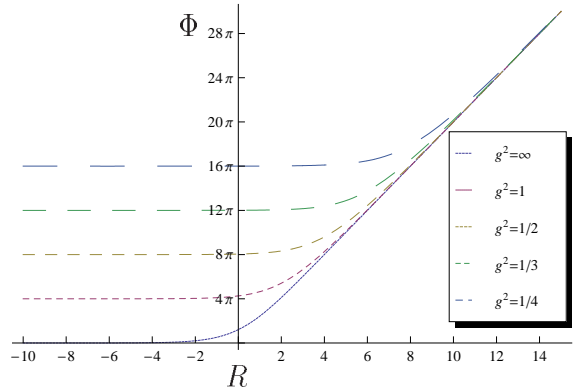}
\caption{
The profiles of $\Phi \equiv \Phi_1 - \Phi_2$ as the function of $R \equiv X^1 - X^2$ obtained by numerical calculations for $\NF=3$, $c=1$, $m_1=1,\,m_2=0,\,m_3=-1$. The function $\Phi$ approaches to the constant value $|B^1_g| + |B^2_g| = 4 \pi \Delta m/g^2$ as the parameter $R$ becomes small.
}
\label{fig:Phi}
\end{center}
\end{figure}

\end{document}